%% file: frahoui_norma_ir_rev2.tex
\documentclass{aa}
\usepackage{epsfig}
\usepackage{graphics}
\usepackage[varg]{txfonts}
\usepackage{natbib,twoopt}
\usepackage[breaklinks=true]{hyperref} 
\usepackage{rotating}

\begin{document}

\newcommand{\intl}{\textit{INTEGRAL}}
\newcommand{\rxte}{\textit{RXTE}}
\newcommand{\xmm}{\textit{XMM-Newton}}
\newcommand{\chandra}{\textit{Chandra}}
\newcommand{\spitzer}{\textit{Spitzer}}
\newcommand\iontiny[2]{#1$\;${\tiny \scshape{#2}}}

\def\nh{$N_{\rm H}$}
\def\chis{$\chi^2$}
\def\chisr{$\chi_{\rm r}^2$}
\def\msun{$M_{\odot}$}
\def\rsun{$R_{\odot}$}
\def\mdot{$\dot{M}$}
\def\lsun{$L_{\odot}$ }
\def\msunyr{$M_{\odot}$~yr$^{-1}$}
\def\ergs{erg~s$^{-1}$}
\def\ergcms{erg~cm$^{-2}$~s$^{-1}$}
\def\kms{km~s$^{-1}$}
\def\acm2{atoms~cm$^{-2}$}
\def\wm2{W~m$^{-2}$}
\def\mic{$\mu$m}
\def\cm2{cm$^{-2}$}
\def\se1{s$^{-1}$}
\def\nhe{N_{\rm H}}
\def\Av{$A_{\rm V}$}
\def\Al{$A_{\rm \lambda}$}
\def\Ave{A_{\rm V}}
\def\Ale{A_{\rm \lambda}}
\def\Ak{$A_{\rm K}$}
\def\Aj{$A_{\rm J}$}
\def\Ake{A_{\rm K}}
\def\Aje{A_{\rm J}}
\def\Ahi{$A_{\textrm{\ion{H}{i}}}$}
\def\Ahd{$A_{\rm H_2}$}
\def\Ax{$A_{\rm x}$}
\def\Ahie{A_{\textrm{\ion{H}{i}}}}
\def\Ahde{A_{\rm H_2}}
\def\Axe{A_{\rm x}}

\title{Near-infrared spectroscopy of 20 new \chandra\ sources in the Norma Arm}

\author{F.~Rahoui\inst{\ref{inst1},\ref{inst2}}
\and
J.~A.~Tomsick\inst{\ref{inst3}}
\and
F.~M.~Fornasini\inst{\ref{inst3},\ref{inst4}}
\and
A.~Bodaghee\inst{\ref{inst3},\ref{inst5}}
\and
F.~E.~Bauer\inst{\ref{inst6},\ref{inst7}, \ref{inst8}}}

\institute{European Southern Observatory,
  Karl-Schwarzschild-Str. 2, D-85748 Garching bei M\"unchen, Germany~\email{frahoui@eso.org}\label{inst1}
  \and
 Harvard University, Department of Astronomy, 60 Garden street, Cambridge, MA 02138, USA\label{inst2}
  \and
  Space Sciences Laboratory, 7 Gauss Way, University of California, Berkeley, CA, 94720-7450, USA\label{inst3}
  \and
  Astronomy Department, University of California, 601 Campbell Hall, Berkeley, CA 94720, USA\label{inst4}
  \and
  Department of Chemistry, Physics, and Astronomy, Georgia College \& State University CBX 82, Milledgeville, GA 31061, USA\label{inst5}
\and
   Instituto de Astrof\'isica, Facultad de F\'isica, Pontifica Universidad Cat\'olica de Chile, 306, Santiago 22, Chile\label{inst6}
   \and
   Millennium Institute of Astrophysics\label{inst7}
   \and
   Space Science Institute, 4750 Walnut Street, Suite 205, Boulder, CO 80301, USA\label{inst8}
  }

\date{Received / Accepted}

\abstract{
We report on CTIO/NEWFIRM and CTIO/OSIRIS photometric and
spectroscopic observations of 20 new 
X-ray (0.5--10~keV) emitters discovered in
the Norma Arm Region Chandra Survey (NARCS). NEWFIRM photometry was
obtained to pinpoint the near-infrared counterparts of NARCS sources,
while OSIRIS spectroscopy was used to help identify 20 sources
with possible high mass X-ray binary properties. We find that (1) two
sources are WN8 Wolf-Rayet stars, maybe in colliding wind binaries, part
of the massive star cluster Mercer 81; (2) two are emission-line
stars, possibly in X-ray binaries, that exhibit near- and mid-infrared 
excesses either due to free-free emission from the decretion discs of
Be stars or warm dust in the stellar winds of peculiar massive stars
such as B[e] supergiants or luminous blue variables; (3) one is a
B8-A3 IV-V star that could be in a quiescent high mass X-ray binary
system; (4) two are cataclysmic variables including one intermediate
polar; (5) three may be neutron star symbiotic binaries; (6) five
are most likely white dwarf symbiotic binaries; and (7) five 
  exhibit properties more consistent with isolated giant/dwarf
stars. The possible detection of one to three high mass X-ray binaries
is in good agreement with our predictions. However, our study
illustrates the difficulty of clearly differentiating quiescent or
intermediate X-ray luminosity systems from isolated massive stars,
which may lead to an underestimation of the number of known high mass
X-ray binaries.}



\maketitle

\section{Introduction}

The arms of spiral galaxies have long been known to be regions of
intense stellar formation, a consequence of the formation of
the galaxies themselves. For instance, \citet{1964Lin} and
\citet{1969Lin} proposed, in the framework of the density wave theory,
that spiral arms are formed through the propagation of linear
discontinuities of the average Galactic gravitational field, which
enhances their local density. When fast-rotating clouds of gas and
dust encounter a slower density wave, their intrinsic density may
therefore increase such that they meet the Jeans criterion and collapse,
triggering star formation. Even though this explanation is still
controversial, it is at least observationally clear that giant
molecular clouds (GMCs) and HII regions are preferentially located in
the  arms of the Milky Way \citep{1987Caswell, 2001Dame}, where the 
number of short-lived O/B massive stars is consequently very
high. Interestingly, \citet{2012Sana} showed that at least
70\% of massive stars interact with a nearby massive companion or, in 
other words, that most O/B stars are found in binary systems. Even if
the authors argue that a third of them will end up in binary
mergers, a significant population of high mass X-ray binaries (HMXBs),
consisting of a neutron star (NS) or a black hole (BH) accreting
from a massive O/B star, must therefore be distributed in the spiral
arms of the Milky Way, a statement supported by two recent studies
reported in \citet{2012Bodaghee} and \citet{2013Coleiro}. 

The last decade has seen a renewed scientific interest in HMXBs, 
following the launch of the \textit{INTErnational Gamma-Ray
  Astrophysics Laboratory} \citep[\intl,][]{2003Winkler} in October
2002. Indeed, \intl\  performed a 20-100~keV survey of the Galactic
plane which led to the detection of more than 700 hard X-ray sources
\citep{2010Bird}, including about 400 new or previously poorly studied
\intl\ Gamma-Ray (IGR) sources. The subsequent multiwavelength follow-ups
\citep[see e.g.][]{2008Tomsick, 2008Chaty, 2008Rahouia,
  2008Rahouib, 2012Tomsick, 2013Coleirob} showed that a significant
fraction of these new sources were actually HMXBs in which a NS fed on
the stellar winds of a blue supergiant through
Bondi-Hoyle processes, known as NS-supergiant X-ray binaries
(NS-SGXBs). Not only did these results increase the fraction of NS-SGXBs
from 10\% of all known HMXBs to about 40\% today, but these new
NS-SGXBs also appear to exhibit extreme behaviours that were not
predicted. Indeed, some are very obscured persistent sources suffering
from a huge intrinsic extinction, with $N_{\rm H}$ as high as
$10^{24}$~\acm2\ \citep[such as IGR~J16318$-$4848,][]{2003Matt},
and others are transient sources, completely undetectable in
quiescence but randomly exhibiting X-ray flares that can be as bright
as the Crab nebula for time periods as short as a few hours; these are
the so-called supergiant fast X-ray transients
\citep[SFXTs;][]{2006Negueruelaa}.

Although the origins of such properties are still a matter of debate
\citep[accretion from clumps of material in the stellar wind or magnetic
barrier, see e.g.][]{2007Walter, 
  2008Bozzo}, these NS-SGXBs have opened up a new window on HMXB
evolution and, as progenitors of NS-NS and NS-BH binaries, the
formation of short gamma-ray bursts. Discovering and 
studying as many of them as possible is therefore crucial and the Norma
arm, as the most active star-forming region in the Milky Way, is the best
candidate for this purpose. In 2011, we therefore performed a
\chandra\ ACIS survey of a $2^{\circ}\times$0\fdg8 area of the
Norma arm to detect new $0.5-10$~keV sources with hard X-ray
spectra, in particular low luminosity HMXBs fainter than the
$10^{34}$~\ergs\ minimum level usually encountered \citep[see][for a
recent review]{2011Reig}. The results of the NARCS survey are
comprehensively presented in \citet{2014Fornasini} and we refer to
that work for more details. In this paper, we report on the
near-infrared (near-IR) photometric and spectroscopic observations
which were carried out to pinpoint the near-IR counterparts of 
the newly discovered hard X-ray sources and identify those we
believed to be HMXB candidates. Section~2 describes the data and the
reduction process. The results are presented in Section~3. We discuss
them and conclude in Section~4.
\input{./Figures/fc_ir.figures}
\input{./Figures/quant.figures}

\section{Observations}

To optimize the detection of new HMXBs, our \chandra\ survey of the
Norma arm was complemented with a near-IR
photometric mapping of the same region, followed by near-IR
spectroscopy of the most promising counterparts. The latter were
selected and ranked on the basis of their fulfillment of six criteria:
(1) X-ray brightness high enough to place good constraints on the
source's X-ray spectral parameters; (2) a high column density ($\nhe >
10^{22}$~\acm2) to exclude foreground sources; (3) a hard X-ray
spectrum with a photon index $\Gamma<2$, which is typical for HMXBs;
(4) X-ray variability, which is often seen in HMXBs; (5) a counterpart
with J-K colour $> 1.5$, indicating that it is also subject to
significant absorption; and (6) the reliability of the near-IR
counterpart.  

Allocating 0, 0.5, or 1 when a criterium was not met, partially met, or
fully met, respectively, the number of criteria fulfilled ranged from
1.5 to 6 for the selected sources. The column density and X-ray photon
index of the sources were measured through X-ray spectral fitting of
the \chandra\ spectra (averaged when a source was present in several
pointings) with {\tt XSPEC} \citep{1996Arnaud}. We principally made
use of the absorbed power-law model {\sc tbabs$\times$pegpwlw}
although, when a spectrum was found to be very soft ($\Gamma \ge 4$),
a thermal plasma model {\sc tbabs$\times$vapec} was also used to obtain
a reliable estimate of the flux. Furthermore, to investigate the
  presence of an intrinsic component of absorption, we also measured
  the total Galactic column density along the line of sight (LOS) of
  each source using the relation
  $N_{\textrm{\ion{H}{i}}+\textrm{H}_{\rm 2}}{\rm (ISM)}$=$N_{\rm
    \ion{H}{i}}$(ISM)+2$N_{\textrm{H}_{\rm 2}}$(ISM). We assessed the
  neutral hydrogen column density $N_{\rm \ion{H}{i}}$(ISM) from the
  Leiden/Argentine/Bonn Survey \citep {2005Kalberla}. In addition, the
  molecular hydrogen column density $N_{\textrm{H}_{\rm 2}}$(ISM) was 
  estimated from the MWA CO survey \citep{1989Bronfman} using the
  $N_{\textrm{H}_{\rm 2}}/I_{\rm CO}$ factor derived in
  \citet{2001Dame}. 

Finally, the reliabilities $R$ of the near-IR
  counterparts of the \chandra\ sources were estimated following the
  analytic treatment introduced in \citet{1992Sutherland}, which is
  based on the positional uncertainties of the X-ray and near-IR
  positions, the distances between the X-ray and near-IR counterparts,
  and the density of near-IR sources in the vicinity of the
  X-ray positions. In this framework, $R$ is the probability for
  a near-IR source of being the real counterpart of an X-ray
  source. Summing up all the $1-R$ values, we therefore expect about
  0.4 spurious X-ray/near-IR associations in our sample. 
 
\subsection{Near-infrared photometry}
\input{./Tables/counter_mag.table}
\input{./Tables/irmag.table}
\input{./Tables/irmag2.table}
\input{./Tables/irmag3.table}
We performed the observations on 2011 July 19 with the NOAO Extremely
Wide Field Infrared Mosaic (NEWFIRM) mounted on the 4~m Blanco
telescope at the Cerro Tololo Inter-America Observatory (CTIO). Each
NEWFIRM image is divided into four extensions and covers a
28\arcmin$\times$28\arcmin\ field of view with a 0\farcs4 plate
scale; this allowed us to observe about 60\% of the
$2^{\circ}\times$0\fdg8 Norma region mapped with \textit{Chandra}
in the $J$, $H$, and $Ks$ bands. For each pointing, the exposure time
in each filter was set to 150~s divided into ten dithered four-extension
frames for median sky construction. Conditions were clear and the
seeing was less than 1\farcs2 during the whole night.

We reduced the data with the dedicated IRAF \citep{1993Tody} package
\textsc{nfextern}. It is composed of several routines that perform
the common tasks for near-IR imaging, tailored for wide-field
mosaics. The first-pass reduction consisted in bad pixel removal, dark
subtraction, linearity correction, flatfielding, and median sky
subtraction. The astrometry of each cleaned four-extension frame was
calibrated to the World Coordinate System (WCS) by deriving the
astrometric transformations through comparison with the 2~Micron All
Sky Survey catalogue \citep[2MASS,][]{2006Skrutskie}. The four
extensions were then combined into a single image
before median stacking. We finally flux-calibrated all the images by
deriving the photometric solutions (zero-point magnitudes, atmospheric
extinction, and colour terms) through relative photometry using the
2MASS catalogue as a reference. While not as robust as the use of
photometric standard stars, we estimate that the measured magnitudes
are accurate to about 0.1~mag in each filter, although statistical
errors are lower.

As already mentioned, our NEWFIRM observations covered roughly 60\% of
the \chandra\ field. We completed and complemented the near-IR
survey using the Vista Variables in the Via Lactae
\citep[VVV,][]{2010Minniti}, 2MASS \citep{2006Skrutskie}, and Deep
Near Infrared Survey of the Southern Sky \citep[DENIS,][]{1999Epchtein}
database\footnote{http://irsa.ipac.caltech.edu/applications/BabyGator/
  (2MASS and DENIS) and
  http://horus.roe.ac.uk:8080/vdfs/Vregion\_form.jsp (VVV)} for the
sources we could not observe or which were saturated.
\input{./Tables/lines_hm_new.table}

\subsection{Near-infrared spectroscopy}

Based on the aforementioned criteria, we selected 45 sources and we
observed most of them from 2012 June 1 to 4 with the Ohio State
Infrared Imager/Spectrometer (OSIRIS) mounted on the 4m~SOAR telescope
at CTIO. We made use of the cross-dispersed mode with a slit of
1\arcsec\ in width, which allowed us to obtain simultaneous $J$, $H$, and $K$-band
spectra with an R$\,\,\sim1200$ spectral resolution using the standard
ABBA nodding technique for background subtraction. Each night,
standard stars were observed in similar conditions for telluric
absorption lines removal. We reduced the data with the IRAF routines
of the \textsc{echelle} package. The basic steps consisted in bad
pixel correction, dark subtraction, linearity correction,
flatfielding, sky subtraction, and spectral extraction along the
dispersion axis. The spectra were then wavelength-calibrated through
comparison with those of an Argon lamp taken before and after each
exposure.
\input{./Figures/sp_ir_hmxb_em.figures}
\input{./Figures/sed_4_16.figures}

Weather and atmospheric conditions were unfortunately not
  optimal, with thin clouds most of the nights and a seeing higher
  than 1\farcs3. Thus, the S/N of all the spectra are lower than we
  expected. Consequently, the $J$-band spectra we obtained are too
  noisy, and we do not use them in this study. Furthermore, the faintest
  sources in our sample were hard to observe in the $H$ and
  $K$ bands. This is the reason why we only report on 20 sources
  (see Figure~\ref{irfc} for their $Ks$-band counterparts); near-IR
  spectroscopy of the rest of our sample is on-going and will be
  presented in another paper.

\section{Results and analysis}

To strengthen the identification of the X-ray emitters, we made use of
the quantile analysis techniques developed in \citet{2004Hong}, and we
refer to \citet{2014Fornasini} for a complete description. Roughly,
it allows the derivation of two energy-binning
independent quantities $Q_{\rm x}$ (which measures the hardness of a
spectrum) and $Q_{\rm y}$ (which depends on how broad or narrow a spectrum
is). By placing a source in a $Q_{\rm x}$ vs. $Q_{\rm y}$ diagram,
it is therefore possible to get a rough measurement of its X-ray
spectral characteristics. The quantile diagram for
the 20 sources in our sample is displayed in
Figure~\ref{quant}. We divided it into five groups labelled from A to
E. Again, comprehensive information can be found in \citet{2014Fornasini}, 
but we can summarize their characteristics as follows: 

$\bullet$ {\it group A}: low column density sources, most
likely X-ray active low mass stars and interacting binaries. 

$\bullet$ {\it group B}: similar to group A with a significant
population of cataclysmic variables (CVs). Owing to their low
  column densities, sources in groups A and B are likely foreground
  objects and their unabsorbed luminosities will thus be
  estimated for a typical 1~kpc distance, when unknown.

$\bullet$ {\it group C}: intermediate column density consistent with a
3-5~kpc distance. Likely populated with CVs, LMXBs, and HMXBs. A
5~kpc distance is used for unabsorbed luminosity calculation.

$\bullet$ {\it group D}: high column density sources, probably located
in the Norma arm or beyond. Hard spectra typical of intermediate
polars (IPs) or HMXBs. However, some of the sources without
  near-IR counterparts may be Type II active galactic nuclei (AGN).
 
$\bullet$ {\it group E}: highly absorbed sources with softer spectra
than group D, likely containing a significant number of isolated
high mass stars and colliding wind binaries (CWBs). Sources without
near-IR counterparts are likely Type I AGN. The unabsorbed
luminosities of sources in Groups D and E are assessed for a  10~kpc
distance, typical of the far Norma arm.
\newline
\input{./Figures/sp_ir_hmxb_em2.figures}

Table~\ref{counter} lists the X-ray characteristics of the 20
  sources (X-ray coordinates, quantile groups, X-ray and
total ISM column densities \nh\ and $N_{\rm \ion{H}{i}+H_2}$,
photon indices $\Gamma$ or plasma temperatures $kT$, unabsorbed
0.5-10~keV luminosities) as well as the reliability of 
the near-IR counterparts. The NEWFIRM and VVV near-IR
counterparts are listed in Table~\ref{irmag} and those from 2MASS and
DENIS are listed in Table~\ref{irmag2}. We note that all the
magnitudes are given in their original photometric systems. Finally,
we also retrieved the GLIMPSE \citep{2003Benjamin, 2009Churchwell} and
{\it WISE} \citep{2010Wright} mid-infrared (mid-IR) counterparts of
most of the 20 sources (see Table~\ref{irmag3}).  

We identified the companion stars of the \chandra\ sources from the
near-IR spectra, relying on several near-IR spectroscopic atlases to
isolate important emission and/or absorption features. We specifically
made use of \citet{1996Wallace, 1997Wallace}, \citet{1998Meyer},
\citet{2000ForsterS}, \citet{2004Ivanov}, and \citet{2009Rayner} for
cool stars, as well as \citet{1996Hanson, 1998Hanson, 2005Hanson},
\citet{1996Morris}, \citet{1997Blum}, \citet{2000Clark},
\citet{2001Steele}, and \citet{2005Repolust} for hot stars.

The 20 sources in our sample are divided into two near-IR
spectral groups, six with spectra dominated by hydrogen-like species,
in emission for all but one (spectra are displayed in Figures~\ref{irspem},
\ref{irspem2}, \ref{irsp1179}, and \ref{irsp747} and features are listed in
Table~\ref{hmlines}), and 14 with spectra typical of cool
stars (spectra are displayed in Figures~\ref{irsp}, \ref{irsp2},
\ref{irsp3}, \ref{irsp4}, and \ref{irsp5} and the important features
are listed Table~\ref{lmlines}).

\subsection{Sources with near-IR spectra dominated by hydrogen-like
  species} 

\subsubsection{Sources 239 and 1326}

Sources 239 and 1326 were detected as relatively bright X-ray
point sources with \chandra. They both suffer from a high column
density \nh$\sim1.0-1.5\times10^{23}$~\acm2, possibly slightly in
excess of the total galactic column density along their LOS
$N_{\textrm{\ion{H}{i}}+\textrm{H}_{\rm 2}}{\rm
  (ISM)}\sim(7-9)\times10^{22}$~\acm2,  but uncertainties are 
high. Furthermore, both are located at almost the same spot in the
quantile diagram (Group D, Figure~\ref{quant}), a hint that they
belong to the same subclass of X-ray emitters. Nonetheless, source 239 is
likely constant (30\% probability) and source 1326 is likely not (0.02\%
probability of being constant), while source 1326 may also be harder
with $\Gamma=1.2_{-2.0}^{+2.7}$ compared to $\Gamma=2.7_{-1.5}^{+2.0}$
for source 239. 

In the near-IR domain, both sources are constant. Alternative
$J$, $H$, and $Ks$ magnitudes are $15.65\pm0.01$, $13.48\pm0.01$, and
$12.16\pm0.01$ for source 239 (from VVV), and $14.55\pm0.04$,
$12.73\pm0.04$, and $11.60\pm0.04$ for source 1326 (from
2MASS). Their near-IR spectra are also almost identical (see 
Figure~\ref{irspem}); we detect strong and narrow emission lines
(roughly 300-500~\kms\ wide) from the whole Brackett series,
\ion{He}{i}~$\lambda20586$~\AA, \ion{Fe}{ii}~$\lambda15778\textrm{,
}\lambda16878$~\AA, as well as very faint emission of
\ion{Na}{i}~$\lambda22090$~\AA. Overall, these spectra are consistent
with those of early Be III/V stars presented in \citet{2000Clark} and
\citet{2001Steele}. However, we also detect, in both spectra, several
emission lines centred at 15842~\AA, 15964~\AA, 17110~\AA, 17455~\AA, 
and 20456~\AA\ that could be identified with $[$\ion{Fe}{ii}$]$. As
stressed in \citet{1996Morris} and \citet{1998Lamers}, the presence
of such features would be more consistent with narrow-emission-line
massive stars with strongly irradiated outflows, such as B[e] 
supergiants (sgB[e]) or luminous blue variables (although, by
definition, we expect LBVs to be strongly variable). These stars are
known to exhibit a mid-IR excess due to a complex
circumstellar environment in which warm dust is produced. To
investigate the presence of such an excess, we built the spectral
energy distributions (SEDs) of  sources 239 and 1326 using their
near-IR magnitudes (completed with those in $Y$ and $Z$ bands from
VVV) as well as their \spitzer/IRAC fluxes at 3.6, 4.5, 5.8, and
8.0~\mic. We then fitted these SEDs first with a single absorbed
20000~K black body mimicking the stellar emission of a blue star, then
adding a complementary black body to take into account a possible
excess, using Galactic extinction laws from
  \citet{1999Fitzpatrick} in the optical/near-IR and \citet{2006Chiar}
  in the mid-IR. For both sources, the best fits were obtained with
two black bodies, the reduced \chis\ being half the value of the single
black body case (see Figure~\ref{sed_4_16} for plots and best-fit
parameters). We stress that the fits were also performed with stellar
temperatures as free parameters, and it is then possible to describe
the SEDs with single black bodies; however, their best-fit temperatures
reach the minimum allowed value of 3000~K, which is not realistic. We
also fitted the SEDs fixing the stellar temperatures to lower and
higher values (10000~K, 15000~K,
25000~K, 30000~K, and 35000~K) and the mid-IR excess is still
present. Thus, despite the simplistic use of spherical black bodies
\citep[see e.g.][]{2012Chaty}, we believe that sources 239 and
1326 do exhibit near- and mid-IR excesses. Whether the latter are
due to warm dust in sgB[e]/LBV stellar winds or free-free emission
from the decretion disc of Be stars is a matter of debate. We note
that the derived \Av\ values are consistent with a minimum 10~kpc
distance for the two sources \citep[using the three-dimensional
extinction law from][]{2006Marshall}. If the stars do reside at this
distance, they must  have radii larger than 20~\rsun\ -- whatever the
stellar temperature we consider in the fits -- which are more
consistent with early-B supergiants than giant or main-sequence stars
\citep{1996Vacca, 2008Searle}. Furthermore, the unabsorbed 0.5-10~keV
luminosities at 10~kpc are about $10^{33}$~erg~s$^{-1}$ for the two
sources, which is possible for isolated O/B supergiants but at least 1
or 2 orders of magnitude larger than what is expected from isolated Be
III/V stars \citep[see e.g.][]{1996Bergofer, 2000Cohen}. It is therefore
reasonable to speculate that if sources 239 and 1326 are
Be stars, they must belong to HMXBs. Alternatively, they may be
isolated sgB[e] or LBV stars, even though their X-ray photon indices
and luminosities may favour the quiescent X-ray binary hypothesis.

\subsubsection{Sources 1278 and 1279}

In the X-ray domain, sources 1278 and 1279 are very similar. They
are located very close to each other in the quantile diagram and both are 
very absorbed and relatively bright, with
\nh$\sim(1.3-1.5)\times10^{23}$~\acm2\ and unabsorbed 0.5-10~keV
luminosities of approximatively $1.4\times10^{33}$~erg~s$^{-1}$ and
$3.5\times10^{33}$~erg~s$^{-1}$ at 10~kpc, respectively. 
Despite large uncertainties in the assessment of their column
densities, they very likely exhibit an excess absorption with respect
to the total galactic extinction along their LOS
$N_{\textrm{\ion{H}{i}}+\textrm{H}_{\rm 2}}{\rm (ISM)}\sim7.8\times10^{22}$~\acm2;
while this is consistent with wind accretion in HMXBs, their X-ray spectra are 
better fitted with an thermal plasma model ($kT=2.0_{-0.8}^{+2.0}$~keV and 
$kT=2.6_{-0.8}^{+2.1}$~keV, respectively), which is more
typical of isolated massive stars. 
\input{./Figures/mercer.figures}

The similarity of their X-ray properties concurs with the
fact that sources 1278 and 1279 belong to the same young
massive star cluster Mercer~81 \citep{2005Mercer},
discovered during the GLIMPSE survey of the Galactic plane with
\spitzer. Mercer 81 was recently studied through near-IR photometry with
\textit{HST} and near-IR spectroscopy with ESO/ISAAC and the results
of these observations are reported in \citet{2012Davies} and
\citet{2013Fuente}. The authors derived a $11\pm2$~kpc distance and
showed that Mercer 81 was mainly populated with nitrogen-rich Wolf-Rayet
and blue supergiant stars; in particular, sources 1278 and 1279
\citep[labelled 8 and 2 in these papers, see Figure~2
in][]{2012Davies} were identified as late WN Wolf-Rayet stars
\citep[see Figure~3 in][]{2013Fuente}. We concur with these results.
Our OSIRIS spectra, displayed in Figure~\ref{irspem2},
are similar to theirs, and our derived
  \ion{He}{ii}~$\lambda21891$~\AA/\ion{H}{i}~$\lambda21661$~\AA\
  ratios -- i.e. about 0.23 and 0.19 for sources 1278 and 1279,
  respectively -- are consistent with a WN8 classification
  \citep[expected values between 0.1 and 0.4, see Table~5 
  in][]{2006Crowther3}. Moreover, Figure~\ref{mc81} displays a
0.5-10~keV \chandra\ image of Mercer 81 on which we superimposed the
near-IR positions of sources 1278 and 1279 (green circles) as well
as those of the remaining eight sources identified as Wolf-Rayet or
supergiant stars (yellow circles). It is clear that only sources
1278 and 1279 are detected with \chandra, although all sources
suffer from the same extinction; this indicates that sources 1278
and 1279 are more effective X-ray emitters than other Wolf-Rayet
stars in Mercer 81. To check this hypothesis, we can follow
  \citet{2010Mauerhan} to estimate the X-ray to bolometric 
  luminosity ratios $L_{\rm X}/L_{\rm Bol}$ for the two sources and
  compare them to the expected $10^{-7}$ value for isolated Wolf-Rayet
  stars. Using the observed $(J-Ks)$ and $(H-Ks)$ colours, the intrinsic
  $(J-Ks)_{0}$ and $(H-Ks)_0$ colours for WN8 stars given in
  \citet{2006Crowther3}, i.e. 0.13 and 0.11, respectively, as well as the
  relations $A_{\rm Ks}=(1.44\pm0.01)\times E(H-Ks)$ and $A_{\rm
    Ks}=(0.494\pm0.006)\times E(J-Ks)$ from \citet{2006Nishiyama}, we
  derive an average $Ks$-band extinction $A_{\rm Ks}=3.11\pm0.03$ for
  each source. For a 10~kpc distance, this results in the absolute
  $Ks$-band magnitudes $-6.85\pm0.04$ and $-8.27\pm0.06$ 
  for source 1278 and 1279, respectively. Using the $Ks$-band
  bolometric correction for WN8 stars given in \citet{2006Crowther3},
  i.e. $-3.4$, we thus find that the total bolometric luminosities are $L_{\rm
    Bol}\approx10^6L_\odot\approx3.9\times10^{39}$~erg~s$^{-1}$ and
  $L_{\rm
    Bol}\approx3.8\times10^6L_\odot\approx1.5\times10^{40}$~erg~s$^{-1}$,
  for sources 1278 and 1279, 
  respectively. This yields $L_{\rm X}/L_{\rm Bol}$ ratios of about 
$3.6\times10^{-7}$ and $2.5\times10^{-7}$ for sources 1278 and
1279, respectively. These values are slightly in excess of the
canonical $10^{-7}$, and this may indicate that the two sources are
CWBs, producing X-rays in the shocks of
their colliding winds. Another explanation is that both sources are
HMXBs in which the compact object accretes material from the strong
stellar winds of Wolf-Rayet stars, but as already mentioned, the fact
that their X-ray spectra are better described by thermal  plasma
models disfavours this possibility. 

\subsubsection{Source 1168}

\input{./Figures/sp_ir_s1179.figures}

We detected source 1168 with \chandra\ as a bright and variable hard
X-ray point source ($\Gamma=1.8_{-0.4}^{+0.4}$) with a
column density much lower than that of ISM along its entire LOS
(\nh$\sim2.9\times10^{21}$~\acm2\ vs.
$N_{\textrm{\ion{H}{i}}+\textrm{H}_{\rm 2}}{\rm (ISM)}\sim6.2\times10^{22}$~\acm2). The 
near-IR spectrum displayed in Figure~\ref{irsp1179} is typical of massive
stars. In the \textit{H} band, we clearly detect absorption features
of the Brackett series centred at 15561~\AA, 15885~\AA, 16114~\AA,
16412~\AA, 16811~\AA, and 17367~\AA. The absence of Brackett
lines below 15500~\AA\ is also consistent with a IV/V luminosity
class \citep{1998Meyer}, while the strength of the detected \ion{H}{i} lines
and the lack of \ion{He}{i} are characteristic of
late-B/early-A stars \citep{1997Blum}. In the \textit{K} band,
Br~$\gamma$ and \ion{Mg}{ii}~21376~\AA, signatures of stellar 
winds, are present but the strength of Br~$\gamma$ and the absence of
\ion{He}{i} features point towards a IV/V class with a spectral type
later than B8 \citep{1996Hanson}; we therefore conclude that source
1168 is a nearby B8-A3IV/V star. Figure~\ref{sed_1179} displays its
optical to mid-IR SED built using archival fluxes from the VVV and
GLIMPSE surveys and fitted with a 11000~K temperature black body. Using
the derived \Av\ value and the three-dimensional law of
\citet{2006Marshall}, we estimate its distance to be around 3.5~kpc,
which results in a stellar radius $R_\ast \approx 4.9$\rsun,
  consistent with the typical radii of late-B/early-A subgiant stars
  listed in \citet{2001Pasi}. Moreover, such a distance gives a
0.5-10~keV unabsorbed luminosity of about
$1.8\times10^{32}$~erg~s$^{-1}$, i.e. at least one order of
magnitude too high for an isolated B8-A3IV/V star \citep{1996Bergofer}
and we tentatively classify source 1168 as a quiescent HMXB.
\input{./Figures/sed_1179.figures}
\input{./Figures/sp_ir_s747.figures}

\subsubsection{Source 750}

 Source 750 is one of the brightest X-ray sources that we detected with
\chandra. It is variable and has a hard spectrum
($\Gamma=1.2_{-0.1}^{+0.1}$) with  a \nh$\sim1.2\times10^{21}$~\acm2
column density;  once corrected from absorption, its 0.5-10~keV
luminosity at 1~kpc is about $1.2\times10^{32}$~erg~s$^{-1}$. It
  is the hardest source of group B and its location in the
  quantile diagram is clearly separated from the other sources of the
  group, hinting at a distinct nature. More importantly, its X-ray light
curve exhibits a 7150~s period \citep{2014Fornasini} more likely
related to the orbit or the spin period of a white dwarf
(WD). Overall, the X-ray behaviour of source 750 is consistent with
that of an intermediate polar (IP), in which a WD accretes from a
main-sequence star via a truncated accretion disc \citep[see][for a
review]{2006Kuulkers}.
\input{./Tables/lines_lm.table}

In the near-IR domain, the source is variable, with $Ks=14.27\pm0.01$
from VVV and $Ks > 15.19$ from 2MASS, pointing towards a
contribution from an accretion stream. Moreover, its near-IR spectrum,
displayed in Figure~\ref{irsp747}, is clearly dominated by the whole
Brackett series as well as \ion{He}{i} $\lambda17007\textrm{~\AA\ and
}\lambda20586$~\AA, all in emission. The lines are particularly strong
in \textit{K}, with equivalent widths of about 18~\AA\ and 43~\AA\ for
\ion{He}{i} and Br~$\gamma$, respectively. Again, this spectrum is
consistent with an accretion stream and is actually very similar to
those of IPs presented in \citet{1997Dhillon} and
\citet{2007Harrison}; we therefore believe that source 750 is very
likely an IP.
 
\subsection{Sources with typical near-IR spectra of cool stars}
\input{./Figures/sp_ir.figures}
\input{./Figures/sp_ir2.figures}
\input{./Figures/sp_ir3.figures}
\input{./Figures/sp_ir4.figures}
\input{./Figures/sp_ir5.figures}
\input{./Tables/class_lm.table}

 In the \textit{H} and \textit{K} bands, cool stars are
 characterized by the presence of many absorption lines of
 \ion{Fe}{i}, \ion{Mg}{i}, \ion{Si}{i}, \ion{Ca}{i}, and \ion{Na}{i},
 but the detection of the CO~(2,0) and CO~(3,1) overtones in
 absorption  beyond 22900~\AA\ is a clear indicator of their
 nature. The near-IR  counterparts of the 14 sources presented
 in this section all  exhibit at least one of these CO features and
 this is the reason why we classify them as cool stars. However,
 deriving a more  accurate spectral classification is difficult
 because (1) the S/N  ratio of our spectra are not high enough to
 perform a quantitative comparison with existing near-IR libraries and
 (2) a possible veiling  by the near-IR continuum of the component
 responsible for the soft  X-ray emission would likely alter the
 equivalent width measurements. We therefore list tentative spectral
 types in Table~\ref{class} obtained by comparing the relative
 strength of the CO features at 16190~\AA\ and 22957~\AA\ as well as
 that of Br{\small $\gamma$}, the last being weak or absent for K/M
 stars. We nevertheless stress that this classification should be
 taken with caution due to the aforementioned limitations.

Determining the possible origin of the soft
 X-ray emission and thus the nature of the X-ray emitters is also not 
 straightforward. On the one hand, the X-ray emission could
 stem from accreting binaries such as LMXBs, CVs, IPs, or symbiotic
 binaries (SBs). On the other hand, it could also come from foreground
 isolated low mass stars (iLMS) or coronally active binaries (ABs). Although
 discriminating between all these possibilities is
 difficult, the presence of near-IR emission lines, in particular
 \ion{H}{i}, and/or near-IR variability can be a strong indicator of the
 presence of an accretion stream. The ISM extinction along the
 LOS and the X-ray luminosity also provide valuable information;
 for example iLMS are likely foreground objects and their X-ray
 luminosity is expected to be lower than $10^{30}$~\ergs\ \citep[see
 e.g.][]{2009Gudel}. Finally, the positions of the sources in the $Q_{\rm X}$ vs.
 $Q_{\rm Y}$ diagram hint at common X-ray spectral properties, 
 and therefore the following sections present the sources
 grouped by similar location in this diagram.

\subsubsection{Groups A/B: sources 694, 1059, 1203, 1358 1374, and 1379 }

The six sources are weakly extincted, with
$\nhe~<~10^{22}$~\acm2, indicating that they are foreground
objects. However, sources 694, 1203, 1358, 1374, and 1379
exhibit similar photon indices ($\Gamma=2.4_{-0.7}^{+0.9}$,  
$\Gamma=2.4\pm0.4$, $\Gamma=2.5_{-0.5}^{+0.6}$,
$\Gamma=1.5_{-0.3}^{+0.5}$, and $\Gamma=2.3_{-0.2}^{+0.4}$,
respectively), while source 1059 is a much softer and better
described by an absorbed thermal plasma model, with
$kT=0.5\pm0.2$~keV. Source 694 is the faintest, with a 0.5-10~keV
luminosity at 1~kpc $\sim6\times10^{30}$~\ergs\, compared to
$\sim(1.1-5)\times10^{31}$~\ergs\ for the other five sources. 

These X-ray properties are consistent with those expected from all
X-ray emitting low mass systems, including iLMSs and ABs, and do not
specifically hint at the presence of accretion discs. However, we note
that the near-IR spectrum of source 694 is the only one that
convincingly exhibits emission lines. This source is also the only one
with variable near-IR emission, with $J$, $H$, and $Ks$ magnitudes of
$12.53\pm0.05$, $11.75\pm0.07$, and $10.26\pm0.06$ in NEWFIRM;
$13.29\pm0.01$, $12.70\pm0.01$, and $12.26\pm0.01$ in VVV; 
and $12.96\pm0.04$, $12.15\pm0.03$, and $11.89\pm0.04$ in 2MASS. We
therefore believe that source 694 is a quiescent accreting
binary. The other five sources are likely isolated giant or main-sequence 
stars, although we stress that sources 1203 and 1379 may
  exhibit a variable emission at 3.5~\mic\, with GLIMPSE and {\it
    WISE} fluxes of about $131.4\pm6.0$~mJy vs. $155.4\pm3.9$~mJy and
  $60.68\pm2.85$~mJy vs. $72.89\pm1.81$~mJy, respectively. We
also note that the optical/near-IR counterpart of source 1059 is
classified as an M4V star in \citet{2011Lepine}, in agreement
  with our own classification as an early main-sequence M star.

\subsubsection{Group C: sources 158, 437, 881, and 1128}

The four sources are more extincted that the previous group
of sources, with \nh\ values clustered in the range
$(1-4.5)\times10^{22}$~\acm2, lower than the total ISM extinction
along their LOS; this extinction indicates that these sources
are located in the near Norma arm. They also have harder spectra than
the sources discussed in Sect.~3.2.1; sources 158, 437, and
1128 have a similar spectral index ($\Gamma=1.5_{-1.0}^{+1.2}$,
$\Gamma=1.7_{-1.0}^{+1.2}$ and $\Gamma=2.2\pm0.6$, respectively) while
the spectrum of source 881 is even flatter, with
$\Gamma=0.8\pm0.4$. Their 0.5-10~keV unabsorbed luminosities at 5~kpc 
($2.1\times10^{32}$~\ergs, $1.7\times10^{32}$~\ergs,
$1.1\times10^{33}$~\ergs, and $5.1\times10^{32}$~\ergs, respectively)
are inconsistent with iLMSs or ABs and thus the four sources are likely
accreting binaries. However, the presence of emission lines in their
near-IR spectra is scarce, maybe because the accretion stream is less
active. Furthermore, source 881 -- the brightest X-ray
emitter of the four -- is convincingly variable in the near-IR, with
2MASS magnitudes significantly fainter than the DENIS ones ($J$:
$6.89\pm0.02$ vs. $6.38\pm0.09$; $Ks$: $4.72\pm0.02$ vs.
$3.87\pm0.16$), as well as in the mid-IR, the GLIMPSE fluxes at
  3.5~\mic\ and 4.5~\mic\ being lower that those from {\it WISE} at
the same wavelengths (3.5~\mic: $4486\pm378$~mJy vs. $5681\pm493$~mJy;
4.5~\mic: $2552\pm114$~mJy vs. $4213\pm220$~mJy). The lack of 
significant near-IR variability for the remaining sources may
indicate that if accretion is occurring, it is weak. 
  However, we note that source 1128 likely has variable mid-IR
  emission, with higher GLIMPSE fluxes at 3.5~\mic\ and 4.5~\mic\ with
respect to {\it WISE} (3.5~\mic: $18.68\pm0.08$~mJy vs. $12.91\pm0.02$~mJy;
4.5~\mic: $11.15\pm0.04$~mJy vs. $7.31\pm0.41$~mJy).

\subsubsection{Groups D and E: sources 38, 78, 396, and 1408}

Sources 38, 396, and 1408 suffer from
an extinction likely higher than that along their LOS,
having $\nhe\ge7.8\times10^{22}$~\acm2. They exhibit a similar photon
index ($\Gamma=2.6_{-1.3}^{+1.4}$, $\Gamma=2.8_{-1.0}^{+1.1}$, and
$\Gamma=2.1_{-1.2}^{+1.6}$, respectively) and are relatively bright
X-ray emitters, with unabsorbed 0.5-10~keV luminosities assuming a
10~kpc distance of $3.7\times10^{33}$~\ergs, $1.3\times10^{33}$~\ergs,
and $1.7\times10^{33}$~\ergs, respectively. The properties of source 78
are slightly different, as it exhibits a lower column density along 
its LOS of $\nhe\sim2.4\times10^{22}$~\acm2 and is a lot
harder, with $\Gamma=0.5\pm0.4$  and a 0.5-10~keV luminosity at
5~kpc of about $3\times10^{33}$~\ergs. That said, the X-ray properties
of the four sources are inconsistent with those expected from
iLMSs or ABs, and their  X-ray emission likely stems from accretion. 

The near-IR properties are in agreement with the latter hypothesis. The
four sources exhibit emission lines in their spectra, hinting 
at an accretion stream presence. In addition, sources 78, 396,
and 1408 are variable. Indeed, the near-IR magnitudes of
source 78 decreased from VVV to NEWFIRM ($H$: $15.77\pm0.04$ vs.
$15.45\pm0.10$; $Ks$: $15.38\pm0.02$ vs. $14.99\pm0.09$). Moreover,
source 396 is undetected with DENIS and its 2MASS magnitudes are
slightly higher than those of NEWFIRM and VVV ($J>15.69$,
$H=13.30\pm0.03$, and $Ks=12.20\pm0.02$ vs. $J=15.49\pm0.01$, 
$H=13.19\pm0.01$, and $Ks=12.0\pm0.01$). Likewise, the 2MASS $J$
magnitude of source 1408 is lower than that measured in DENIS
($8.24\pm0.03$ vs. $8.53\pm0.08$). Its GLIMPSE fluxes at 3.5~\mic\
  is also significantly lower than that from {\it WISE} ($3120\pm78$~mJy vs.
  $1593\pm84$~mJy). Based on their X-ray and infrared
behaviours and properties, it is therefore reasonable to conclude that
all four sources are accreting binaries hosting giant stars.

\section{Discussion and conclusion}

We have conducted near-IR photometric and spectroscopic
observations of 20 soft X-ray sources discovered during a
\chandra\ survey of a $2^{\circ}\times$0\fdg8 region of the Norma
arm. Our main goal was to detect new low-luminosity HMXBs, thought to
be principally located in very active star-forming regions. We identify 
(1) two massive emission-line stars, possibly in HMXBs, exhibiting
near- and mid-IR excesses consistent with either free-free emission
from the decretion discs of Be stars or warm dust in the stellar winds
of peculiar massive stars such as sgB[e] or LBVs; (2) two WN8 Wolf
Rayets, maybe in CWBs and located in the Mercer 81 massive star cluster;
(3) a foreground B8-A3 IV/V star, likely in an HMXB; (4) one IP, the
near-IR spectrum of which is dominated by the accretion disc; and (5) a
foreground isolated M4V dwarf. Among the 13 remaining sources,
four are likely isolated late-type giant or main-sequence stars, and
nine are low mass accreting binaries.

\subsection{The nature of the low mass accreting systems}

Even if the outcome of the ongoing near-IR spectroscopy of additional
NARCS sources may increase the number of detected massive
stars, it is clear that our sample is dominated by low mass accreting
binaries. However, it is difficult to be more specific about the nature
of the primaries (WDs, NSs, or BHs) or that of the systems themselves
(LMXBs, CVs, or SBs). Based on the presence of many emission lines in
their near-IR spectra, their relatively high 0.5-10~keV luminosity and
the red giant nature of their companion stars, we can
speculate that sources 38, 396, and 1408 are quiescent LMXBs in
the outer Norma arm. It is, however, interesting that they all suffer
from an intrinsic extinction in excess of the ISM along their LOS (especially
source 38) and that they are located at a similar position in the
quantile diagram as candidate HMXBs and CWBs. It is therefore possible
that sources 38, 396, and 1408 actually are SBs in which the
compact objects accrete material from the stellar winds of their late
giant companion stars in an eccentric orbit. Whether the compact
object is a WD or a NS is a matter of debate, but we stress that their
X-ray and near-IR properties (in particular their relative X-ray
hardness) could be consistent with the new and rare class of symbiotic
X-ray binaries (SyXRBs), in which
a magnetized NS accretes matter from the wind of a red giant
\citep[see][]{2007Masetti, 2008Corbet, 2013DeWitt}. Source 694 in
groups A/B as well as sources 158, 437, 881, and 78 in groups C/D
are also possible SBs but likely belong to two different
classes. Indeed, source 694 has all the characteristics of a
standard quiescent SB with a WD primary, i.e. a giant companion, weak
extinction, and low 0.5-10~keV luminosity. In contrast, sources 78,
158, 437, and 881 exhibit X-ray luminosities of
$(1-30)\times10^{32}$~\ergs, are relatively extincted, and have a hard
spectrum. This behaviour may be more consistent with that expected from
hard spectrum SBs, in which a WD accretes from a red
giant through an accretion disc \citep[dubbed as class $\delta$;
see][]{2013Luna}.  Finally, source 1128 has all the properties
of a quiescent CV in the inner Norma arm, and likely hosts a WD
accreting from a main-sequence companion.

\subsection{The nature of the high mass systems}

The low number of massive stars in our sample was expected because of
their short lifetime. Nonetheless, the possible detection of
up to three new HMXBs (sources 239, 1168, and 1326) in a region
covering about one square degree is consistent with our prediction of
three to four HMXBs per square degree with unabsorbed fluxes
larger than $5\times10^{-14}$~\ergcms \citep{2014Fornasini}. One
could argue that none of these sources is an HMXB, in particular
because their unabsorbed luminosities are at least one order of
magnitude lower than expected. However, these X-ray 
luminosities also seem too high to be those of isolated stars. A
possible way to reconcile these discrepancies is that sources 239,
1326, and 1168 are quiescent HMXBs. We
stress that SFXTs, which are transient sources, exhibit luminosities as low as  
$10^{32}$~erg~s$^{-1}$ when in quiescence. Furthermore, such low
luminosities have been observed in some Be X-ray binaries \citep[see
e.g. V0332+53,][]{2002Campana} and some $\gamma$-Cas analogs which
could be -- although this explanation is still controversial -- WD-Be
binaries \citep{2013Nebot}; such systems may represent up to 70\% of all
compact object-Be star associations \citep{2001Raguzova}. More
  recently, \citet{2014Casares} also confirmed that the Be star MWC~656 was
the companion of an X-ray quiescent BH, resulting in the first
identification of a BH-BeXB. Population synthesis models point towards
the extreme rarity of such systems, which are also thought to be very
faint in the X-ray domain due to accretion disc truncation. This 
illustrates one of the major limitations of soft X-ray surveys in 
characterizing new low-luminosity HMXBs, as a significant fraction of 
them appear similar to CWBs or slightly more energetic isolated
massive stars. It is therefore reasonable to believe that a
significant population of quiescent HMXBs, some with BH
  primaries, exist in the Galaxy but remain undetected and/or
misidentified. A possible way to address this issue is to observe
candidate low-luminosity HMXBs through optical and near-IR
  high-resolution spectroscopy for radial velocity measurements. 
  Alternatively, hard X-ray spectroscopy may allow us to characterize
  their emission in this spectral domain more accurately, which should be
power-law-like and harder than that of isolated massive stars if even
weak accretion occurs. In particular, hard X-ray observations of
the five X-ray sources with massive stellar counterparts we identified
in this study should enable us to verify their HMXB, CWB, or isolated
massive star nature.

\begin{acknowledgements}
We thank the anonymous referee for her/his comments that helped
to improve this work. FMF acknowledges support from a National Science
Foundation Graduate Research Fellowship. FEB acknowledges support from
Basal-CATA PFB-06/2007, CONICYT-Chile FONDECYT 1141218, ``EMBIGGEN''
Anillo ACT1101, and Project IC120009 ``Millennium Institute of
Astrophysics (MAS)'' funded by the Iniciativa Cient\'{\i}fica Milenio del
Ministerio de Econom\'{\i}a, Fomento y Turismo. We acknowledge the use
of data products from observations made with ESO Telescopes at the La
Silla or Paranal Observatories under ESO programme ID~179.B-2002. The
VVV Survey is supported by ESO, by BASAL Center for Astrophysics and
Associated Technologies PFB-06, by FONDAP Center for Astrophysics
15010003, and by the Iniciativa Cient\'{\i}fica Milenio del
Ministerio de Econom\'{\i}a, Fomento y Turismo through grant
IC~12009, awarded to the ``Millennium Institute of
Astrophysics (MAS)''.
This publication makes use of data products from the Two Micron All
Sky Survey, which is a joint project of the University of
Massachusetts and the Infrared Processing and Analysis
Center/California Institute of Technology, funded by the National
Aeronautics and Space Administration and the National Science
Foundation.The DENIS project has been partly funded by the SCIENCE and
the HCM plans of the European Commission under grants CT920791 and
CT940627. It is supported by INSU, MEN and CNRS in France, by the
State of Baden-W\"urttemberg in Germany, by DGICYT in Spain, by CNR in
Italy, by FFwFBWF in Austria, by FAPESP in Brazil, by OTKA grants
F-4239 and F-013990 in Hungary, and by the ESO C\&EE grant A-04-046.
This work is partly based on observations made with the Spitzer Space
Telescope, which is operated by the Jet Propulsion Laboratory,
California Institute of Technology under a contract with the National
Aeronautics and Space Administration. This publication makes use of
data products from the Wide-field Infrared Survey Explorer, which is a
joint project of the University of California, Los Angeles, and the
Jet Propulsion Laboratory/California Institute of Technology, funded
by the National Aeronautics and Space Administration.
\end{acknowledgements}

\bibliographystyle{aa}
\bibliography{./mybib}

\end{document}

%% file: Figures/fc_ir.figures
\begin{figure*}
\begin{center}
\begin{tabular}{cccc}
\includegraphics[width=4cm]{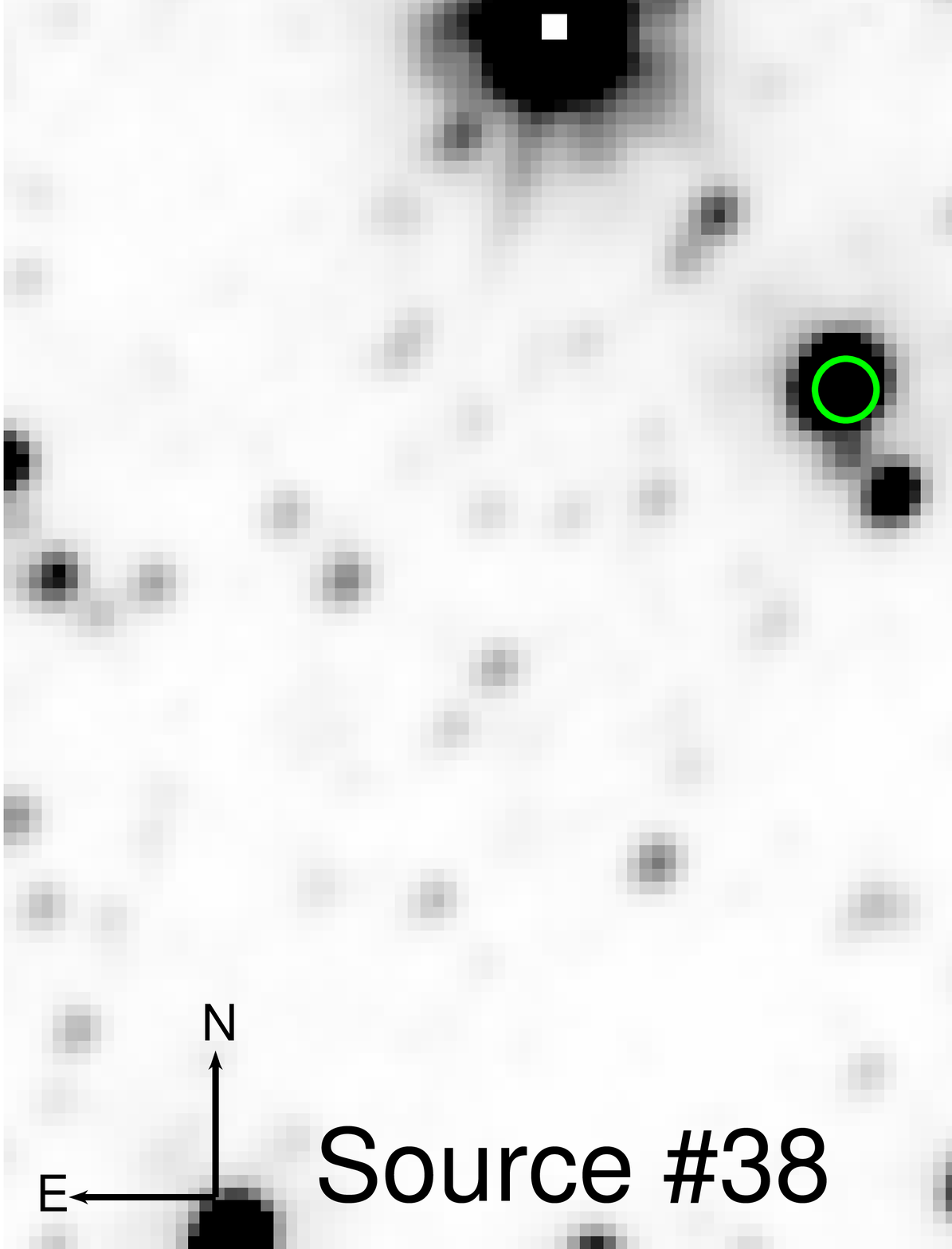}&\includegraphics[width=4cm]{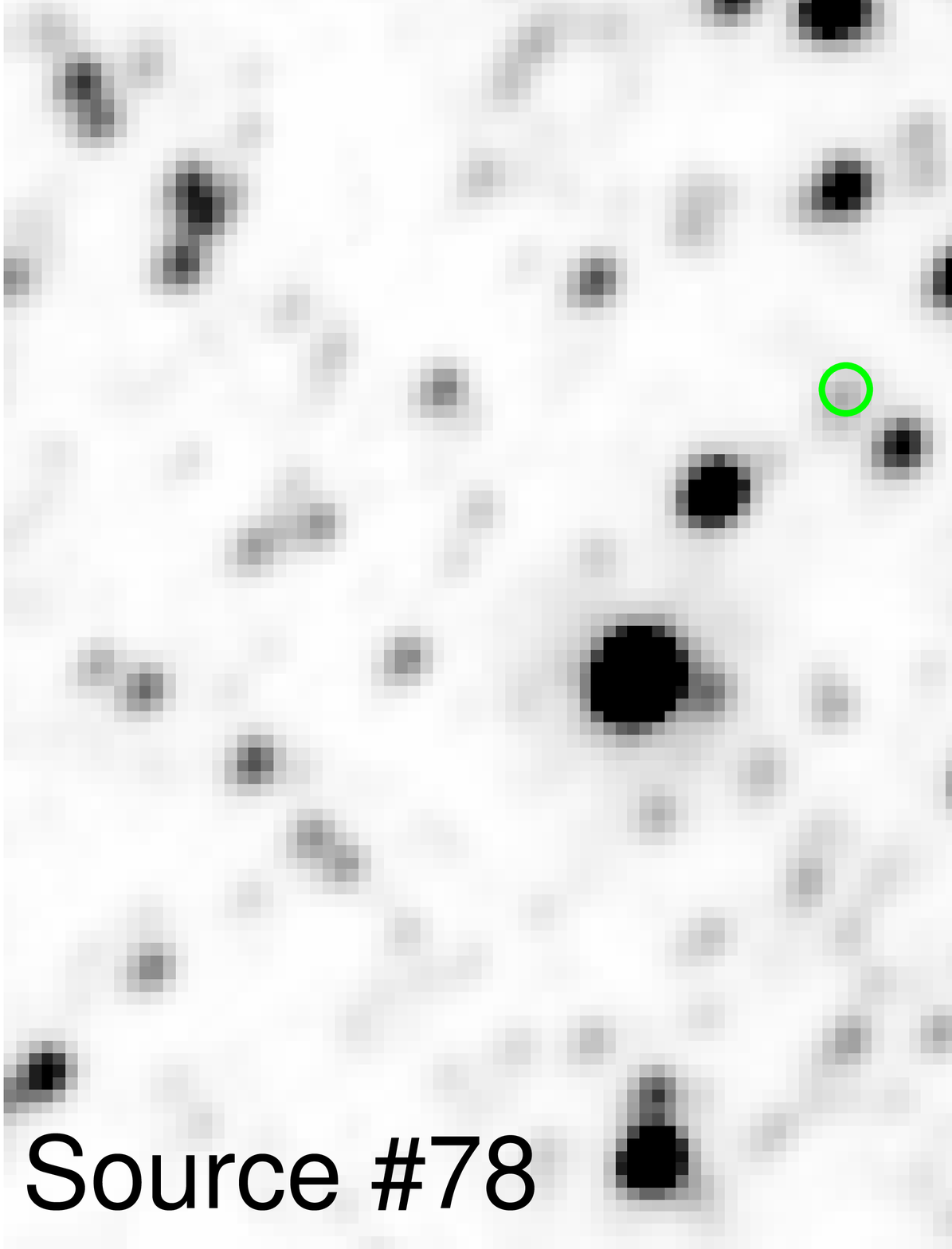}&\includegraphics[width=4cm]{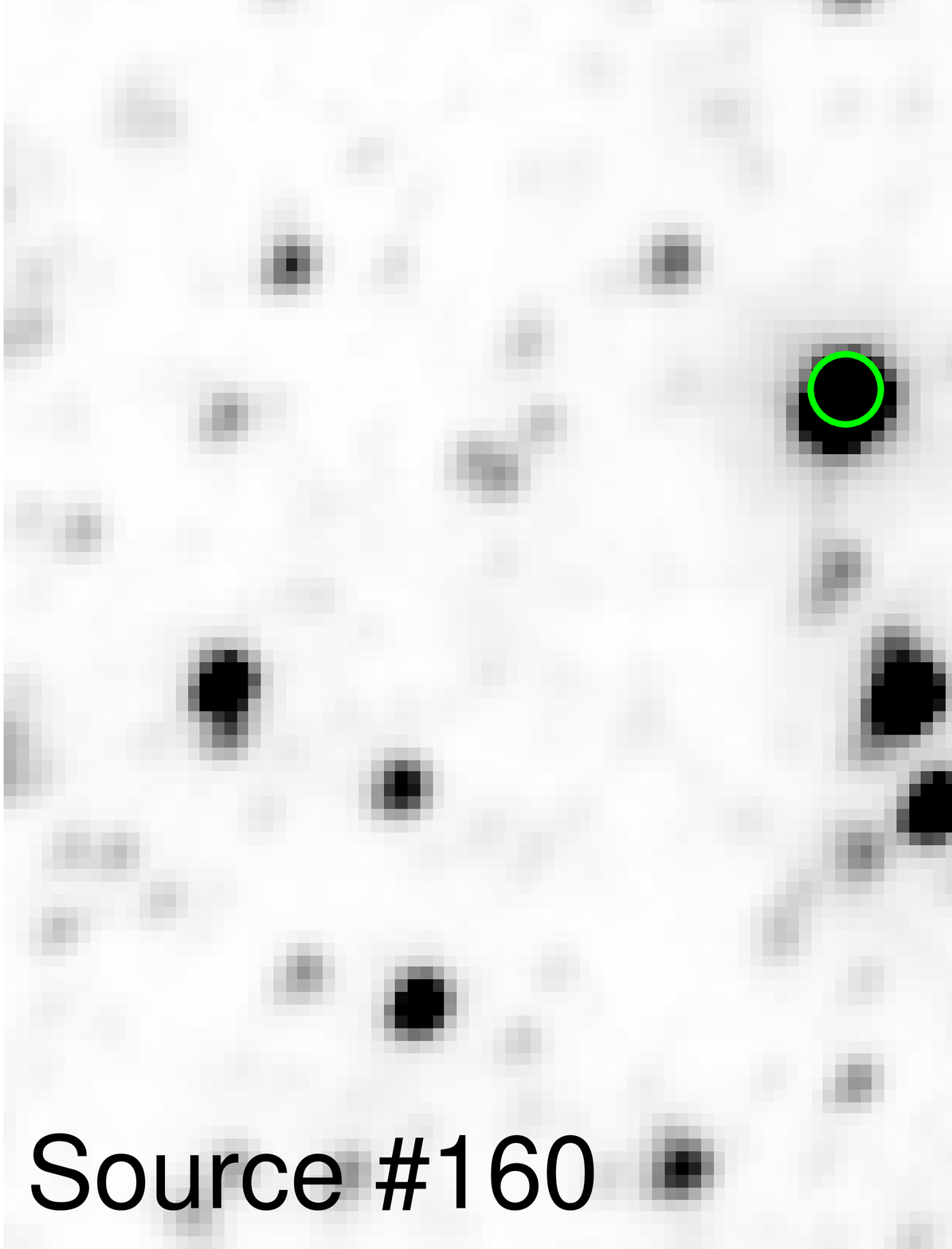}&\includegraphics[width=4cm]{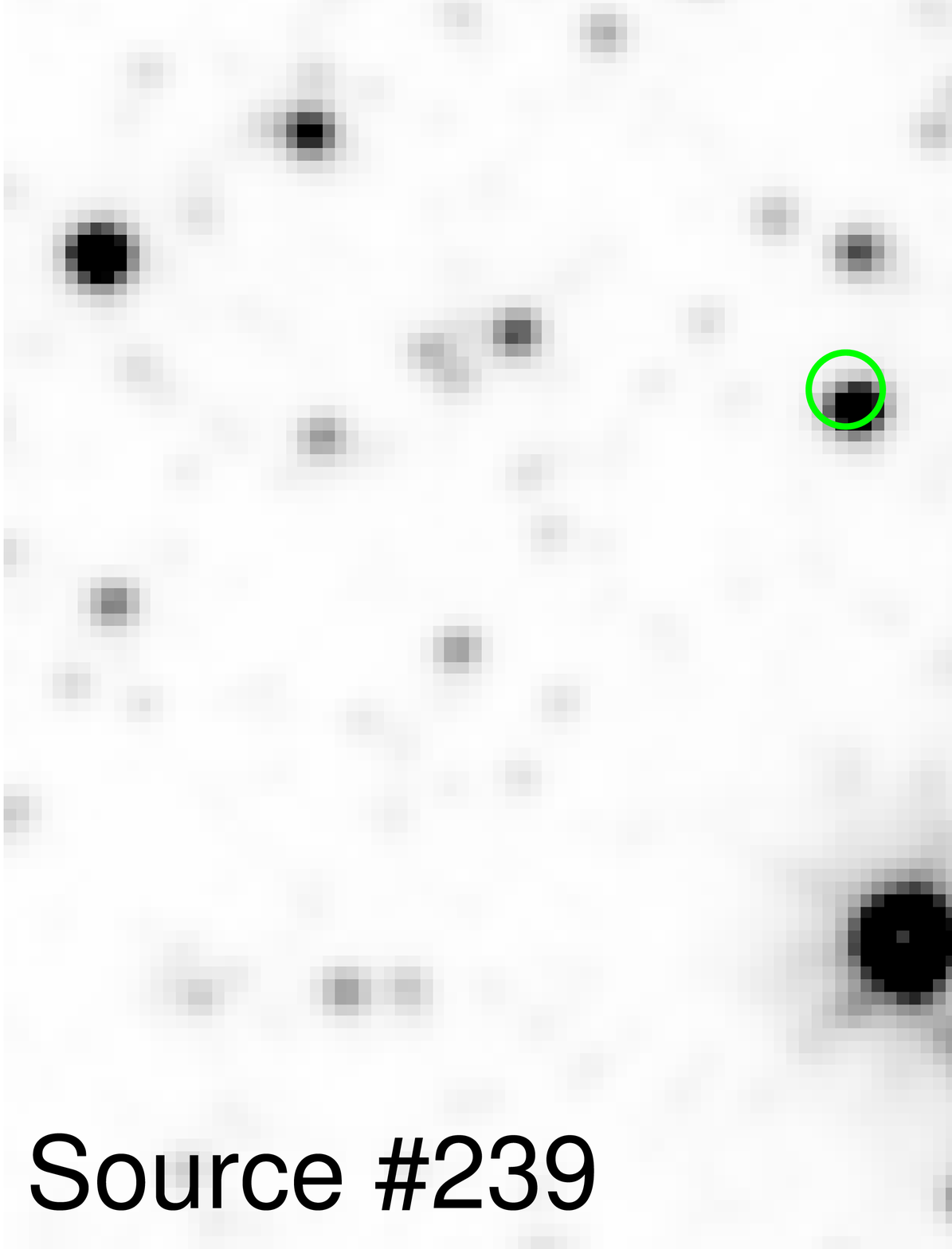}\\
\includegraphics[width=4cm]{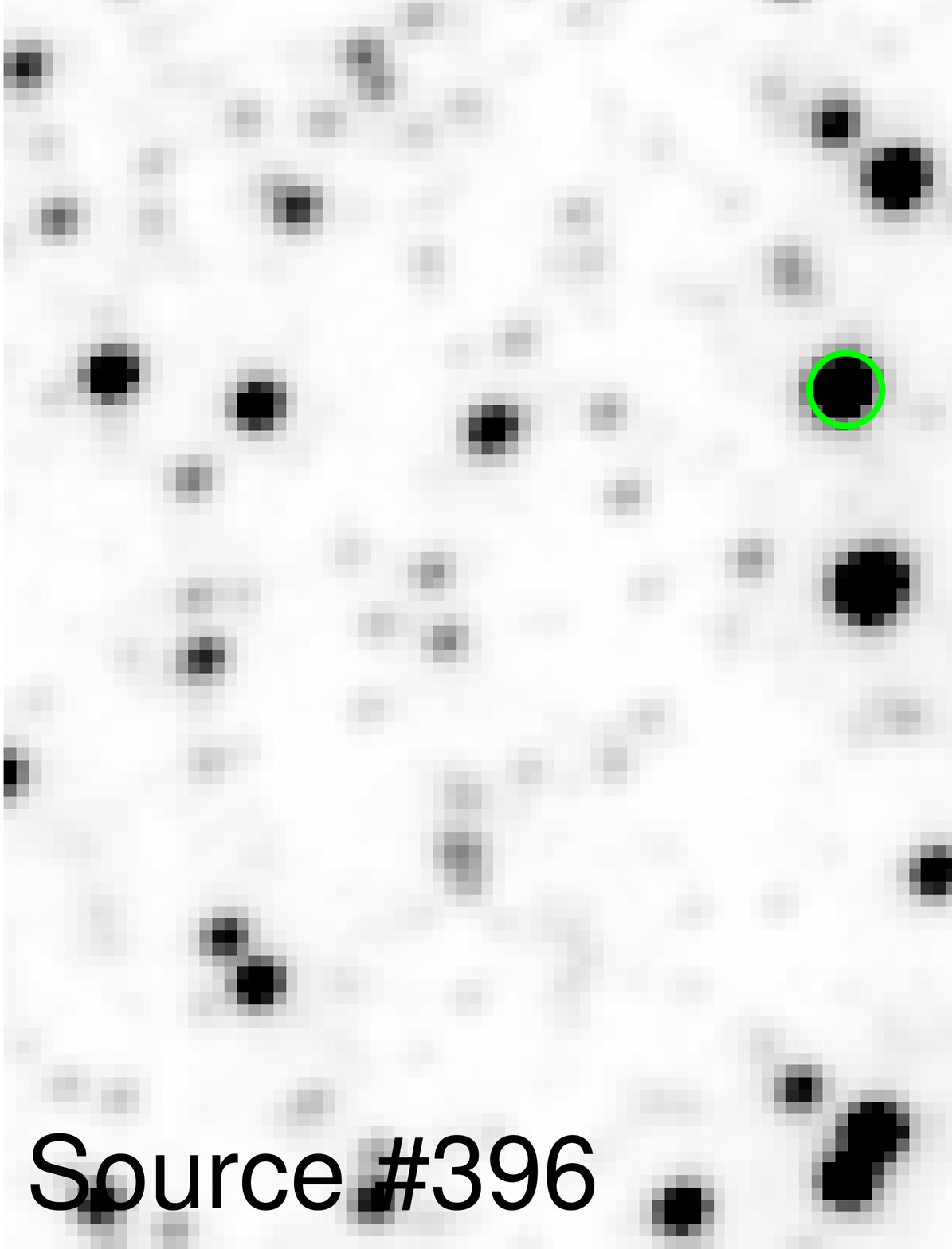}&\includegraphics[width=4cm]{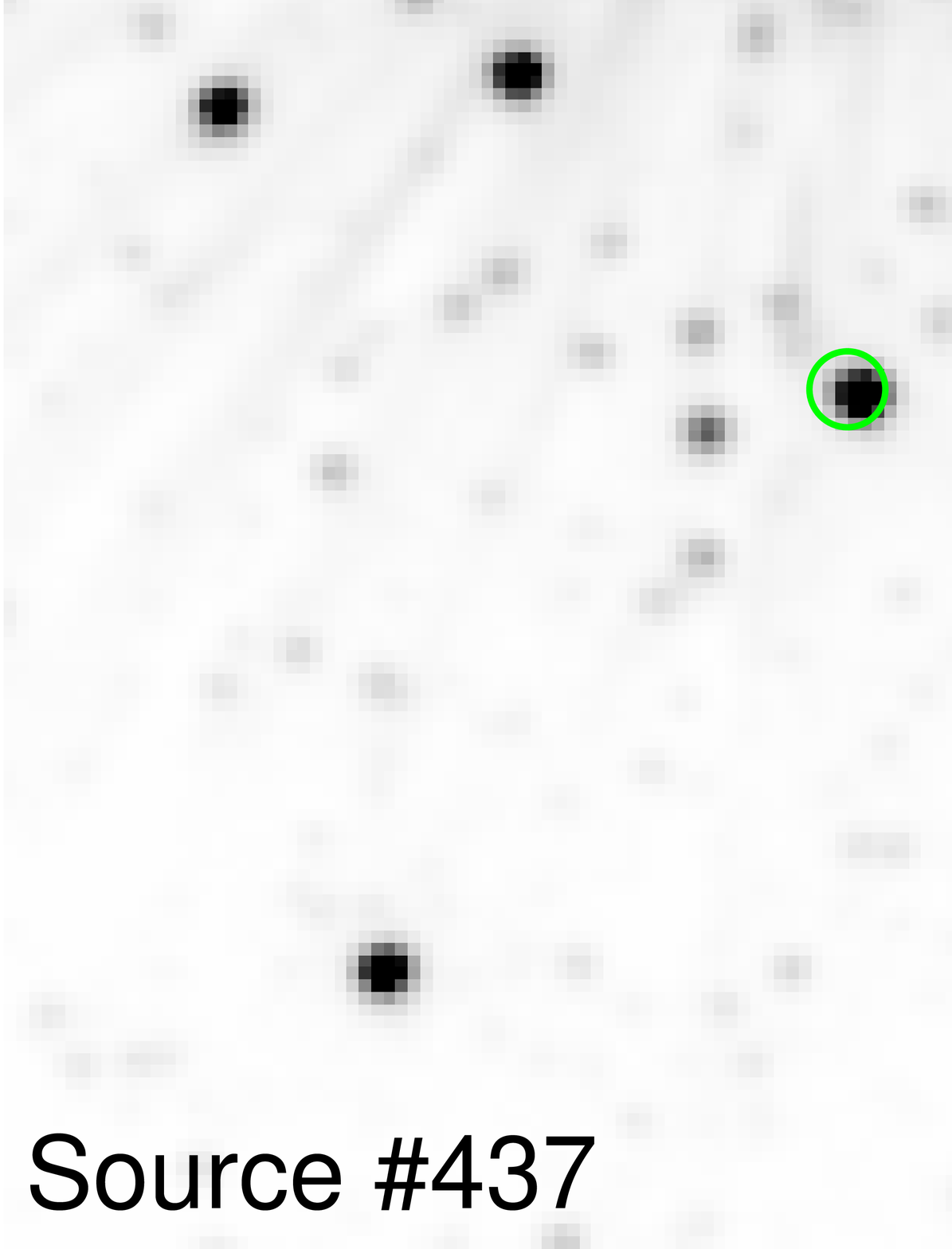}&\includegraphics[width=4cm]{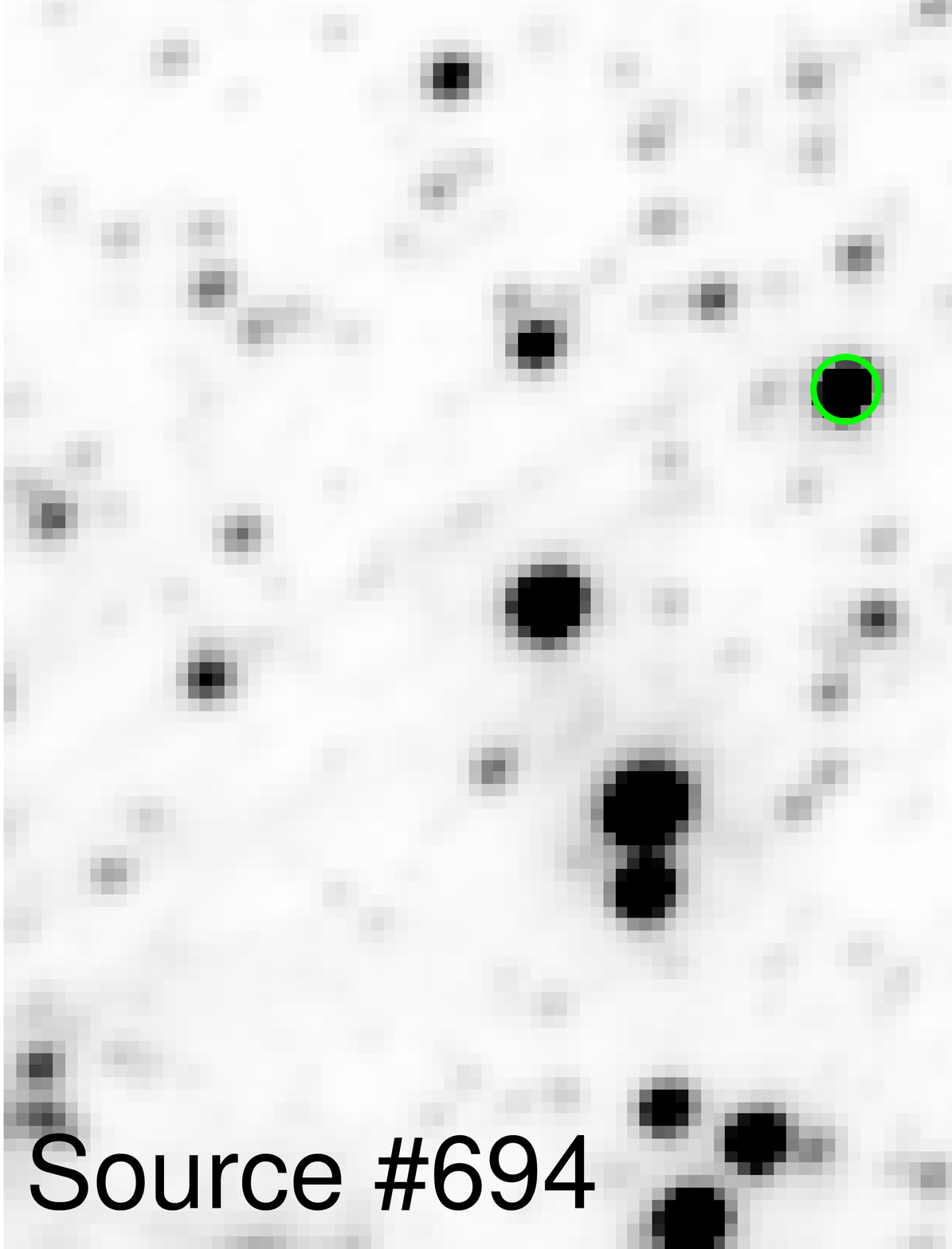}&\includegraphics[width=4cm]{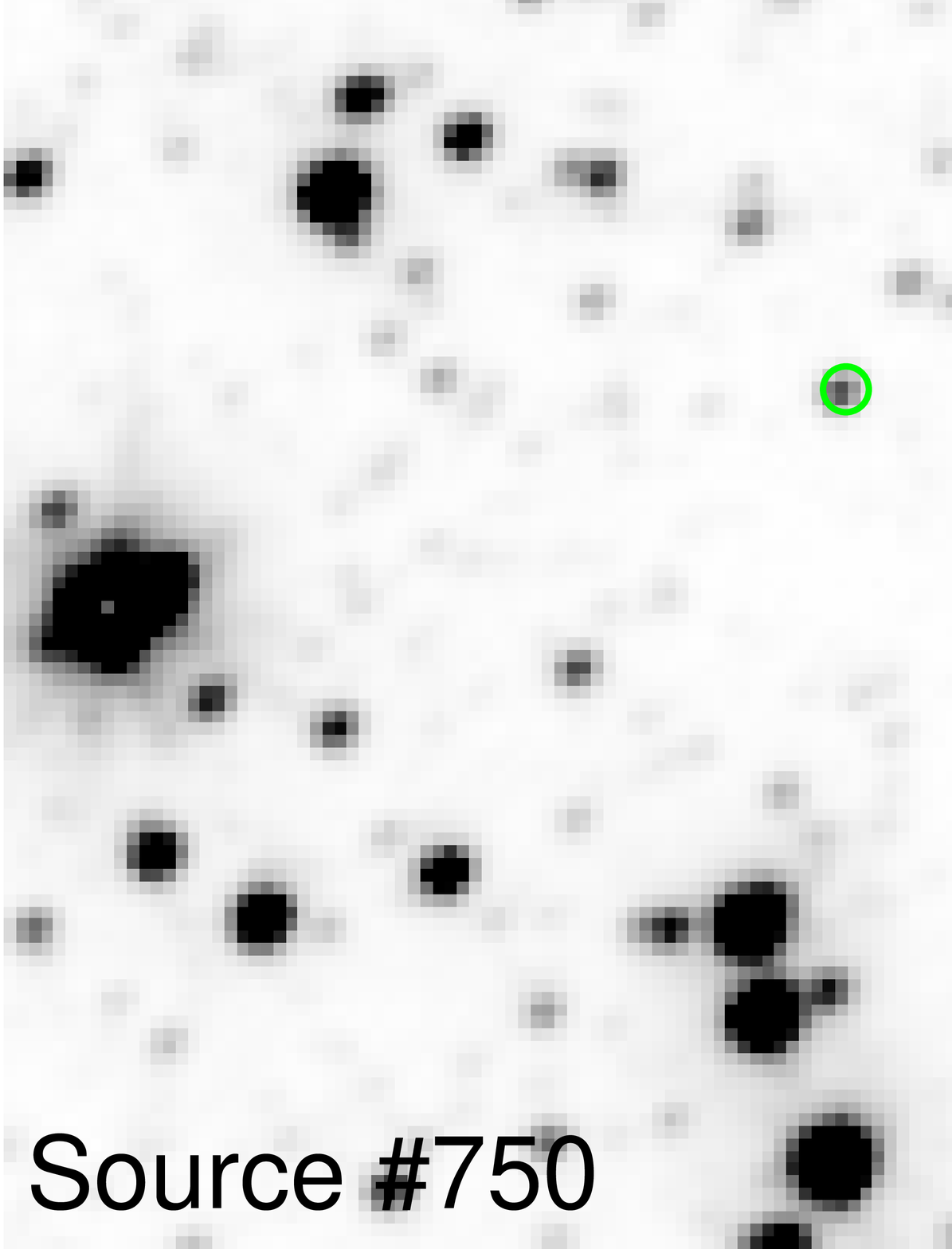}\\
\includegraphics[width=4cm]{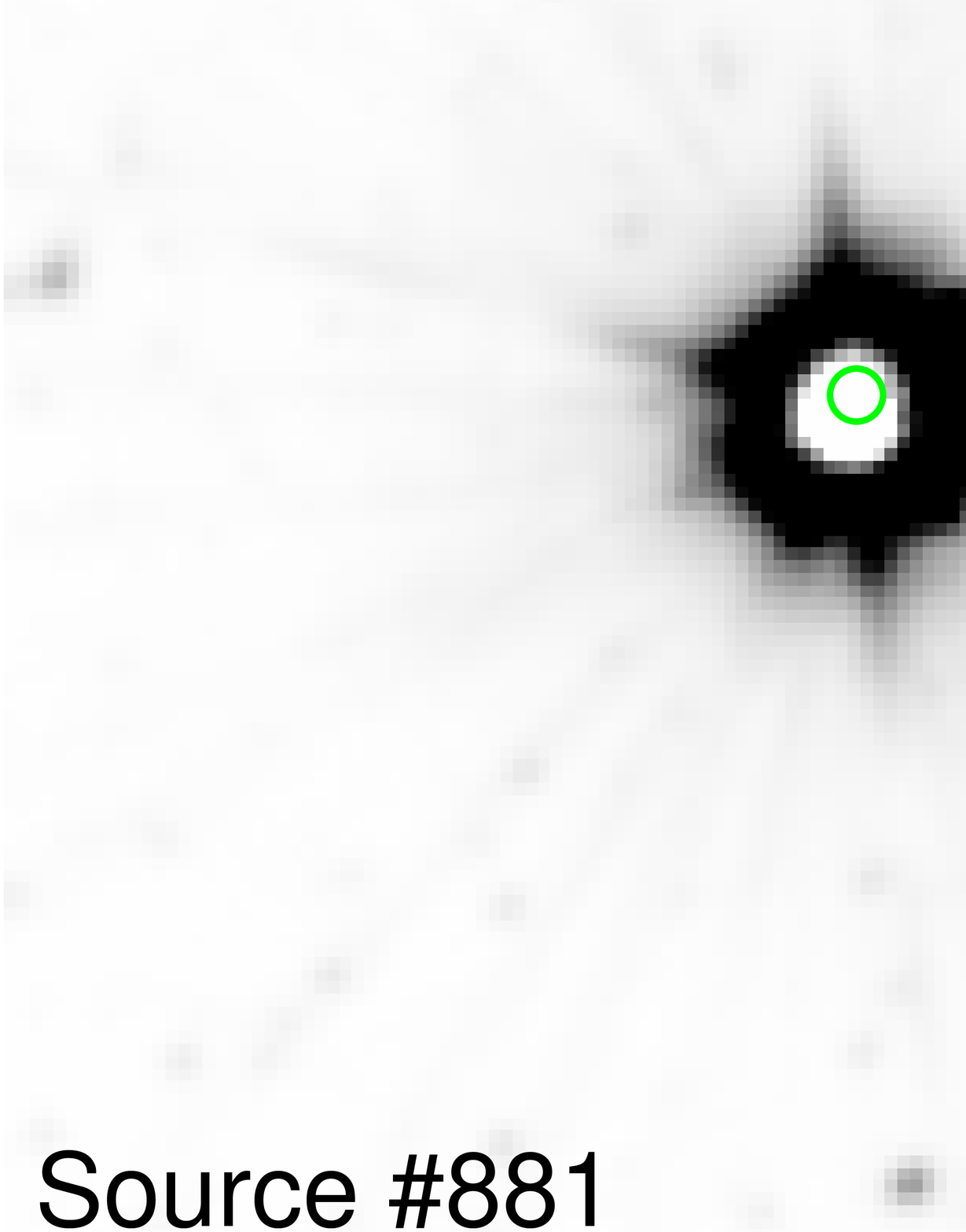}&\includegraphics[width=4cm]{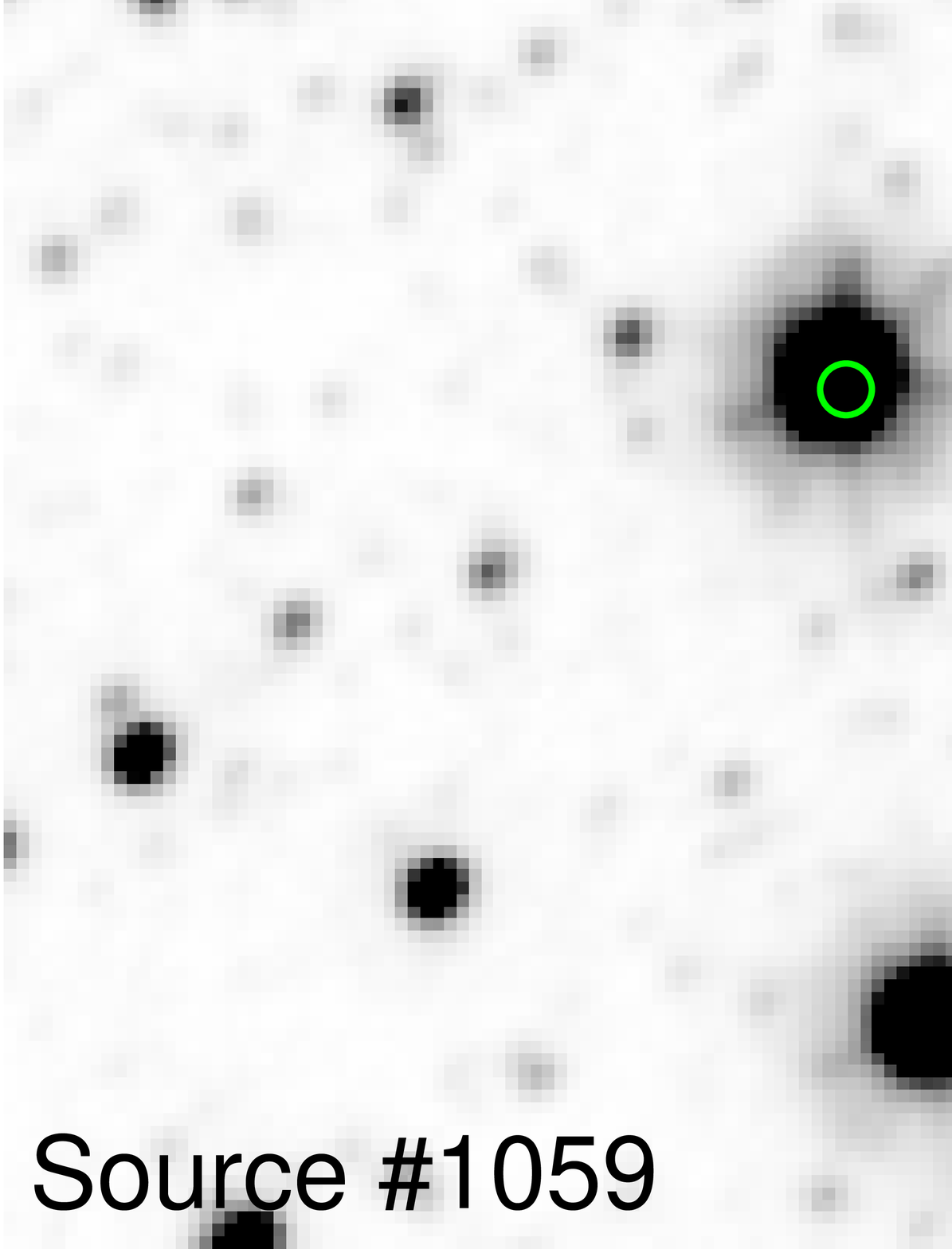}&\includegraphics[width=4cm]{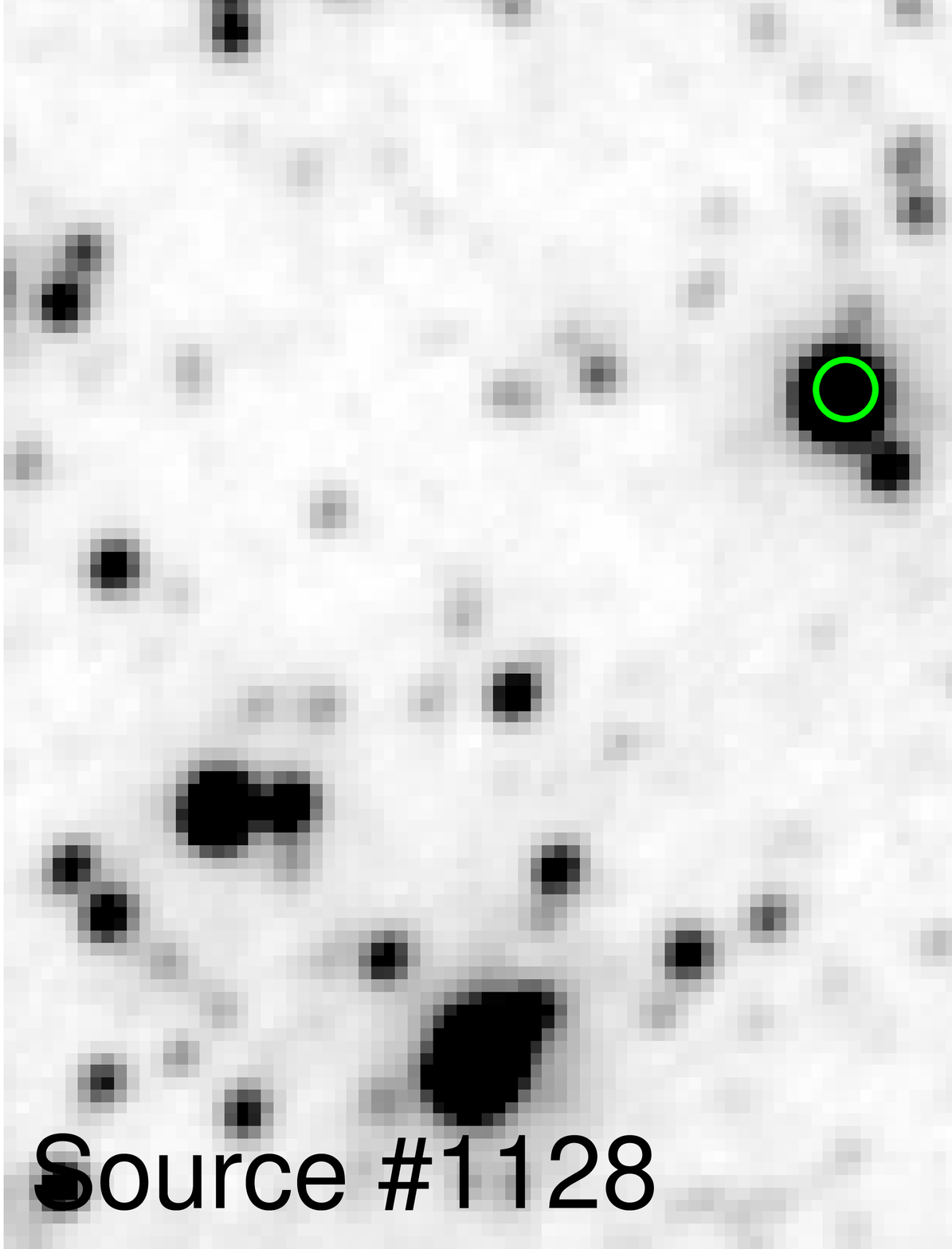}&\includegraphics[width=4cm]{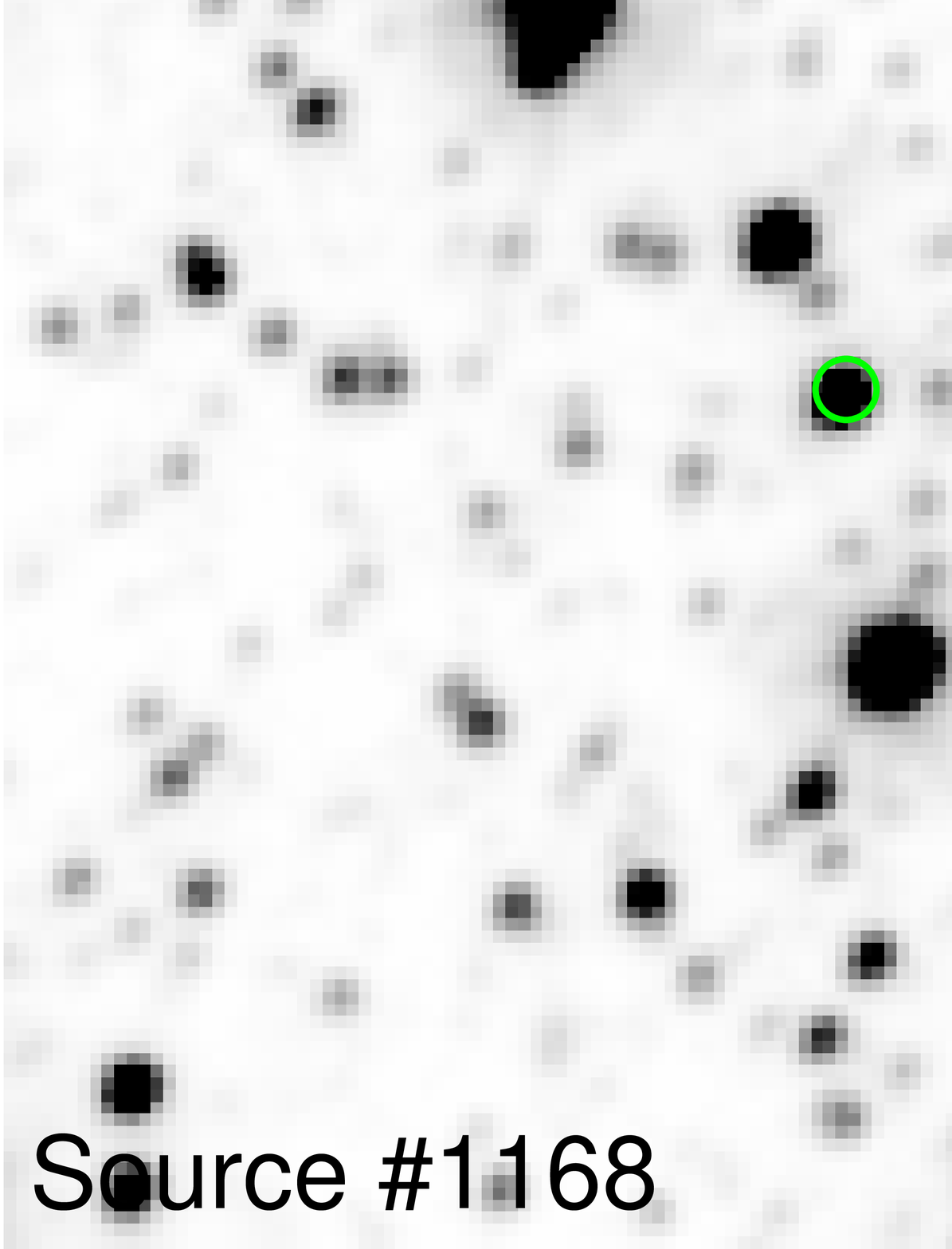}\\
\includegraphics[width=4cm]{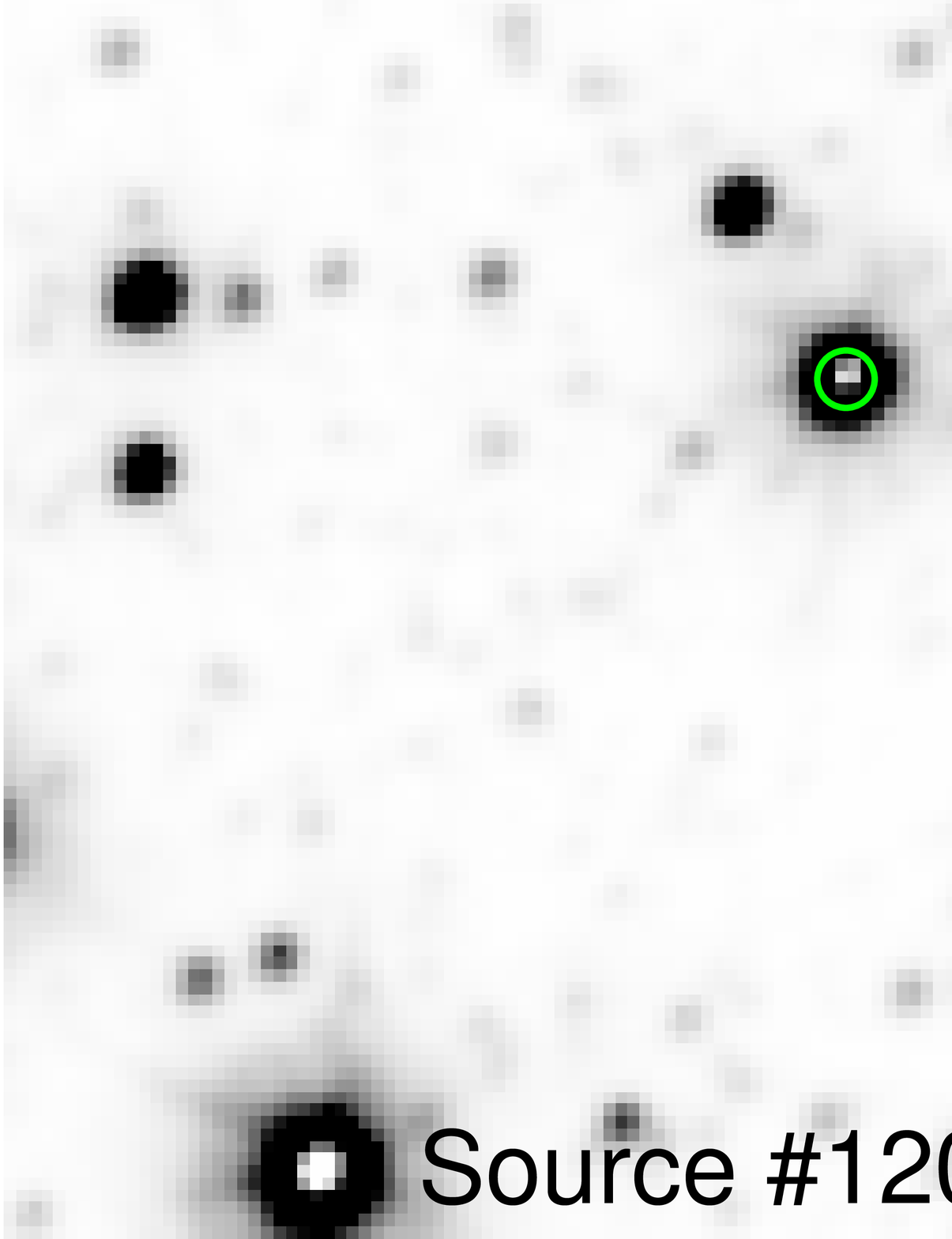}&\includegraphics[width=4cm]{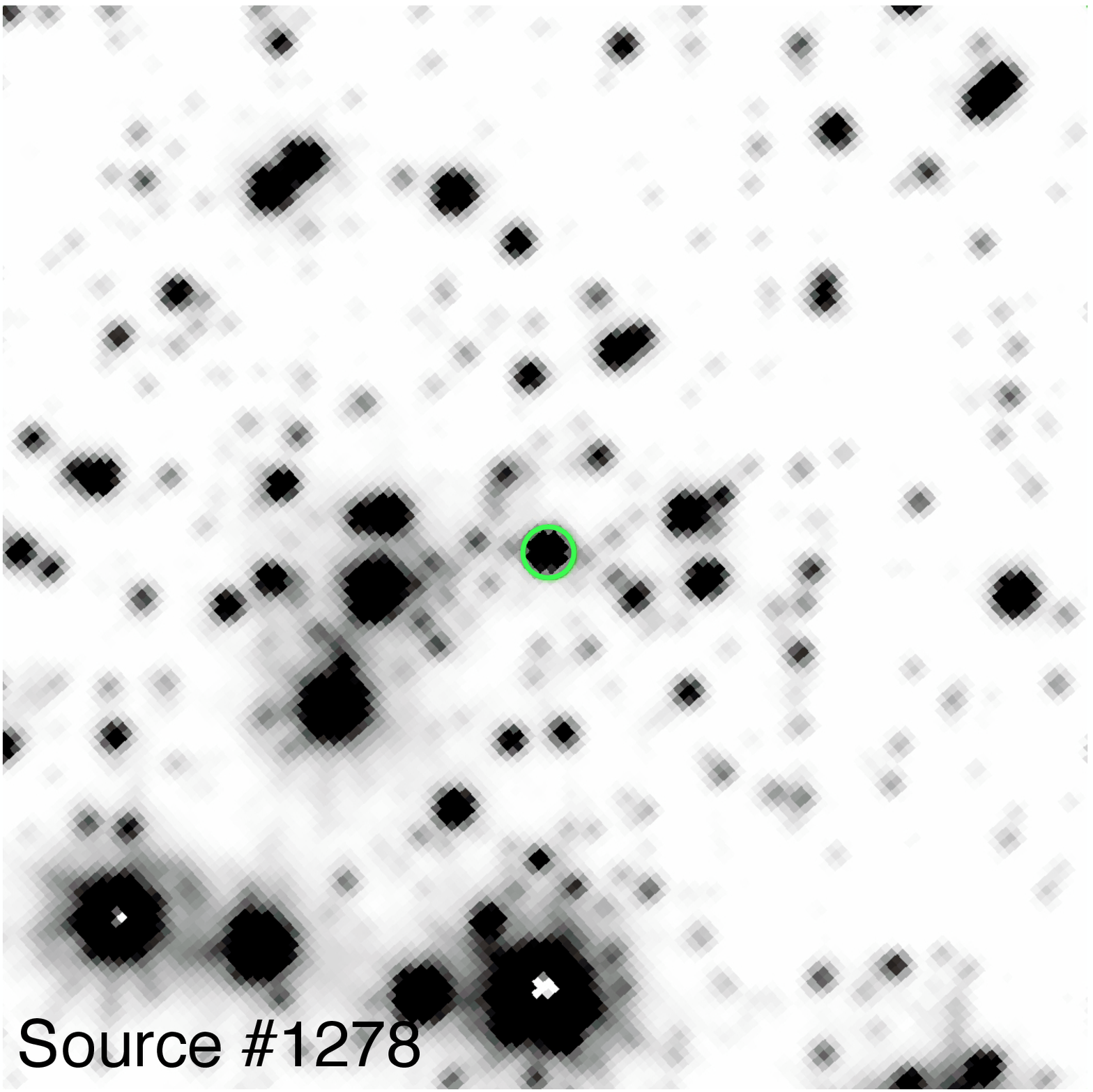}&\includegraphics[width=4cm]{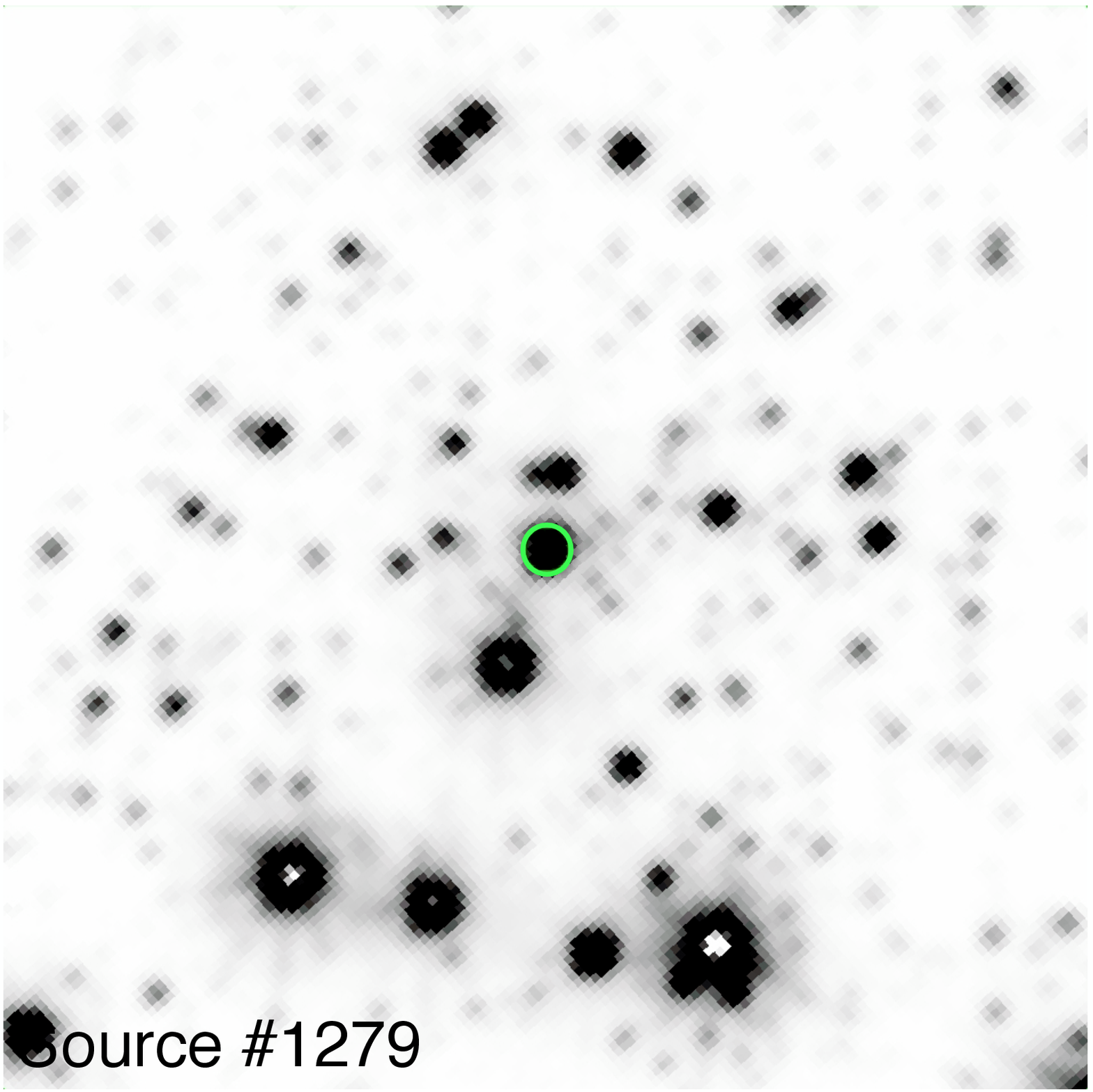}&\includegraphics[width=4cm]{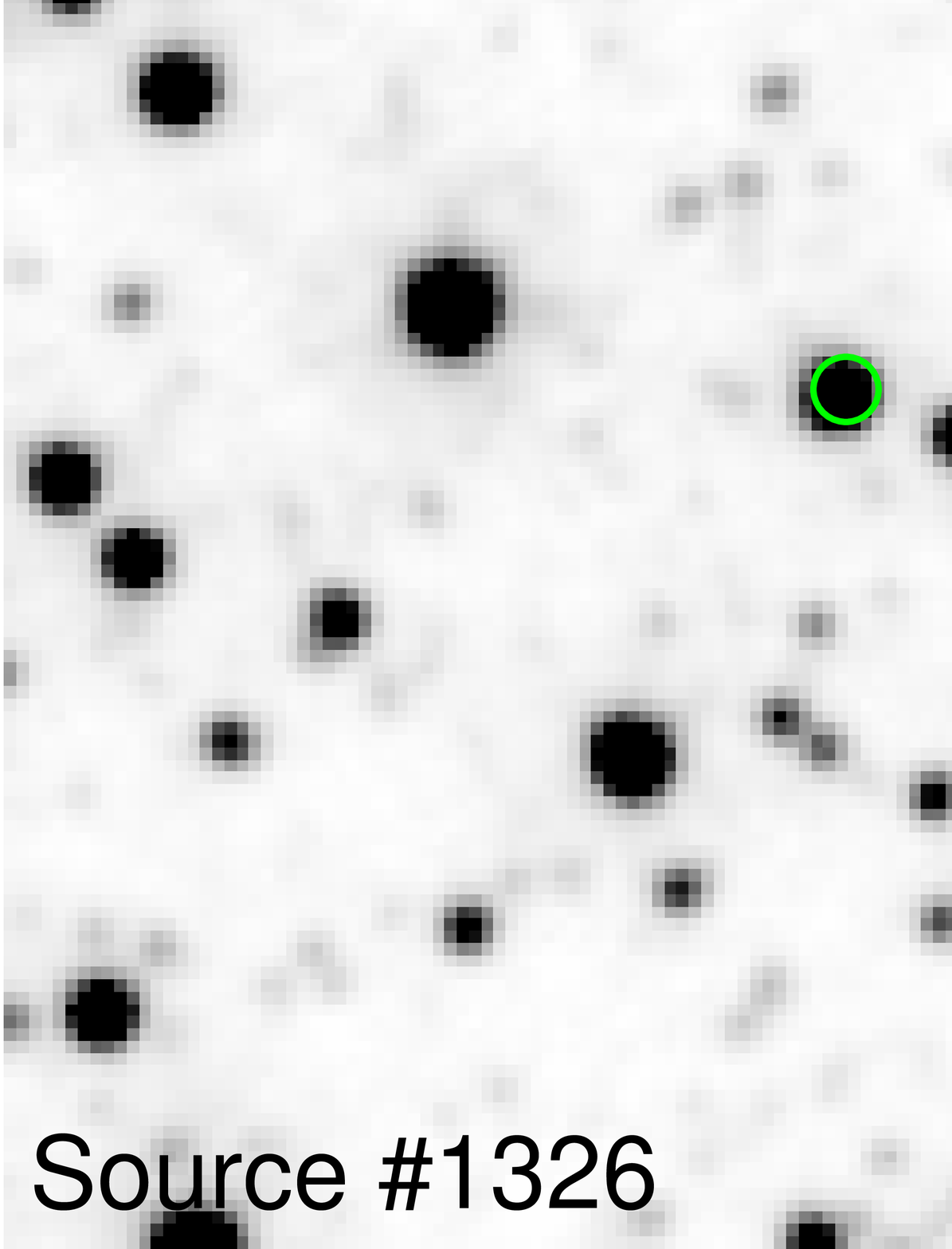}\\
\includegraphics[width=4cm]{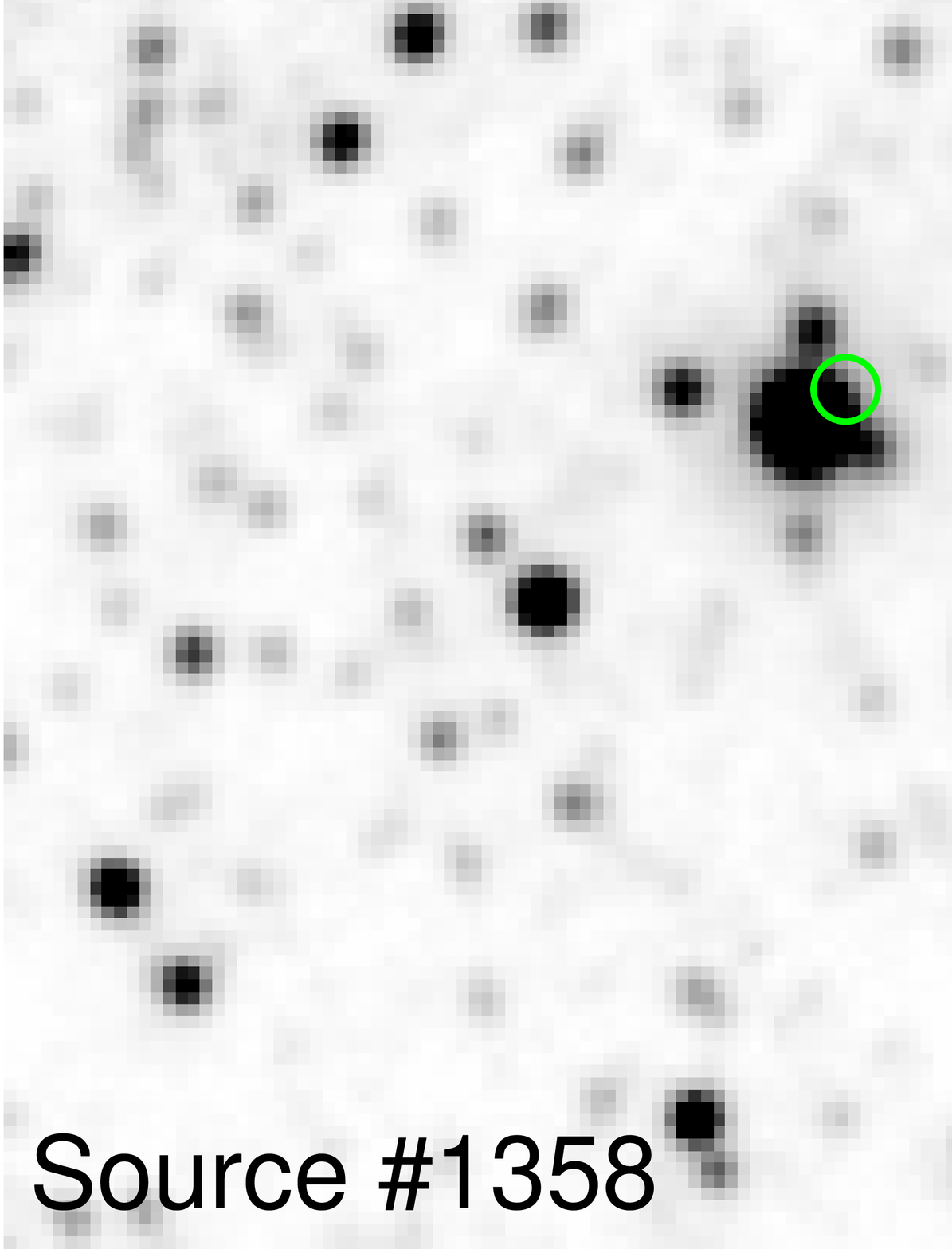}&\includegraphics[width=4cm]{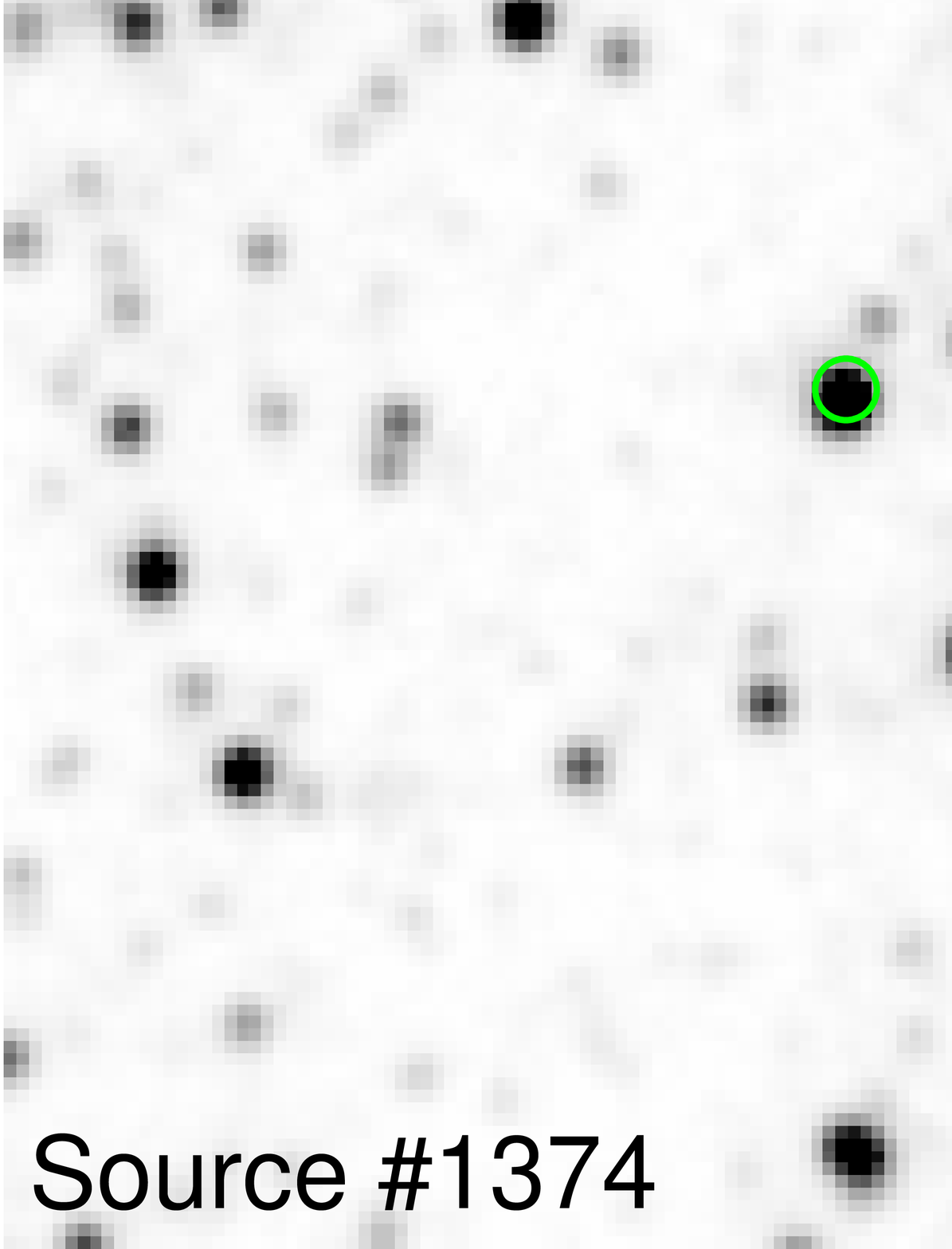}&\includegraphics[width=4cm]{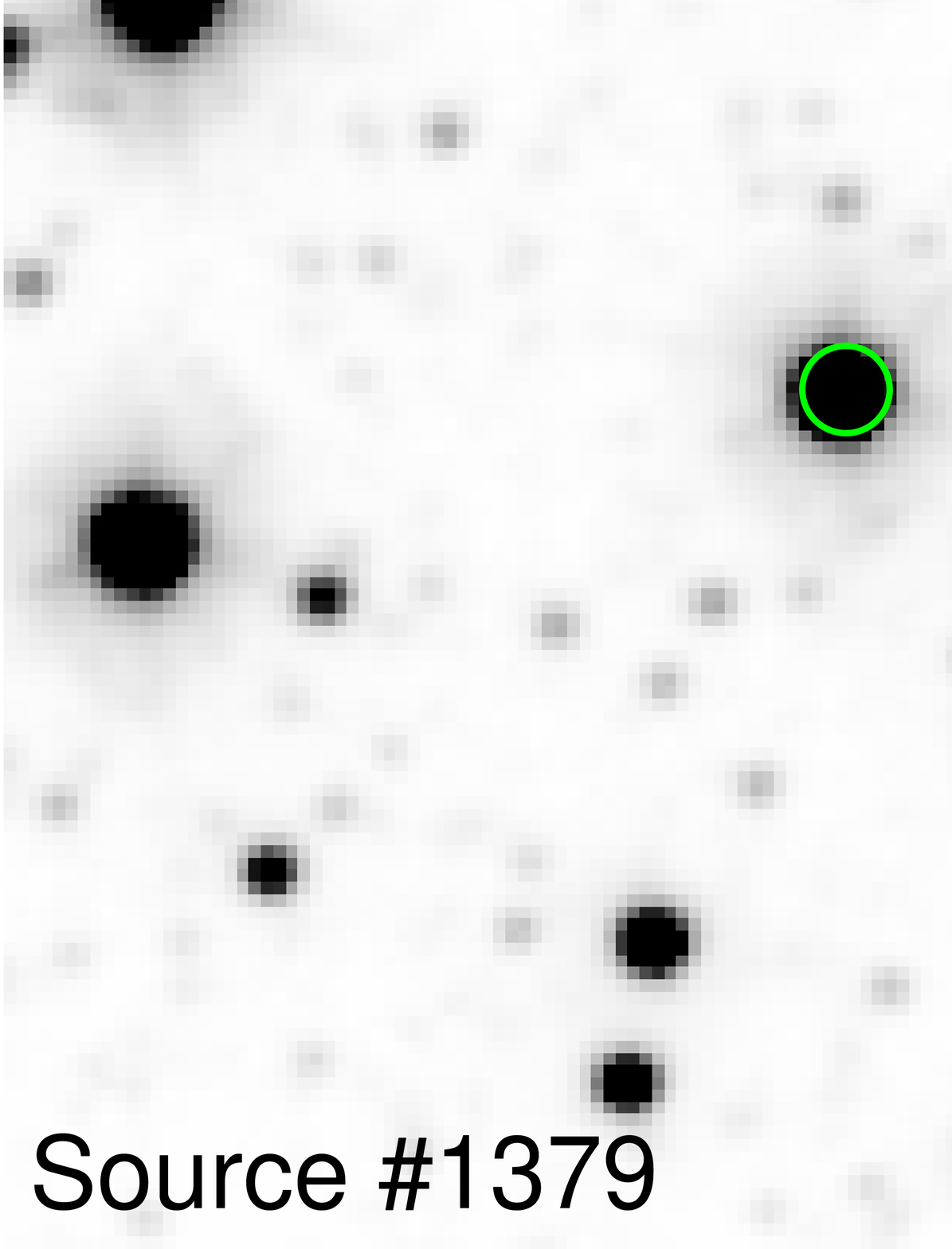}&\includegraphics[width=4cm]{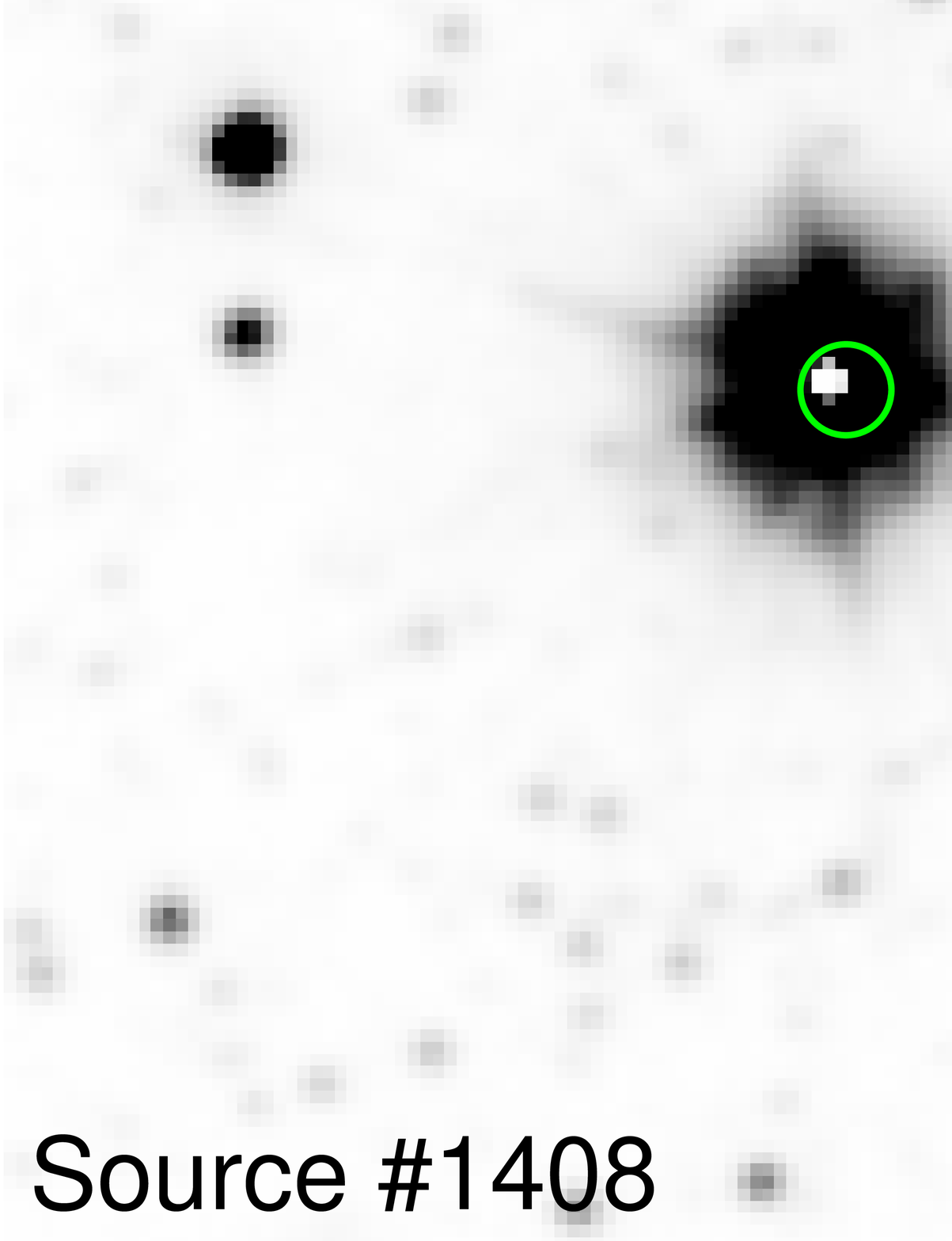}\\
\end{tabular}
\caption{VVV $Ks$-band counterparts of the newly discovered hard X-ray sources we identified through near-IR spectroscopy with OSIRIS. $Chandra$ position error circles (green) are superimposed. Field of view is 47\arcsec$\times$47\arcsec and orientation north (up) and east (left).}
\label{irfc}
\end{center}
\end{figure*}

%% file: Tables/counter_mag.table
\begin{table*}
\caption{\scriptsize List of the 20 new X-ray sources
discovered during our survey of the Norma arm with \chandra, the
companion stars of which we identified through near-IR
spectroscopy.\label{counter}}
\resizebox{0.8\textwidth}{!}{\begin{minipage}{\textwidth}
\begin{tabular}{llccccccccc}
\hline\hline
No.&IAU name&Quantile group&RA (J2000)&DEC (J2000)&$N_{\rm \ion{H}{i}+H_2}$(ISM)&$N_{\rm H}$(X-ray)&$\Gamma$&$kT$&$F_{\rm 0.5-10}$&$R$\\
&CXOU~J&&(hms)&(dms)&($\times10^{22}$cm$^{-2}$)&($\times10^{22}$ cm$^{-2}$)&&keV&($\times10^{-13}$ cgs)&\\
\hline
38&	163329.5-473332		&  D&16 33 29.58	&-47 33 32.8	& 5.9 &$21.3_{-9.0}^{+10.4}$& $2.6_{-1.3}^{+1.4}$&--& 3.05 &0.994\\
78&	163355.1-473804	   	&  D&16 33 55.11	&-47 38 04.9	& 7.0 &$2.4_{-0.9}^{+1.1}$& $0.5\pm0.4$&--& 10.0   	&0.994\\
158&	163443.0-472603	   	&  C/D&16 34 43.05	&-47 26 03.3	& 7.0 &$4.4_{-2.9}^{+1.9}$& $1.5_{-1.0}^{+1.2}$&--& 0.71	&0.990\\
239&	163515.1-472304 	&  D&16 35 15.19	&-47 23 04.0	& 8.9 &$12.7_{-6.1}^{+10.5}$& $2.7_{-1.5}^{+2.0}$&--& 0.78	&0.989\\
396&	163553.5-470037		&  D/E&16 35 53.52	&-47 00 37.1	& 4.9 &$8.3_{-3.1}^{+3.9}$& $2.8_{-1.0}^{+1.1}$&--& 1.26	&0.898\\
437&	163607.0-471434		&  C&16 36 07.03	&-47 14 34.8	& 8.4 &$2.8_{-1.6}^{+2.4}$& $1.7_{-1.0}^{+1.2}$&--& 0.58	&0.984\\
694&	163734.1-464630		&  A/B&16 37 34.11	&-46 46 30.1	& 7.7 &$0.4_{-0.4}^{+0.5}$& $2.4_{-0.7}^{+0.9}$&--& 0.50	&0.983\\
750&	163750.8-465545		&  B&16 37 50.82	&-46 55 45.3	& 9.4 &$0.11_{-0.1}^{+0.1}$& $1.2\pm0.1$&--& 10.3	 	&0.989\\
881&	163832.2-470339		&  C&16 38 32.20	&-47 03 39.2	& 8.5 &$1.3_{-0.6}^{+0.8}$& $0.8\pm0.4$&--& 2.19		&0.883\\
1059&	163925.7-465303		&  A&16 39 25.73	&-46 53 03.6	& 8.0 &$0.3_{-0.2}^{+0.4}$&--& $0.5\pm0.2$& 0.88		&0.989\\
1128&	163944.3-465614		&  C&16 39 44.39	&-46 56 14.9	& 6.4 &$2.6_{-0.9}^{+1.0}$& $2.2\pm0.6$&--& 3.82		&0.994\\
1168&	163955.2-463145		&  B&16 39 55.25	&-46 31 45.8	& 6.2 &$0.3_{-0.2}^{+0.3}$ & $1.8\pm0.4$&--& 1.19		&0.986\\
1203&	164002.4-463200 	&  B&16 40 02.43	&-46 32 00.6	& 6.2 &$0.7\pm0.3$ & $2.4_{-0.4}^{+0.4}$&--& 1.51		&0.981\\
1278&	164029.4-462328 	&  D/E&16 40 28.97	&-46 23 27.6	& 7.8 &$15.2_{-5.2}^{+8.7}$&--& $2.0_{-0.8}^{+2.0}$& 1.18	&0.986\\
1279&	164029.5-462329 	&  D/E&16 40 29.68	&-46 23 29.2	& 7.8 &$13.2_{-3.4}^{+4.4}$&--& $2.6_{-0.8}^{+2.1}$& 2.89	&0.986\\
1326&	164045.5-464607		&  D&16 40 45.54	&-46 46 07.1	& 7.3 &$13.8_{-11.2}^{+20.3}$& $1.2_{-2.0}^{+2.7}$&--& 0.79&0.991\\
1358&	164105.5-465108		&  A/B&16 41 05.54	&-46 51 08.4	& 6.0 &$0.5_{-0.3}^{+0.4}$& $2.5_{-0.5}^{+0.6}$&--& 1.28	&0.993\\
1374&	164122.8-464529 	&  A/B&16 41 22.87	&-46 45 29.4	& 6.8 &$0.4\pm0.3$& $1.5_{-0.3}^{+0.5}$&--& 2.45	 	&0.991\\
1379&	164130.8-463048 	&  A/B&16 41 30.88	&-46 30 48.8	& 7.0 &$0.01_{-0.01}^{+0.24}$& $2.3_{-0.2}^{+0.4}$&--& 4.10&0.982\\
1408&	164204.5-464341 	&  D/E&16 42 04.59	&-46 43 41.3	& 6.3 &$7.8_{-4.2}^{+7.7}$& $2.1_{-1.2}^{+1.6}$&--& 1.71	&0.983\\
\hline
\end{tabular}
\end{minipage}}
\tablefoot{\scriptsize{\bf Column 1}: source number in our catalogue. {\bf Column 2}: 
\chandra\ name. {\bf Column 3}: quantile group. {\bf Columns 4 and 5}: \chandra\ equatorial
coordinates. {\bf Column 6}: total interstellar column density of the source along its line of sight. {\bf Column 7}: X-ray 
column density of the source $N_{\rm H}$. {\bf Column 8}: Power-law photon index $\Gamma$ when a power-law model better 
describes the \chandra\ spectrum. {\bf Column 9}: Plasma temperature $kT$ in keV when a thermal plasma model better 
describes the \chandra\ spectrum. {\bf Column 10}: 0.5-10~keV unabsorbed flux $F_{\rm 0.5-10}$. {\bf Column 11}: 
Reliability $R$ of the near-infrared counterpart.}
\end{table*}

%% file: Tables/irmag.table
\begin{table*}
\caption{\scriptsize NEWFIRM and VVV counterparts of the 20 \chandra\ sources identified through near-IR spectroscopy in this study. \label{irmag}}
\resizebox{0.81\textwidth}{!}{\begin{minipage}{\textwidth}
\begin{tabular}{l|ccc|ccccccc}
\hline\hline
	&	\multicolumn{3}{c|}{NEWFIRM}&\multicolumn{7}{c}{VVV}	\\
\hline
No.&		$J$		&$H$			&$Ks$			&VVV ID.		&Dist.		&$Z$		&$Y$		&$J$		&$H$		&$Ks$		\\	
       	 &	(mag)		&(mag)			&(mag)			&			&(\arcsec)	&(mag)		&(mag)		&(mag)		&(mag)		&(mag)		\\	
\hline
38	 & 	15.29$\pm$0.07	& 11.92$\pm$0.10	& 10.13$\pm$0.06	&515727245968		&0.12		&--		&19.33$\pm$0.15	&15.21$\pm$0.01	&11.99$\pm$0.01	&Sat.	\\		
78	 & 	16.43$\pm$0.07	& 15.45$\pm$0.10	& 14.99$\pm$0.09	&515726911773		&0.06		&17.02$\pm$0.01	&16.74$\pm$0.02	&16.46$\pm$0.01	&15.77$\pm$0.04	&15.38$\pm$0.02	\\	
158	 & 	12.96$\pm$0.03	& 11.39$\pm$0.05	& 10.58$\pm$0.05	&515727009565		&0.06		&16.44$\pm$0.01	&14.88$\pm$0.01	&12.96$\pm$0.01 &11.30$\pm$0.01	&10.58$\pm$0.01	\\	
239	 & 	15.74$\pm$0.07	& 13.60$\pm$0.08	& 12.12$\pm$0.06	&515720865399		&0.13		&--		&18.42$\pm$0.08	&15.65$\pm$0.01	&13.48$\pm$0.01	&12.16$\pm$0.01	\\	
396	 & 	15.48$\pm$0.05	& 13.18$\pm$0.10	& 11.99$\pm$0.07	&515720906266		&0.96		&--		&18.16$\pm$0.06	&15.49$\pm$0.01	&13.19$\pm$0.01	&12.19$\pm$0.01	\\	
437	 & 	14.51$\pm$0.04	& 12.85$\pm$0.06	& 12.01$\pm$0.08	&515720093751		&0.42		&17.53$\pm$0.02 &15.97$\pm$0.01	&14.31$\pm$0.01	&12.82$\pm$0.01	&12.00$\pm$0.01	\\	
694	 & 	12.53$\pm$0.05	& 11.75$\pm$0.07	& 10.26$\pm$0.06	&515720390344		&0.44		&13.56$\pm$0.01	&13.17$\pm$0.01	&13.29$\pm$0.01 &12.70$\pm$0.01 &12.26$\pm$0.01   \\	
750	 & 	No Cov.		  	& No Cov.		& No Cov.	&515720338856		&0.22		&15.08$\pm$0.01	&15.06$\pm$0.01	&14.91$\pm$0.01	&14.64$\pm$0.01	&14.27$\pm$0.01	\\	
881	 & 	No Cov.		  	& No Cov.		& No Cov.	&515725172945		&0.66		&Sat.     	&Sat.     	&Sat.     	&Sat.	      	&Sat.		\\	
1059 	 &	No Cov.		  	& No Cov.		& No Cov.	&515725290815		&0.30		&Sat.     	&Sat.     	&Sat.     	&Sat.	      	&Sat.     	\\	
1128 	 &	No Cov.			& No Cov.		& No Cov.	&515725286315		&0.08		&14.71$\pm$0.01	&13.63$\pm$0.01	&12.53$\pm$0.01	&11.48$\pm$0.01	&10.76$\pm$0.01	\\	
1168 	 &	No Cov.		  	& No Cov.		& No Cov.	&515720689720		&0.26		&12.75$\pm$0.01	&12.63$\pm$0.01	&12.23$\pm$0.01	&12.14$\pm$0.01	&11.83$\pm$0.01	\\	
1203 	 &	No Cov.		  	& No Cov.		& No Cov.	&515720696577		&0.38		&Sat.     	&Sat.     	&Sat.     	&Sat.	      	&Sat.		\\	
1278 	 &	No Cov.		  	& No Cov.		& No Cov.	&515721533405		&0.18		&--		&--		&17.72$\pm$0.06	&13.61$\pm$0.01	&11.26$\pm$0.01	\\	
1279 	 &	No Cov.		  	& No Cov.		& No Cov.	&515721039612		&0.30		&--		&--		&16.02$\pm$0.01	&12.19$\pm$0.04	&Sat.	\\	
1326 	 &	No Cov.		  	& No Cov.		& No Cov.	&515725424702		&0.20		&17.59$\pm$0.02 &16.28$\pm$0.01	&14.56$\pm$0.01	&12.71$\pm$0.01	&11.48$\pm$0.01	\\	
1358 	 &	No Cov.		  	& No Cov.		& No Cov.	&515725732674		&0.13		&Sat.		&11.45$\pm$0.01	&10.96$\pm$0.01	&10.46$\pm$0.01	&10.18$\pm$0.01	\\	
1374 	 &	No Cov.		  	& No Cov.		& No Cov.	&515725467960		&0.17		&14.27$\pm$0.01	&13.84$\pm$0.01 &13.36$\pm$0.01 &12.56$\pm$0.01 &12.38$\pm$0.01   \\	
1379 	 &	No Cov.		  	& No Cov.		& No Cov.	&515725581165		&0.38		&Sat.     	&Sat.     	&Sat.     	&Sat.	      	&Sat.		\\	
1408 	 &	No Cov.		  	& No Cov.		& No Cov.	&515725521664		&0.41		&Sat.     	&Sat.     	&Sat.     	&Sat.	      	&Sat.		\\	
\hline
\end{tabular}
\end{minipage}}
\tablefoot{\scriptsize {\bf Column 1}: Source number in our catalogue. {\bf Columns 2 to 4}: 
NEWFIRM $J$, $H$, and $Ks$ magnitudes. {\bf Column 5}: IDs of the VVV counterparts. {\bf Column 6}: Distance (\arcsec) of the VVV 
counterparts to the corresponding X-ray sources. {\bf Columns 7 to 11}: VVV $Z$, $Y$, $J$, $H$, and $Ks$ magnitudes. ``Sat.'' means that the source is detected but saturated while a ``--'' means that the source is not detected in the corresponding filter.}
\end{table*}

%% file: Tables/irmag2.table
\begin{table*}
\caption{\scriptsize 2MASS and DENIS counterparts of the 20 \chandra\ sources identified through near-IR spectroscopy in this study. \label{irmag2}}
\resizebox{0.82\textwidth}{!}{\begin{minipage}{\textwidth}
\begin{tabular}{l|ccccc|ccccc}
\hline\hline
	&	\multicolumn{5}{c|}{2MASS}		&\multicolumn{5}{c}{DENIS}\\
\hline
No.&IAU name	&Dist.&$J$			&$H$			&$Ks$			&IAU name		&Dist.	&	$I$				&$J$			&$Ks$\\	
       	       &2MASS~J			&(\arcsec)&(mag)			&(mag)			&(mag)			&DENIS J				&(\arcsec)	&	(mag)			&(mag)			&(mag)\\	
\hline
38	 &16332956-4733327	&0.14	&15.26$\pm$0.04	&11.75$\pm$0.03	&9.99$\pm$0.022	&163329.6-473332	&0.29		&	--				&15.30$\pm$0.15	&10.01$\pm$0.05\\
78	 &--    				&--		&--				&--	    			&--	    		       	&--	 				&--			&	--				&--				&--\\
158	 &16344306-4726033	&0.17	&12.63$\pm$0.03	&11.26$\pm$0.02	&10.54$\pm$0.02	&163443.0-472603	&0.31		&	17.38$\pm$0.18	&12.97$\pm$0.07	&10.54$\pm$0.06\\
239	 &16351519-4723042	&0.25	&>15.97			&>13.77 		     	&12.27$\pm$0.04	&--	 				&--   		&	--				&--				&--\\
396	 &16355350-4700381	&1.02	&>15.69			&13.30$\pm$0.03	&12.20$\pm$0.03	&--	 				&--   		&	--				&--				&--\\
437	 &--				&--		&--     		      	&--	    			&--				&163607.0-471434	&0.45		&	--				&14.54$\pm$0.11	&12.04$\pm$0.10\\
694	 &16373414-4646299	&0.40	&12.96$\pm$0.04	&12.15$\pm$0.03	&11.89$\pm$0.04	&163734.1-464630	&0.16		&	14.02$\pm$0.11	&12.59$\pm$0.12	&11.50$\pm$0.12\\
750	 &16375079-4655452	&0.26	&14.69$\pm$0.03	&14.57$\pm$0.06	&>15.19		     	&163750.7-465545	&0.54		&	14.66$\pm$0.11	&14.30$\pm$0.15	&--\\
881	 &16383217-4703397	&0.68	&6.89$\pm$0.02	&5.44$\pm$0.04	&4.72$\pm$0.02	&163832.1-470339	&0.50		&	11.07$\pm$0.03	&6.38$\pm$0.09	&3.87$\pm$0.16\\
1059 &16392581-4653002	&3.50	&9.30$\pm$0.02	&8.69$\pm$0.03	&8.43$\pm$0.02	&163925.7-465300	&3.05		&	0.65$\pm$0.03	&9.32$\pm$0.07	&8.49$\pm$0.05\\
1128 &16394439-4656149	&0.09	&12.42$\pm$0.02	&11.26$\pm$0.02   &10.86$\pm$0.02	&163944.3-465614	&0.13		&	15.40$\pm$0.05	&12.41$\pm$0.07	&10.77$\pm$0.07\\
1168 &16395525-4631454	&0.27	&12.22$\pm$0.03	&11.97$\pm$0.06	&11.83$\pm$0.04	&163955.2-463145	&0.24		&	12.91$\pm$0.02	&12.30$\pm$0.08	&11.68$\pm$0.09\\
1203 &16400244-4632002	&0.40	&9.60$\pm$0.02	&8.84$\pm$0.02	&8.51$\pm$0.02   	&164002.4-463200	&0.38		&	11.19$\pm$0.02	&9.59$\pm$0.06	&8.46 $\pm$0.05\\
1278 &16402896-4623276	&0.04	&>15.68  	      	&13.63$\pm$0.12  	&11.29$\pm$0.07	&164029.0-462327	&1.28		&	17.83$\pm$0.17	&--				&11.13$\pm$0.08\\
1279 &16402966-4623295	&0.36	&>15.16        		&>11.48	    		&9.84$\pm$0.05	&164029.6-462329	&0.75		&	--				&15.81$\pm$0.23	&9.69$\pm$0.06\\
1326 &16404554-4646072	&0.14	&14.55$\pm$0.04	&12.73$\pm$0.04	&11.60$\pm$0.04	&164045.5-464607	&0.46		&	--				&14.42$\pm$0.11	&11.73$\pm$0.08\\
1358 &16410554-4651085	&0.15	&11.04$\pm$0.03	&10.39$\pm$0.02	&10.16$\pm$0.02	&164105.5-465108	&0.18		&	12.22$\pm$0.02	&11.07$\pm$0.07	&10.13$\pm$0.06\\
1374 &16412285-4645292	&0.25	&13.34$\pm$0.02	&12.57$\pm$0.02	&12.33$\pm$0.03	&164122.8-464529	&0.24		&	14.77$\pm$0.04   	&13.44$\pm$0.10   &12.27$\pm$0.11\\
1379 &16413093-4630488	&0.48	&9.76$\pm$0.02	&9.30$\pm$0.02     &9.13$\pm$0.02   	&164130.9-463048	&0.33		&	10.55$\pm$0.03	&9.83$\pm$0.06	&9.14$\pm$0.05\\
1408 &16420462-4643411	&0.41	&8.24$\pm$0.03	&6.73$\pm$0.04	&5.87$\pm$0.03   	&164204.6-464341	&0.33		&	14.13$\pm$0.04   	&8.53$\pm$0.08	&6.01$\pm$0.03\\	
\hline
\end{tabular}
\end{minipage}}
\tablefoot{\scriptsize {\bf Column 1}: Source number in our catalogue. {\bf Column 2}: IAU names of the 2MASS counterparts. 
{\bf Column 3}: Distance (\arcsec) of the 2MASS counterparts to the corresponding X-ray sources. {\bf Columns 4 to 6}: 2MASS $J$, $H$, and $Ks$ magnitudes. {\bf Column 7}: IAU names of the DENIS counterparts. {\bf Column 8}: Distance (\arcsec) of the DENIS 
counterparts to the corresponding X-ray sources. {\bf Columns 9 to 11}: DENIS $I$, $J$, and $Ks$ magnitudes. A ``--'' means that the source is not detected in the corresponding filter.}
\end{table*}

%% file: Tables/irmag3.table
\begin{table*}
\caption{\scriptsize GLIMPSE and {\it WISE} counterparts of the 20 \chandra\ sources identified through near-IR spectroscopy in this study. \label{irmag3}}
\setlength{\tabcolsep}{2pt}
\resizebox{0.78\textwidth}{!}{\begin{minipage}{\textwidth}
\begin{tabular}{l|cccccc|cccccc}
\hline\hline
	&	\multicolumn{6}{c|}{GLIMPSE}&\multicolumn{6}{c}{WISE}\\
\hline
No.&IAU Name&Dist.&$F_{\rm 3.6}$&$F_{\rm 4.5}$&$F_{\rm 5.8}$&$F_{\rm 8.0}$&IAU Name&Dist.&$F_{\rm 3.4}$&$F_{\rm 4.6}$&$F_{\rm 12}$&$F_{\rm 22}$\\	
  &				&(\arcsec)&(mJy)&(mJy)&(mJy)&(mJy)&WISE J&(\arcsec)&(mJy)&(mJy)&(mJy)&(mJy)\\	
\hline
38	 &G336.7280+00.2037&0.28&85.79$\pm$5.25&53.95$\pm$2.36&54.59$\pm$2.46&26.23$\pm$2.19&163329.53-473331.8&1.09&80.44$\pm$2.07&64.35$\pm$1.48&--&--\\
78	 &-- &--&--&--&--&--&--&--&--&--&--&--\\
158	 &G336.9601+00.1363&0.18&27.93$\pm$1.17&17.88$\pm$0.98&12.76$\pm$0.71&7.80$\pm$0.42&163443.04-472603.7&0.43&29.56$\pm$0.92&16.75$\pm$0.86&--&--\\
239	 &G337.0580+00.1032&0.14&13.02$\pm$0.58&11.95$\pm$0.62&10.51$\pm$0.50&--&--&--&--&--&--&--\\
396	 &G337.4072+00.2754&0.15&8.21$\pm$0.20&5.25$\pm$0.30&4.86$\pm$0.22&--&--&--&--&--&--&--	 \\
437	 &G337.2613+00.0906&0.30&6.47$\pm$0.49&4.54$\pm$0.46&2.56$\pm$0.38&--&--&--&--&--&--&--	 \\
694	 &G337.7742+00.2214&0.10&7.20$\pm$0.27&4.59$\pm$0.26&3.08$\pm$0.25&--&--&--&--&--&--&--	 \\
750	 &--&--&--&--&--&--&--&--&--&--&--&--\\
881	 &G337.6719-00.0926&0.21&4486$\pm$378&2552$\pm$114&2420$\pm$40&1348$\pm$23&163832.16-470339.7&0.11&$5681\pm493$&$4213\pm220$&$634.6\pm17.2$&$299.1\pm58.4$\\
1059 &G337.9058-00.0881&1.31&$155.3\pm3.1$&$100.8\pm3.0$&$72.17\pm1.34$&$38.52\pm0.86$&163925.74-465303.2&3.11&$154.9\pm3.2$&$94.36\pm2.12$&$52.03\pm2.02$&$115.4\pm5.4$\\
1128 &G338.2265+00.0846&0.15&18.68$\pm$0.08&11.15$\pm$0.04&8.10$\pm$0.03&5.20$\pm$0.03&163944.37-465615.0&0.22&12.91$\pm$0.02&7.31$\pm$0.41&--&--\\
1168 &G337.9008-00.1634&0.20&5.92$\pm$0.37&4.11$\pm$0.34&2.23$\pm$0.27&--&--&--&--&--&--&--\\
1203 &G338.2371+00.0665&0.20&$131.4\pm6.0$&$73.45\pm3.65$&$52.48\pm1.95$&$29.84\pm0.84$&164002.45-463159.9&0.31&$155.4\pm3.9$&$79.45\pm2.36$&$20.23\pm1.48$&--\\
1278 &G338.3941+00.1040&0.21&36.24$\pm$2.68&47.31$\pm$3.37&36.51$\pm$2.28&2.03$\pm$1.43&--&--&--&--&--	&--	 \\
1279 &G338.3951+00.1022&0.46&$146.6\pm19.3$&$181.6\pm7.9$&$163.9\pm5.6$&$103.7\pm3.6$&--&--&--&--&--&--\\
1326 &G338.1427-00.1817&0.15&16.19$\pm$0.69&15.67$\pm$0.56	&13.86$\pm$0.70&6.76$\pm$0.36&--&--&	--&--&--&--	 \\
1358 &G338.1178-00.2799&0.21&26.30$\pm$2.28&17.62$\pm$0.64	&12.13$\pm$0.70&6.57$\pm$0.17&164105.56-465108.5&0.13&28.49$\pm$0.76&15.15$\pm$0.44&--&--\\
1374 &G338.2212-00.2546&0.33&4.25$\pm$0.28&2.26$\pm$0.31&--&2.51$\pm$0.52&--&--	&--&--&--&--\\
1379 &G338.4200-00.1104&0.23&60.68$\pm$2.85&37.25$\pm$1.75&27.72$\pm$1.13&16.73$\pm$0.64&164130.91-463048.9&0.27&72.89$\pm$1.81&39.82$\pm$1.02&--&--\\
1408 &G338.3224-00.3244&0.16&$3120\pm78$&$2353\pm100$&$2424\pm76$&$1601\pm43$&164204.62-464341.1&0.04&$1593\pm84$&$2113\pm73$&$653.8\pm1.1$&$1662\pm93$\\	
\hline
\end{tabular}
\end{minipage}}
\tablefoot{\scriptsize {\bf Column 1}: Source number in our catalogue. {\bf Column 2}: IAU names of the GLIMPSE counterparts. 
{\bf Column 3}: Distance (\arcsec) of the GLIMPSE counterparts to the corresponding X-ray sources. {\bf Columns 4 to 7}: GLIMPSE 3.6~\mic, 4.5~\mic, 5.8~\mic, and 8.0~\mic\ flux densities (mJy). 
{\bf Column 8}: IAU names of the {\it WISE} counterparts. {\bf Column 9}: Distance (\arcsec) of the {\it WISE} 
counterparts to the corresponding X-ray sources. {\bf Columns 10 to 13}: {\it WISE} {\it W1} (3.4~\mic), {\it W2} (4.6~\mic), {\it W3} (12~\mic), and {\it W4} (22~\mic) flux densities (mJy). A ``--'' means that the source is not detected in the corresponding filter.}
\end{table*}

%% file: Tables/lines_hm_new.table
\begin{table*}
\caption{\scriptsize $H$- and $K$-band {\bf spectral} features of hydrogen-like lines dominated sources.}
\label{hmlines}
\resizebox{0.7\textwidth}{!}{\begin{minipage}{\textwidth}
\begin{tabular}{l|cc|cc|cc|cc|cc|cc}
\hline\hline
&\multicolumn{2}{c|}{Source 239}&\multicolumn{2}{c|}{Source 750}&\multicolumn{2}{c|}{Source 1168}&\multicolumn{2}{c|}{Source 1278}&\multicolumn{2}{c|}{Source 1279}&\multicolumn{2}{c}{Source 1326}\\
&$\mathring{W}^a$&$FWHM^b$&$\mathring{W}$&$FWHM$&$\mathring{W}$&$FWHM$&$\mathring{W}$&$FWHM$&$\mathring{W}$&$FWHM$&$\mathring{W}$&$FWHM$\\
&(\AA)&(\kms)&(\AA)&(\kms)&(\AA)&(\kms)&(\AA)&(\kms)&(\AA)&(\kms)&(\AA)&(\kms)\\
\hline
\ion{H}{i}~15087~\AA	&  $-$		&$-$	& $-$		& $-$		& $-$       & $-$           & $-$        &$-$          &$-$ 		& $-$          &*             & *\\
\ion{H}{i}~15137~\AA	&  $-$		&$-$	& $-$		& $-$		& $-$       & $-$           & $-$        &$-$          &$-$ 		& $-$	   &-4.71$\pm$0.59&348$\pm$76\\
\ion{H}{i}~15196~\AA	&  *			& *		&  *	  		& *			& $-$       & $-$           & $-$        &$-$          &$-$ 		& $-$	   &-4.67$\pm$0.46&259$\pm$73\\
\ion{H}{i}~15265~\AA	&  *			& *		& -3.59$\pm$0.99	& 508$\pm$216		& $-$       & $-$           & $-$        &$-$          &$-$		& $-$	   &-3.82$\pm$0.51&277$\pm$105\\
\ion{H}{i}~15346~\AA	&  -6.21$\pm$0.61	&369$\pm$70  	& 	*	  	& *			& $-$       & $-$ 	    &-4.65$\pm$0.91	& 370$\pm$91& $-$        &$-$           	   &-6.65$\pm$0.57&329$\pm$57\\
\ion{Fe}{ii}~15403~\AA	&  *			&*		& $-$	  	& $-$		& $-$       & $-$ 	    &-4.29$\pm$0.53		& 355$\pm$111	   & $-$        &$-$          &*             &*\\
\ion{H}{i}~15443~\AA	&  -10.61$\pm$0.99	&589$\pm$92	& 	*	  	& *			& $-$       & $-$ 	    & $-$	     &$-$          &$-$			& $-$	   &-4.71$\pm$0.59&304$\pm$84\\
\ion{H}{i}~15561~\AA	&  -8.98$\pm$0.48 	&395$\pm$38  	& -2.60$\pm$1.15	& 374$\pm$224		& 8.37$\pm$1.71 & 1073$\pm$202	    & $-$        &$-$          &$-$ 		& $-$	   &12.28$\pm$0.88&371$\pm$49\\
\ion{H}{i}~15705~\AA	&  -5.31$\pm$0.51 	&253$\pm$61  	& 	*	  	& *			& $-$       & $-$ 	    & $-$        &$-$          &$-$ 		& $-$	   &-7.91$\pm$0.52&344$\pm$46\\
\ion{Fe}{ii}~15778~\AA	&  -1.69$\pm$0.62	&258$\pm$196 	& $-$	  	& $-$		& $-$       & $-$ 	    & $-$        &$-$          &$-$			& $-$	   &-2.62$\pm$0.98&556$\pm$298\\
$[$\ion{Fe}{ii}$]$~15842~\AA&  -5.29$\pm$0.63	&416$\pm$86  	& $-$	  	& $-$		& $-$       & $-$ 	    & $-$        &$-$          &$-$			& $-$	   &-3.57$\pm$0.35&524$\pm$88\\
\ion{H}{i}~15885~\AA	&  -7.65$\pm$0.61	&268$\pm$45  	& 	*	  	& *			& 2.62$\pm$0.42	& 720$\pm$233 	    &-7.73$\pm$0.92		& 975$\pm$142	   & -3.34$\pm$0.66 &841$\pm$211      &-9.09$\pm$0.53&409$\pm$45\\
$[$\ion{Fe}{ii}$]$~15954~\AA&  -1.41$\pm$0.77 	&324$\pm$75  	& $-$	  	& $-$		& $-$ 	& $-$ 	    &-4.12$\pm$0.65		& 446$\pm$111	   & $-$	     &$-$          &$-$       &$-$\\
\ion{He}{i}~16008~\AA	&  $-$		&$-$     	& $-$	  	& $-$		& $-$	& $-$	    &-4.77$\pm$0.85		& 604$\pm$140	   & $-$	     &$-$          &$-$	    &$-$\\
\ion{H}{i}~16114~\AA 	&  -8.15$\pm$0.54	&374$\pm$48  	& -4.64$\pm$1.94	& 384$\pm$224		& 4.53$\pm$0.96	& 1184$\pm$311 	    &*			& *	   	   & -3.98$\pm$0.94 &680$\pm$203      &-5.01$\pm$0.57&313$\pm$79\\
\ion{H}{i}~16412~\AA	&  -7.70$\pm$0.50	&332$\pm$33  	& -2.29$\pm$0.60	& 259$\pm$166		& 4.79$\pm$0.82	& 850$\pm$168 	    &-1.88$\pm$0.77		& 500$\pm$291	  & -5.34$\pm$0.87 &872$\pm$169       &-9.23$\pm$0.71&429$\pm$35\\
\ion{H}{i}~16811~\AA	&  -11.1$\pm$0.79	&507$\pm$40  	& -5.04$\pm$0.65	& 489$\pm$95		& 4.31$\pm$0.70	& 762$\pm$257 	    &*			& *	   	   & -6.11$\pm$1.29 &1077$\pm$251     &-9.03$\pm$0.62&379$\pm$36\\
\ion{Fe}{ii}~16878~\AA	&  -6.00$\pm$0.65	&540$\pm$80  	& $-$	  	& $-$		& $-$	& $-$ 	    &$-$			& $-$	   & $-$	     &$-$          &-5.01$\pm$0.96&590$\pm$170\\
\ion{He}{i}~16919~\AA	&  $-$		&$-$     	& $-$	  	& $-$		& $-$	& $-$ 	    &*			& *	   	   & $-$	     &$-$          &*             &*\\
\ion{He}{i}+$[$\ion{Fe}{ii}$]$~17007~\AA&$-$&$-$	& -1.71$\pm$0.35	& 183$\pm$64		& $-$	& $-$ 	    &-3.32$\pm$0.52		& 297$\pm$87 	   & -11.0$\pm$1.12 &1458$\pm$217     &-1.94$\pm$1.19&504$\pm$362\\
$[$\ion{Fe}{ii}$]$~17110~\AA& *			& *	     	& $-$	  	& $-$		& $-$       & $-$ 	    &$-$			& $-$	   & $-$        &$-$          & *            & *			\\
\ion{H}{i}~17367~\AA	&  -7.56$\pm$0.52	&432$\pm$56  	& -9.26$\pm$0.89	& 799$\pm$92		& 6.89$\pm$1.00	& 1233$\pm$201 	    &-5.72$\pm$1.31		& 726$\pm$189	   & -12.0$\pm$1.11 &1209$\pm$135     &-8.22$\pm$0.75&540$\pm$60\\
$[$\ion{Fe}{ii}$]$~17455~\AA& *			& *	     	& $-$	  	& $-$		& $-$       & $-$ 	    &$-$			& $-$	   & $-$        &$-$          & $-$	    & $-$		\\
$[$\ion{Fe}{ii}$]$~20456~\AA& *			& *	     	& $-$	  	& $-$		& $-$       & $-$ 	    &$-$			& $-$	   & $-$        &$-$          & $-$	    & $-$		\\
\ion{He}{i}~20586~\AA	&  -6.50$\pm$0.93	&514$\pm$105 	& -17.8$\pm$0.80	& 532$\pm$37		& $-$	& $-$ 	    &-10.1$\pm$1.57		& 1105$\pm$171	   & *	             &*                &-7.68$\pm$0.69&304$\pm$57\\
\ion{He}{i}~21127~\AA	&  $-$		&$-$     	& $-$		& $-$		& $-$	& $-$ 	    &-7.71$\pm$0.91		& 593$\pm$98 	   & -11.4$\pm$1.36 &761$\pm$86       &$-$       &$-$\\
\ion{Mg}{ii}~21376~\AA	&  $-$		&$-$     	& $-$		& $-$		& 2.97$\pm$0.53	& 410$\pm$188 	    & $-$	     &$-$ 	       &$-$		& $-$ 	   &$-$       &$-$\\
\ion{H}{i}~21661~\AA	&  -10.1$\pm$0.65	&159$\pm$53  	& -42.6$\pm$1.14	& 750$\pm$19		& 12.4$\pm$1.19	& 1199$\pm$130 	    &-29.4$\pm$1.33		& 1175$\pm$52 	   & -45.1$\pm$1.34 &1318$\pm$56      &-10.9$\pm$0.72&220$\pm$41\\
\ion{He}{ii}~21891~\AA	&  $-$		&$-$     	& $-$		& $-$		& $-$	& $-$ 	    &-6.90$\pm$1.27		& 570$\pm$113	   & -8.57$\pm$1.01 &533$\pm$73       &$-$       &$-$\\
\ion{Na}{i}~22090~\AA	& *			& *	     	& $-$	  	& $-$		& $-$       & $-$ 	    & $-$        &$-$          &$-$			& $-$	   & *            & *			\\
\hline
\end{tabular}
\end{minipage}}
\tablefoot{\scriptsize A ``*'' means that the line is detected but difficult to properly measure, while a ``--'' means that the line is not detected.\\
\tablefoottext{a}{Equivalent width (\AA)}\\
\tablefoottext{b}{Full-width at half-maximum (\kms).}}
\end{table*}

%% file: Figures/sp_ir_hmxb_em.figures
\begin{figure*}
\begin{tabular}{cc}
\includegraphics[width=8.5cm]{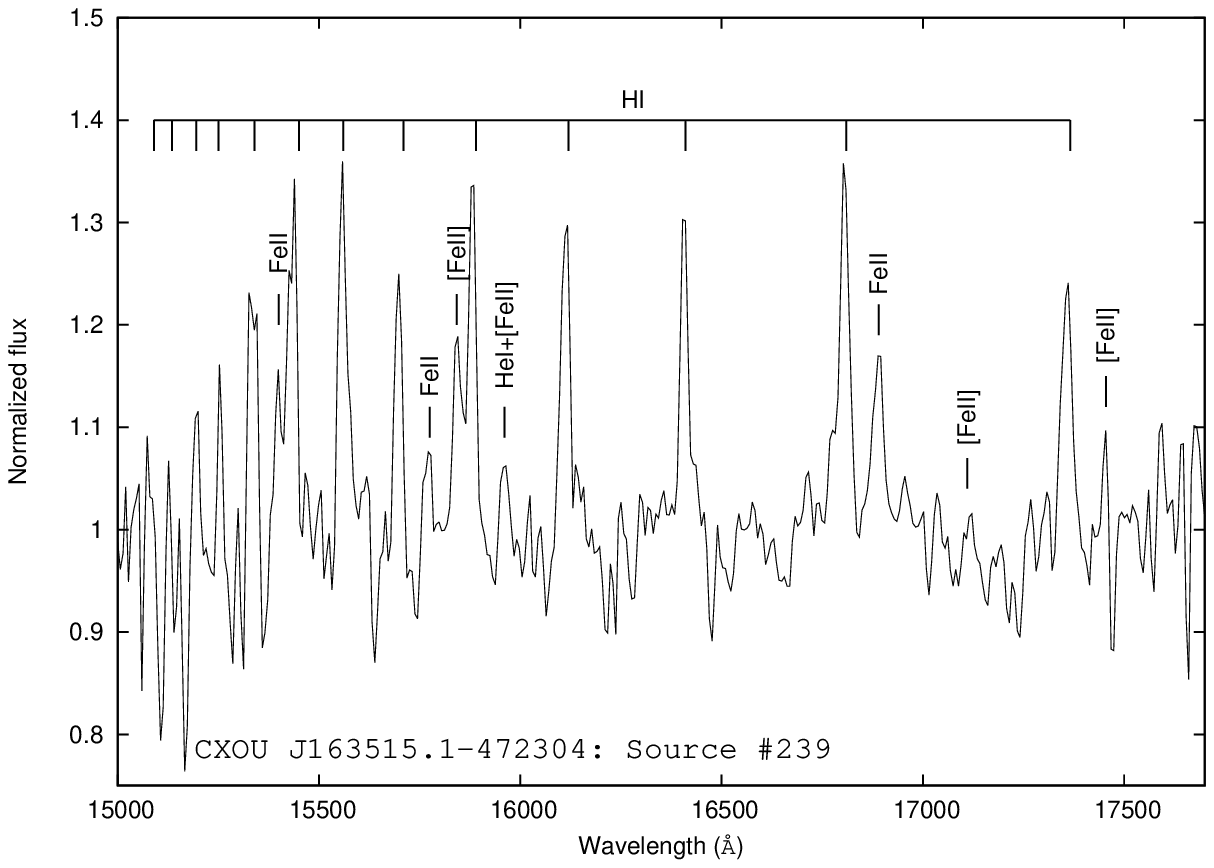}&\includegraphics[width=8.5cm]{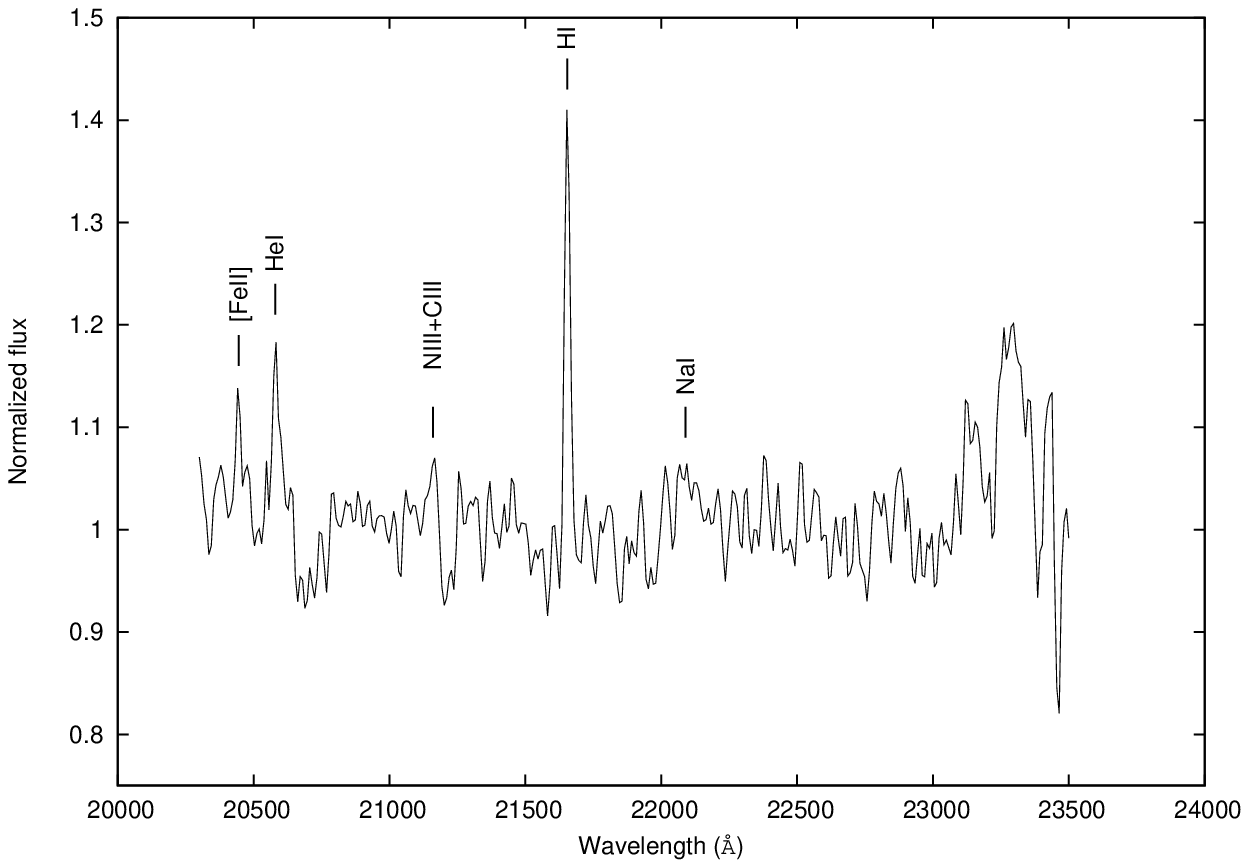}\\
\includegraphics[width=8.5cm]{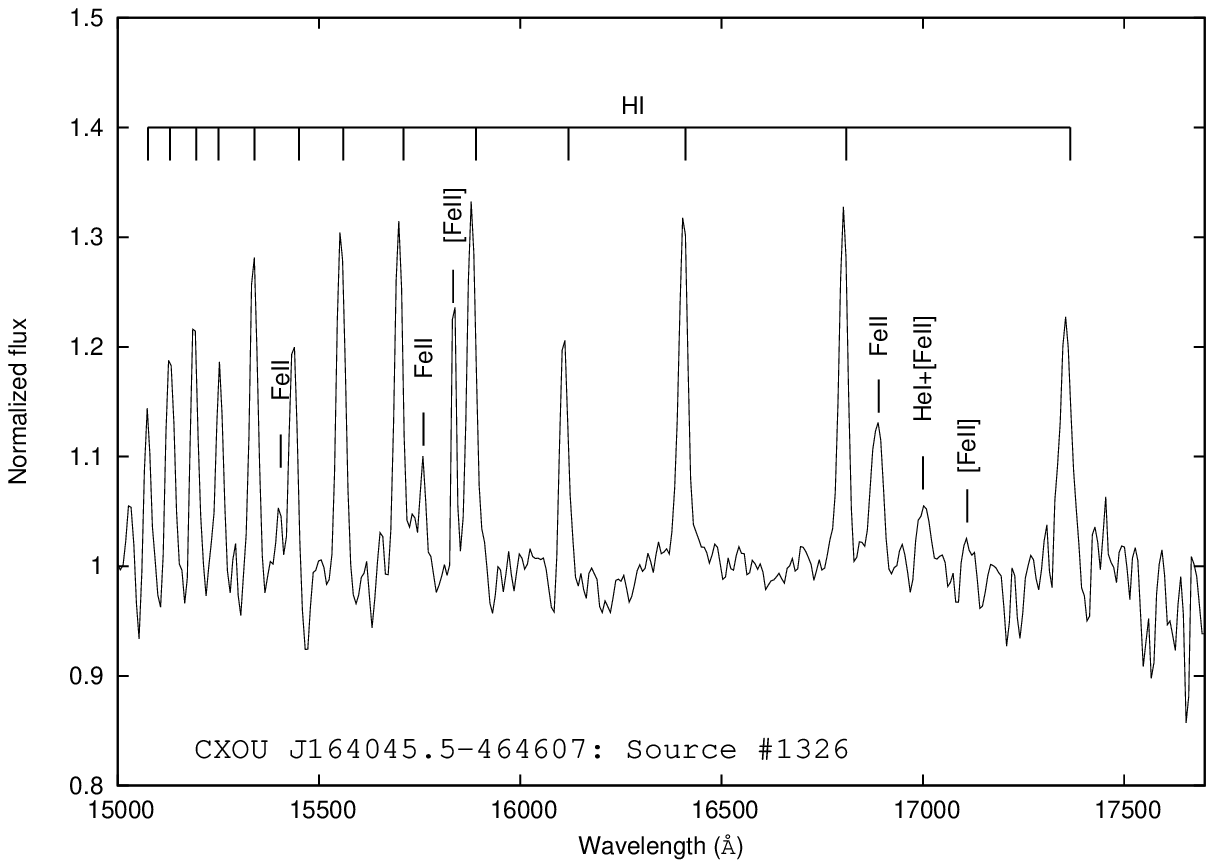}&\includegraphics[width=8.5cm]{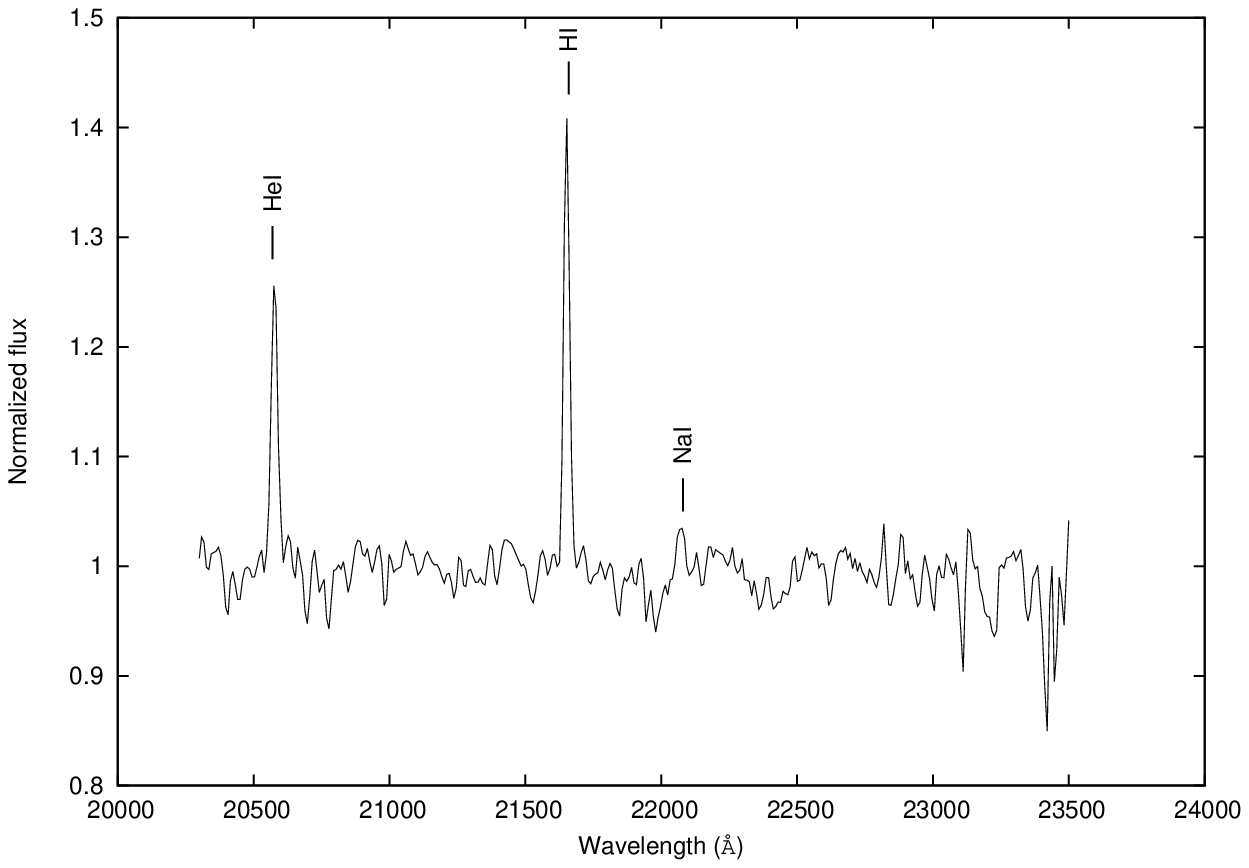}\\

\end{tabular}
\caption{Near-IR \textit{H}-band (left) and \textit{K}-band (right)
spectra of Source 239 (top) and Source 1326 (bottom).}
\label{irspem}
\end{figure*}

%% file: Figures/sed_4_16.figures
\begin{figure*}
\begin{tabular}{cc}
\includegraphics[width=8.5cm]{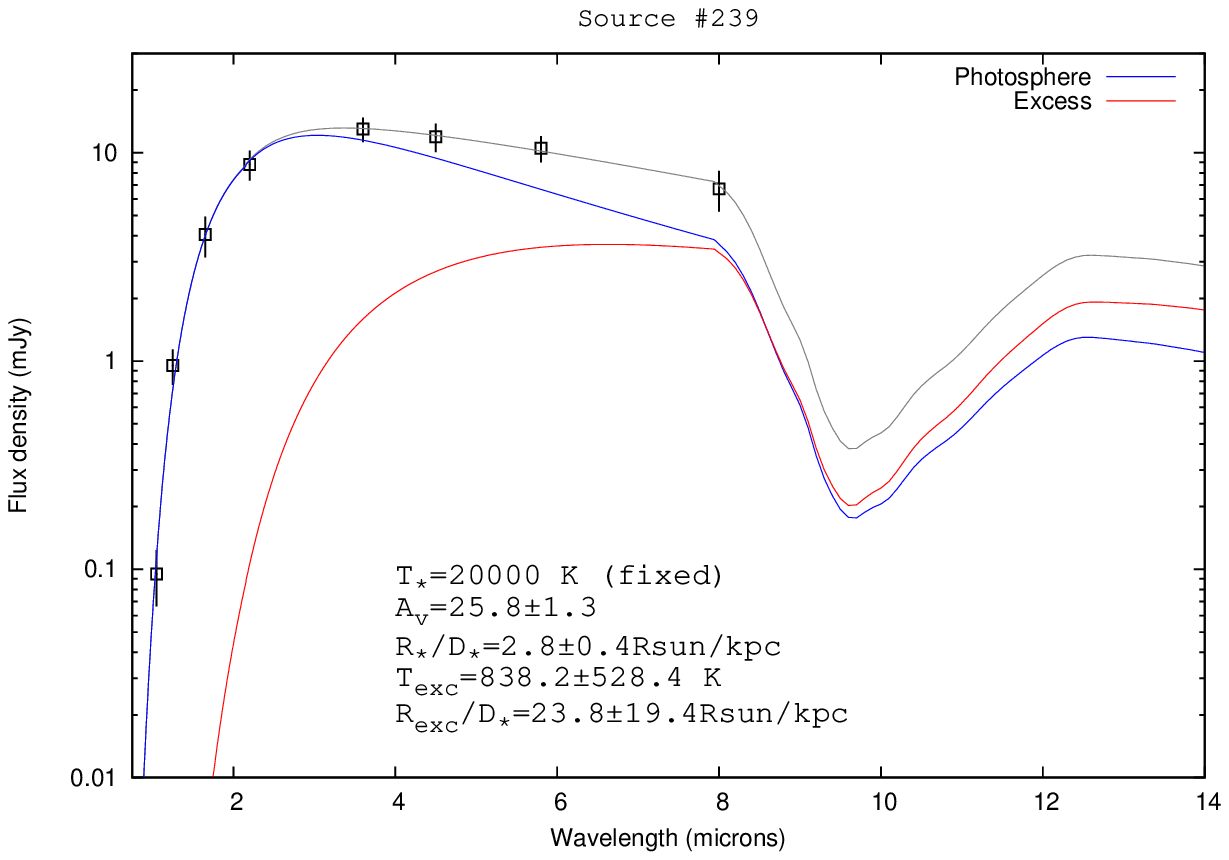}&\includegraphics[width=8.5cm]{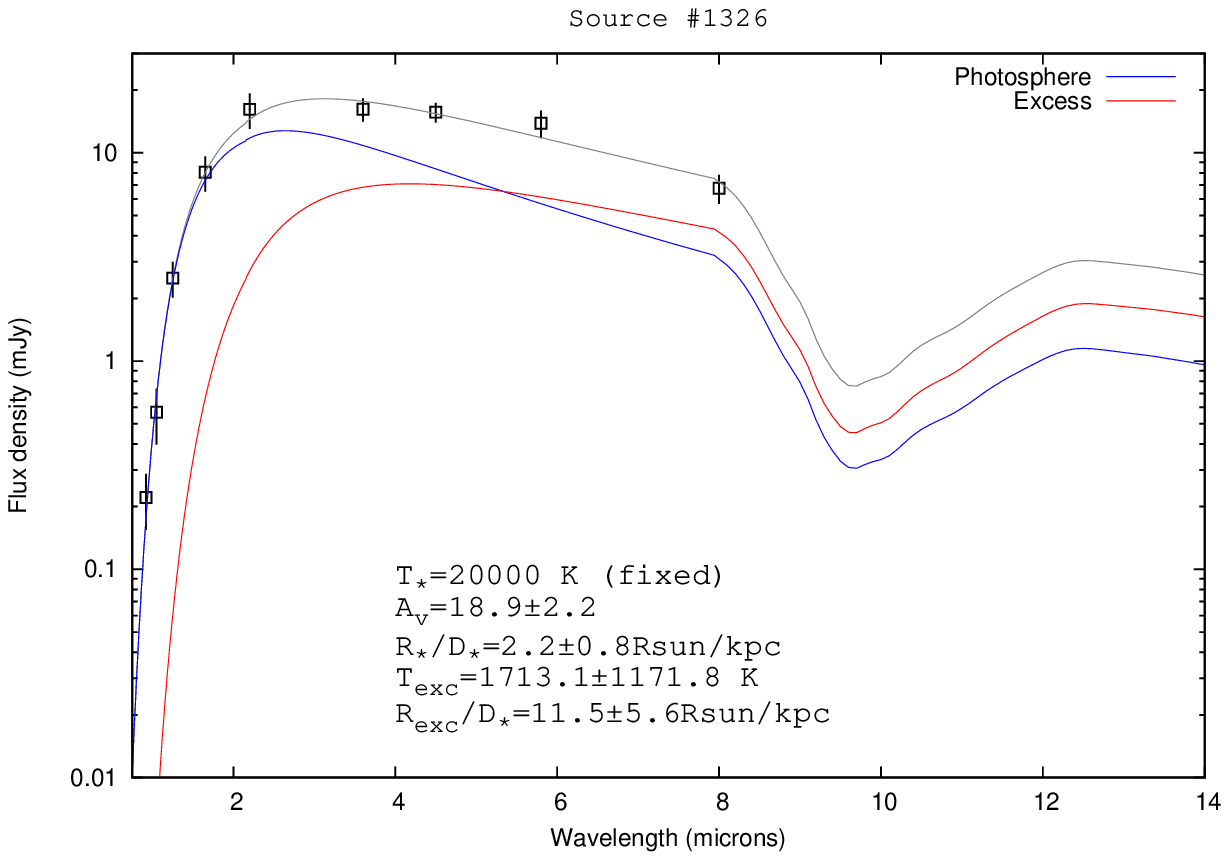}\\
\end{tabular}
\caption{Optical to mid-IR SEDs of sources 239 (left) and 1326 (right)
fitted with two absorbed black bodies. Uncertainties are given at 3$\sigma$.}
\label{sed_4_16}
\end{figure*}

%% file: Figures/sp_ir_hmxb_em2.figures
\begin{figure*}
\begin{tabular}{cc}
\includegraphics[width=8.5cm]{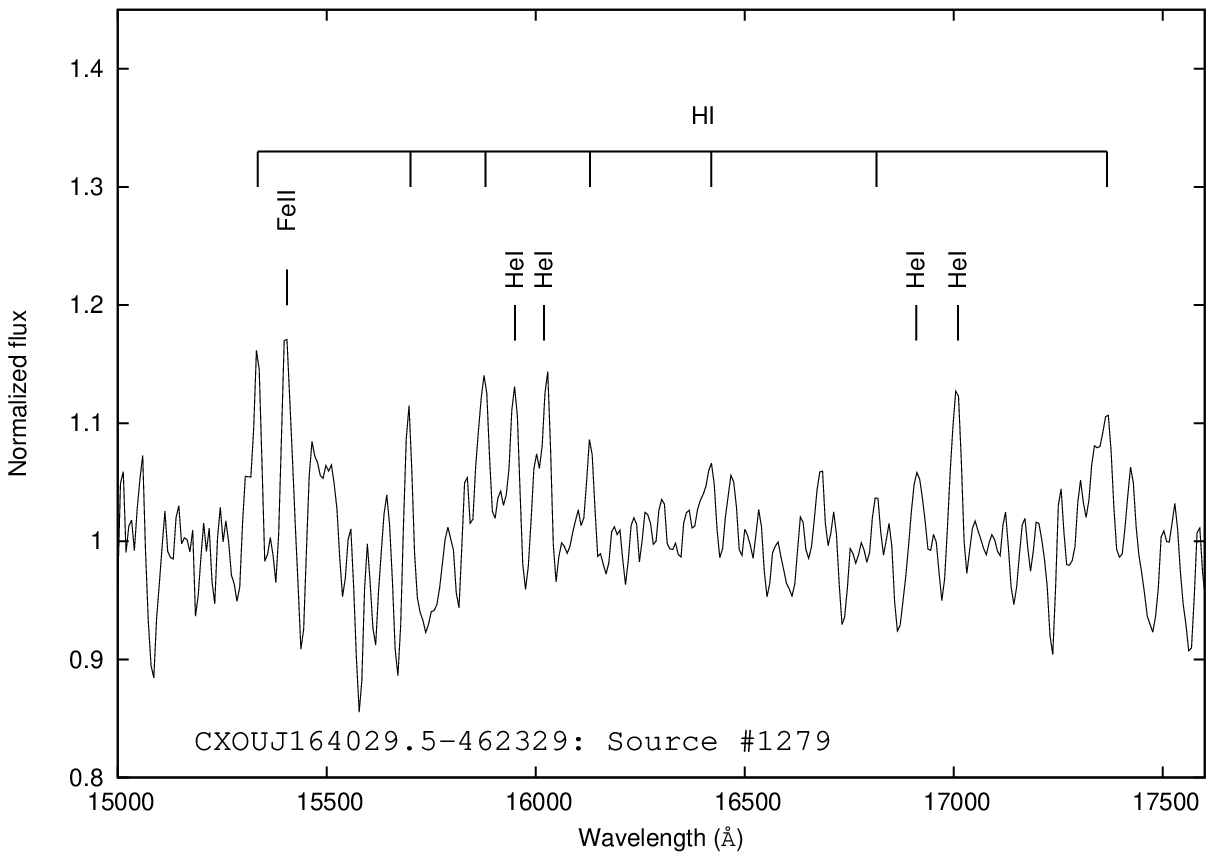}&\includegraphics[width=8.5cm]{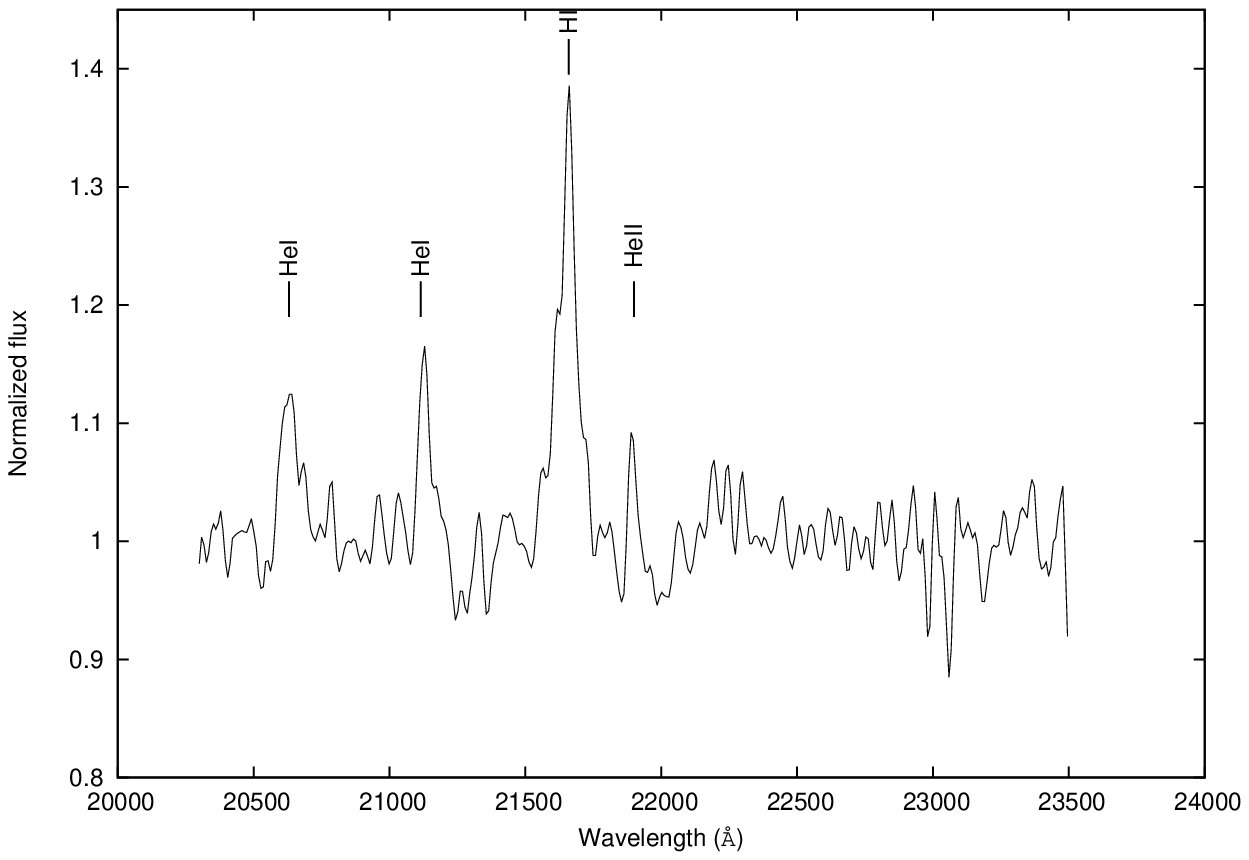}\\
\includegraphics[width=8.5cm]{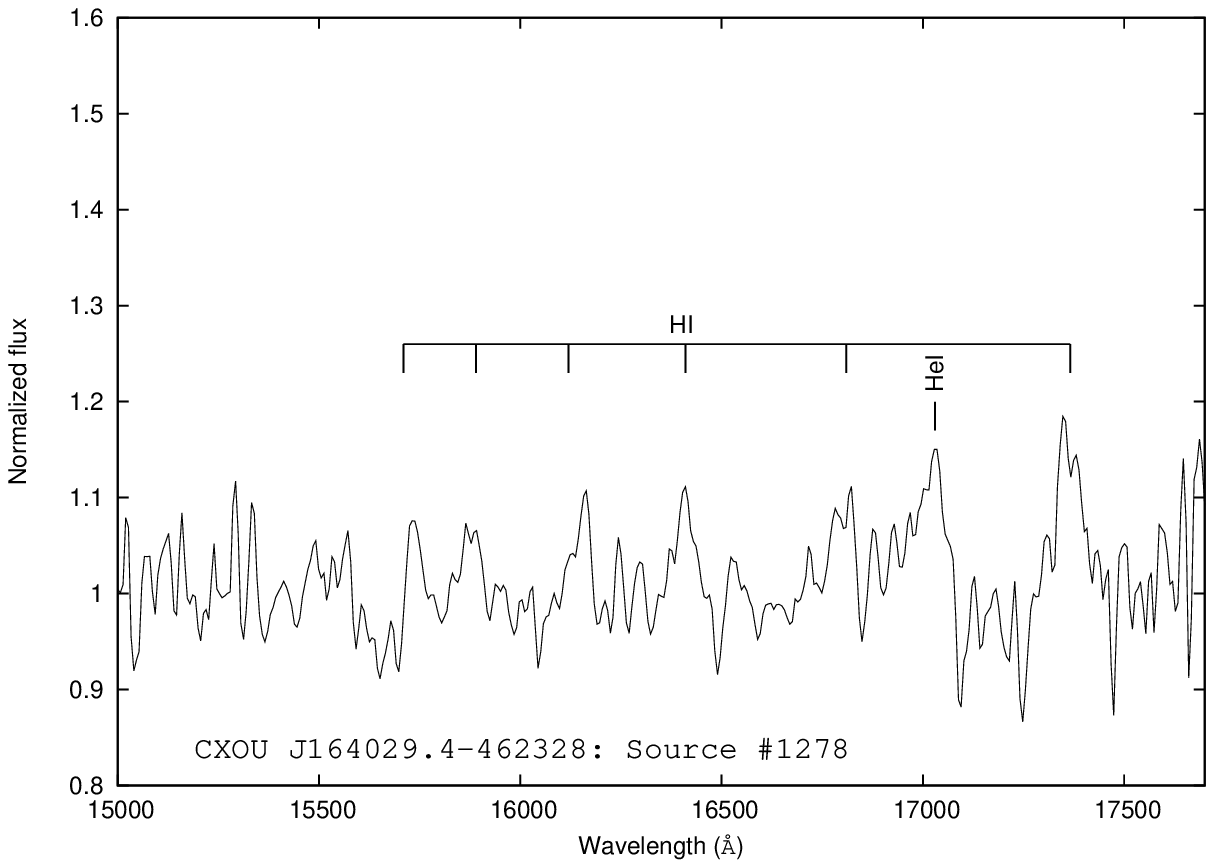}&\includegraphics[width=8.5cm]{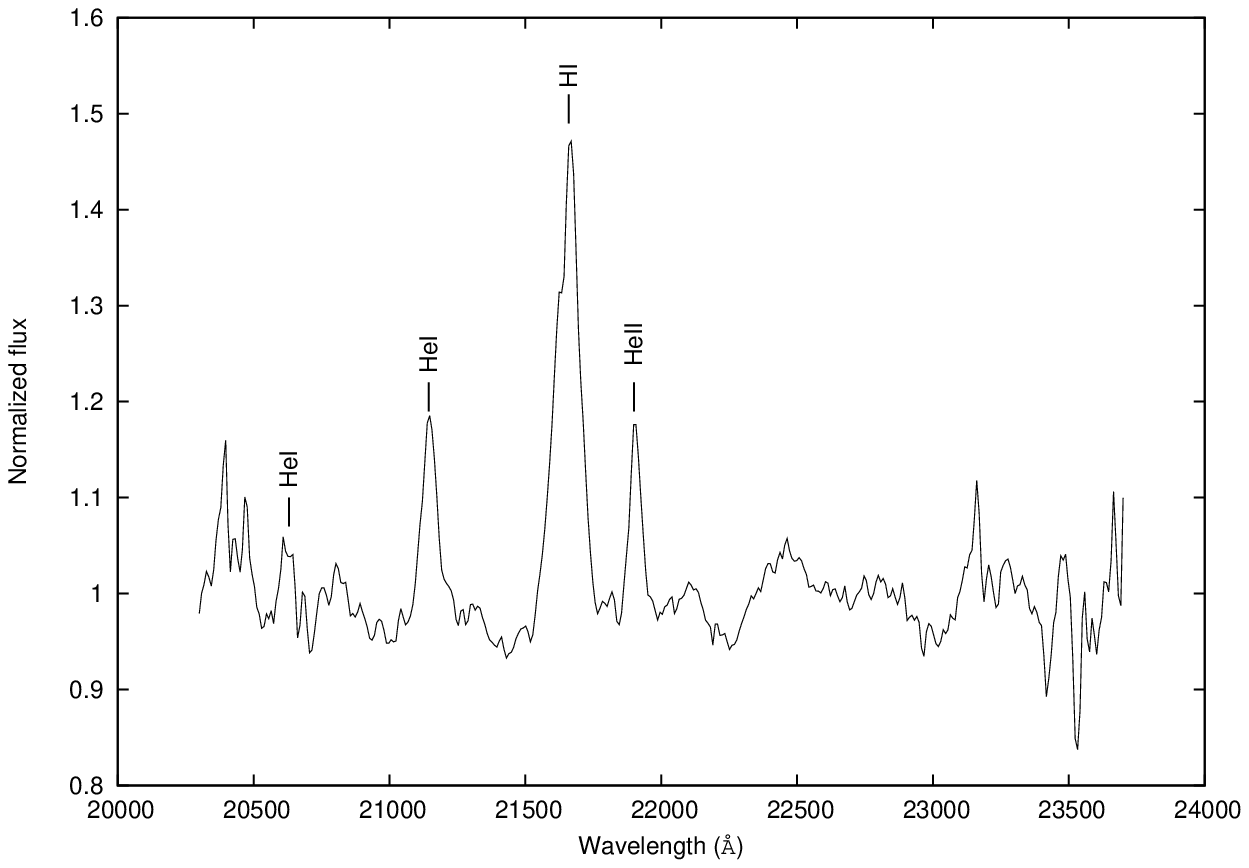}\\
\end{tabular}
\caption{Near-IR \textit{H}-band (left) and \textit{K}-band (right)
spectra of Source 1278 (top) and Source 1279 (bottom).}

\label{irspem2}
\end{figure*}

%% file: Figures/sp_ir_s1179.figures
\begin{figure*}
\begin{tabular}{cc}
\includegraphics[width=8.5cm]{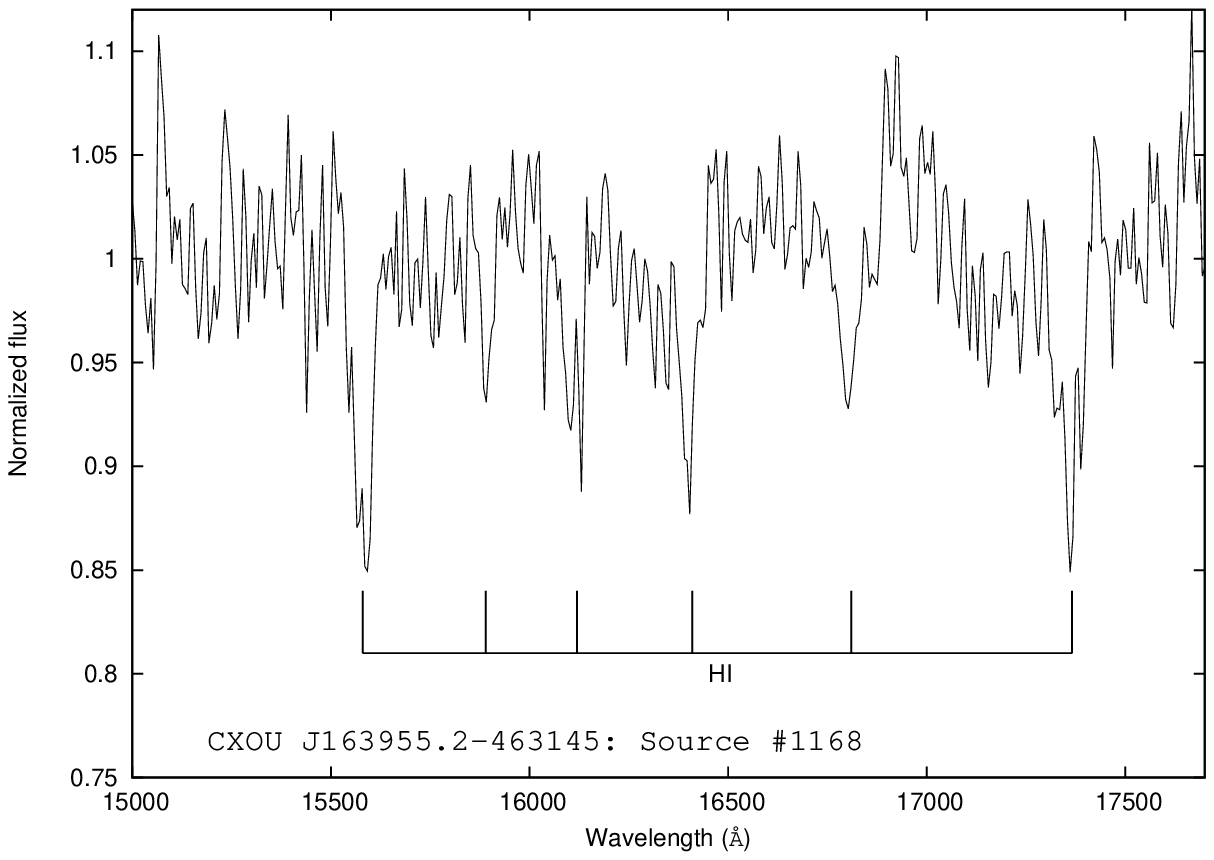}&\includegraphics[width=8.5cm]{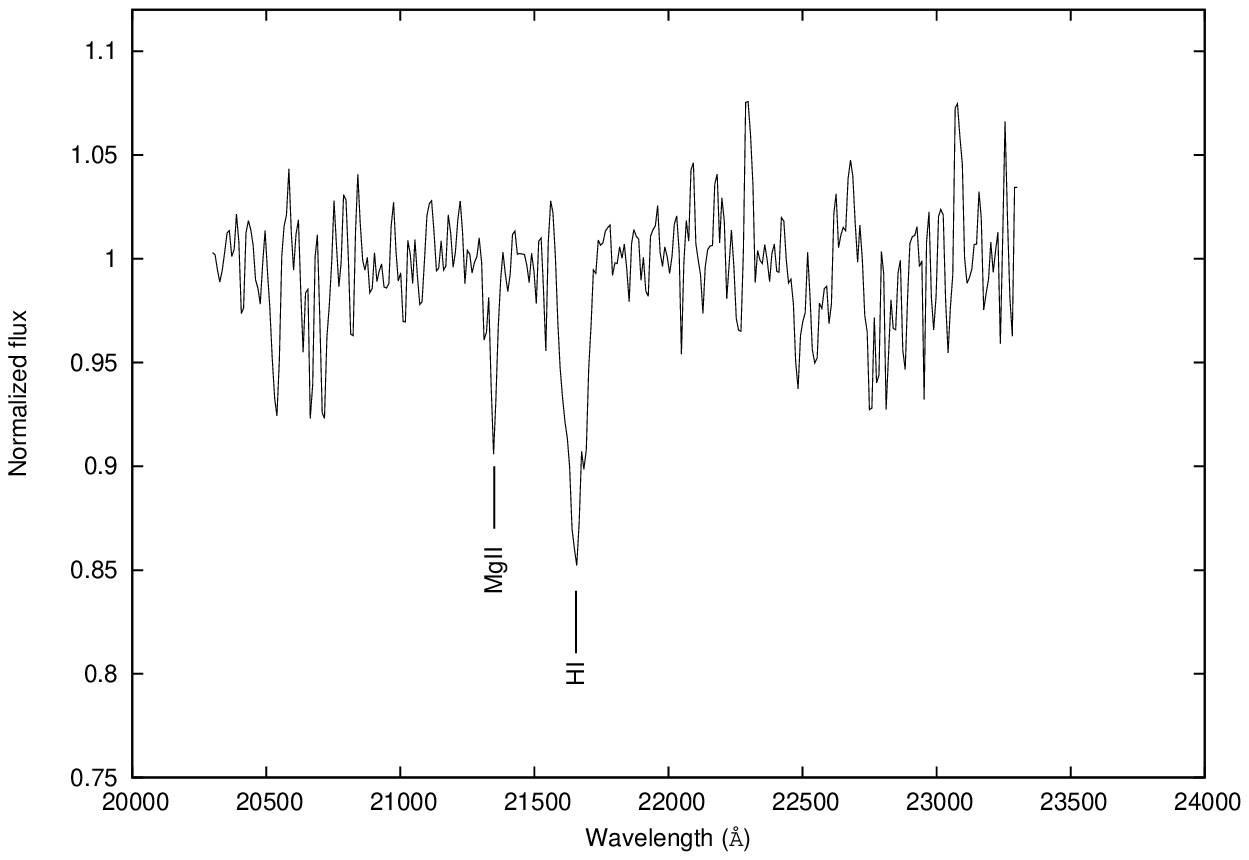}\\
\end{tabular}
\caption{Near-IR \textit{H}-band (left) and \textit{K}-band (right)
spectra of Source 1168.}
\label{irsp1179}
\end{figure*}

%% file: Figures/sp_ir_s747.figures
\begin{figure*}
\begin{tabular}{cc}
\includegraphics[width=8.5cm]{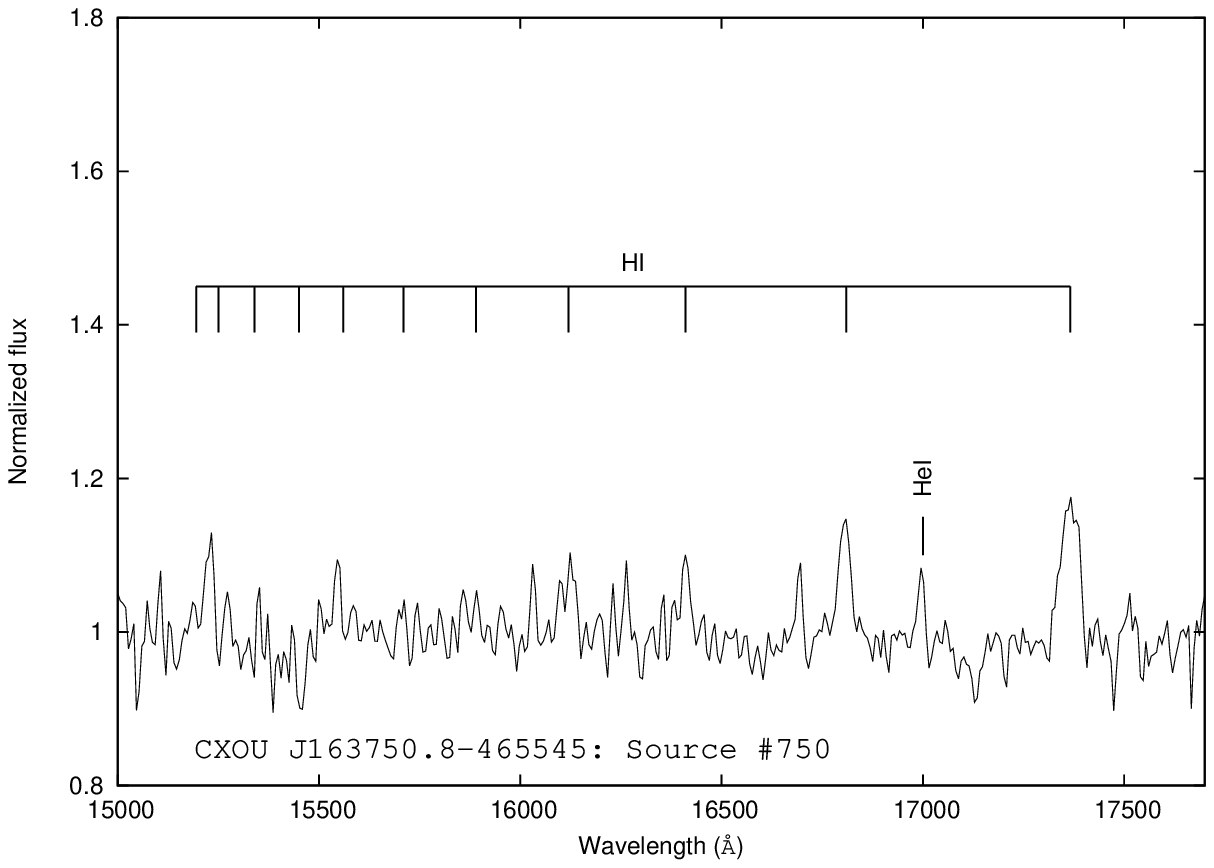}&\includegraphics[width=8.5cm]{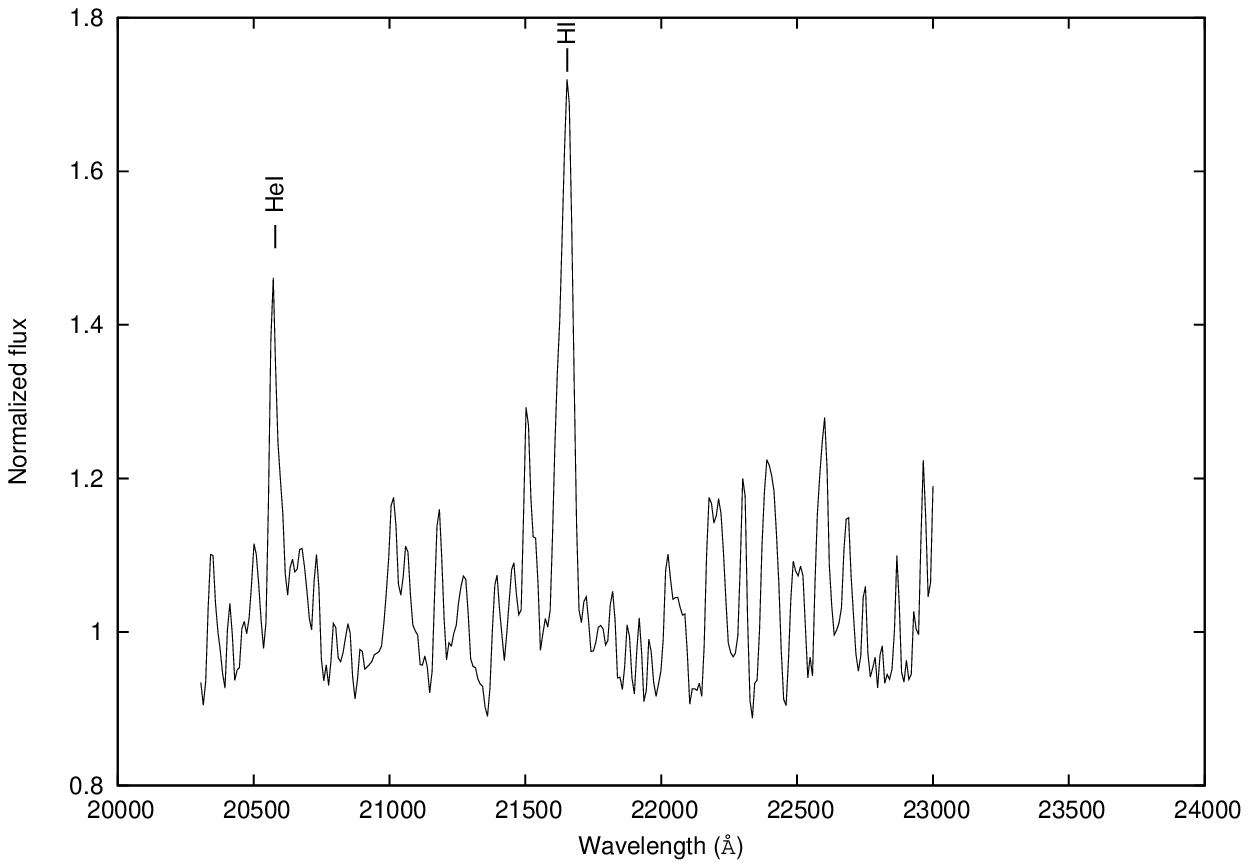}\\
\end{tabular}
\caption{Near-IR \textit{H}-band (left) and \textit{K}-band (right)
spectra of Source 750.}
\label{irsp747}
\end{figure*}

%% file: Tables/lines_lm.table
\begin{table*}
\caption{\scriptsize  Equivalent widths (\AA) of the main absorption lines for cool stars detected
in the $H$ and $K$-band spectra of 14 sources in our sample.}
\label{lmlines}
\resizebox{0.8\textwidth}{!}{\begin{minipage}{\textwidth}
\begin{tabular}{lccccccccccc}
\hline
\hline
No.&\iontiny{Fe}{i}&\iontiny{Si}{i}&CO&\iontiny{Mg}{i}&\iontiny{H}{i}&\iontiny{Na}{i}&\iontiny{Ca}{i}&CO&CO&CO&CO\\
&15830 \AA&15890 \AA&16198 \AA&17115 \AA&21663 \AA&22075 \AA&22637 \AA&22957 \AA&23245 \AA&23464 \AA&23550 \AA\\
\hline
38	&  2.88$\pm$0.51&1.28$\pm$0.40	& 7.35$\pm$0.77	& 1.73$\pm$0.45	& 2.43$\pm$0.95	& 1.57$\pm$0.55 &5.25$\pm$1.00	& 43.45$\pm$1.50& 10.88$\pm$0.55&$-$      &12.06$\pm$0.40\\
78	&  $-$	&2.44$\pm$0.51	& 2.57$\pm$1.12	& $-$	& 3.37$\pm$0.85 & 1.12$\pm$0.40 &$-$ 	& 9.13$\pm$0.72	& 3.09$\pm$0.35 &3.24$\pm$0.61&$-$\\
158	&  1.56$\pm$0.73&1.74$\pm$0.48	& 5.19$\pm$1.94	& 1.31$\pm$0.75	& 4.72$\pm$0.61	& 3.17$\pm$0.66 &$-$ 	& 10.51$\pm$0.75& 4.05$\pm$0.50	&$-$      &$-$\\
396	&  $-$	&2.71$\pm$0.46	& 3.28$\pm$0.63	& 3.43$\pm$0.86	& 1.81$\pm$0.85	& 1.83$\pm$0.76 &$-$	& 8.41$\pm$0.70	& 5.49$\pm$0.73	&$-$      &$-$\\
437	&  $-$	&1.75$\pm$0.62	& 1.58$\pm$0.68	& 1.84$\pm$0.56	& 3.99$\pm$0.69	& $-$ 	&$-$	& 6.50$\pm$0.49	& 6.95$\pm$0.73	&$-$      &$-$\\
694	&  $-$	&1.71$\pm$0.48	& 3.20$\pm$1.04	& 2.01$\pm$0.57	& 4.11$\pm$0.75	& $-$ 	&$-$	& 6.10$\pm$1.01	& $-$	&3.48$\pm$0.81&$-$\\
881	&  $-$	&2.11$\pm$0.57	& 7.16$\pm$0.70	& $-$	& 2.50$\pm$0.68	& 2.79$\pm$0.57 &0.80$\pm$0.40	& 27.91$\pm$2.01& 16.29$\pm$0.94&2.48$\pm$0.52&12.28$\pm$0.88\\
1059	&  $-$	&$-$	& $-$	& $-$	& 0.96$\pm$0.40	& $-$ 	&4.26$\pm$0.82	& 5.79$\pm$1.07	& 4.95$\pm$0.75 &$-$      &$-$\\
1128	&  2.21$\pm$0.56&2.31$\pm$0.77	& 1.46$\pm$0.68	& $-$	& 5.44$\pm$1.05	& $-$ 	&2.18$\pm$0.86	& 4.37$\pm$0.98	& 2.27$\pm$0.68	&$-$      &$-$\\
1203 	&  $-$	&2.64$\pm$0.65	& 2.49$\pm$0.70	& $-$	& $-$	& 1.66$\pm$0.50 &2.13$\pm$0.90	& 13.12$\pm$1.02& 5.90$\pm$0.65	&$-$      &$-$\\
1358	&  $-$	&2.24$\pm$0.50	& 2.50$\pm$0.81	& 1.90$\pm$1.05	& 1.49$\pm$0.75	& 1.41$\pm$0.51 &$-$	& 5.29$\pm$1.04	& $-$	&$-$      &5.73$\pm$0.71\\
1374	&  $-$	&1.88$\pm$0.41	& 2.87$\pm$1.32	& 2.28$\pm$0.52	& 3.60$\pm$0.70	& 2.22$\pm$0.57 &1.24$\pm$0.61	& 4.05$\pm$0.57	& 8.86$\pm$0.72&$-$	      &$-$\\
1379	&  $-$	&1.51$\pm$0.57	& $-$	& 1.62$\pm$0.50	& 4.96$\pm$0.72	& $-$ 	&$-$	& 3.41$\pm$0.97	& $-$	&$-$      &$-$\\
1408	&  $-$	&$-$	& 4.43$\pm$0.61	& $-$	& $-$	& $-$ 	&1.42$\pm$0.41	& 13.95$\pm$0.87& 5.11$\pm$0.48&$-$	      &8.94$\pm$0.45\\
\hline
\end{tabular}
\end{minipage}}
\tablefoot{\scriptsize A ``--'' means that the line is not detected}.
\end{table*}

%% file: Figures/sp_ir.figures
\begin{figure*}
\begin{tabular}{cc}
\includegraphics[width=8.5cm]{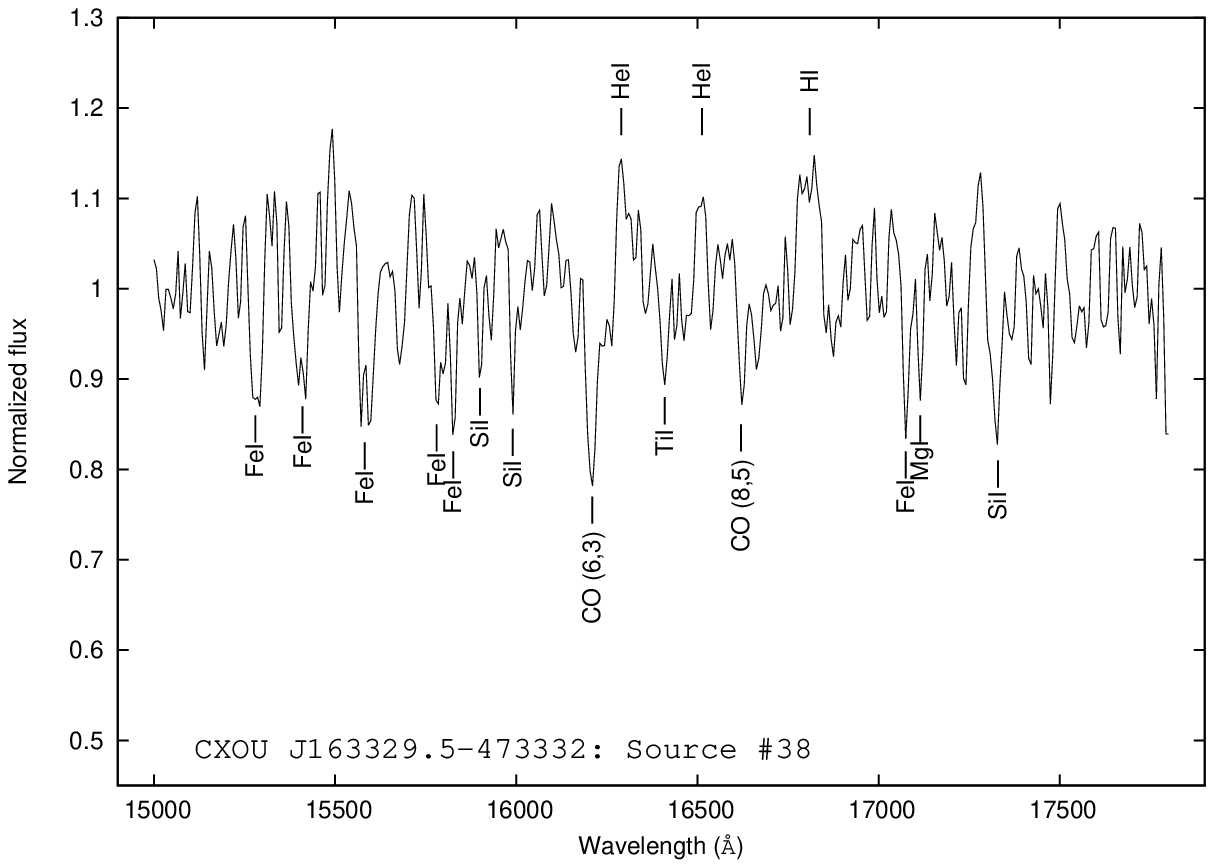}&\includegraphics[width=8.5cm]{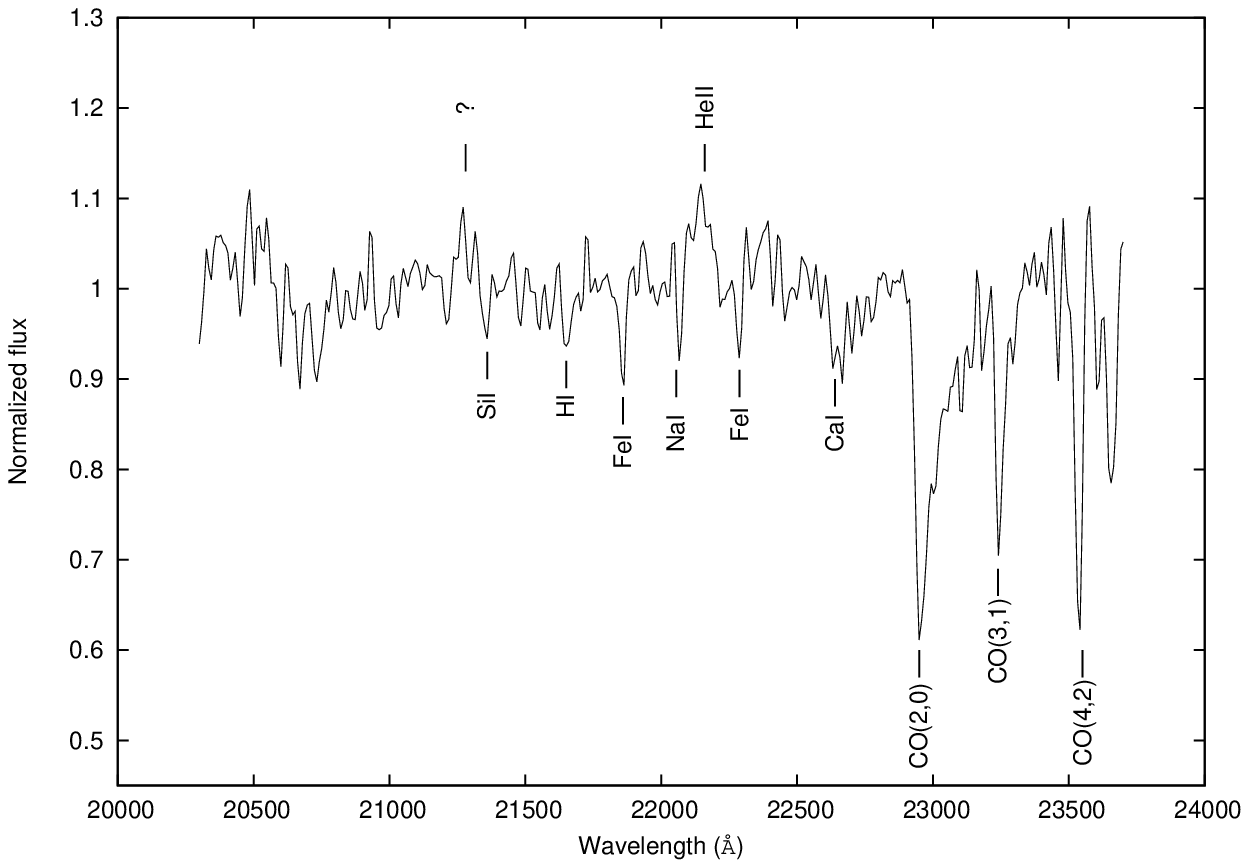}\\
\includegraphics[width=8.5cm]{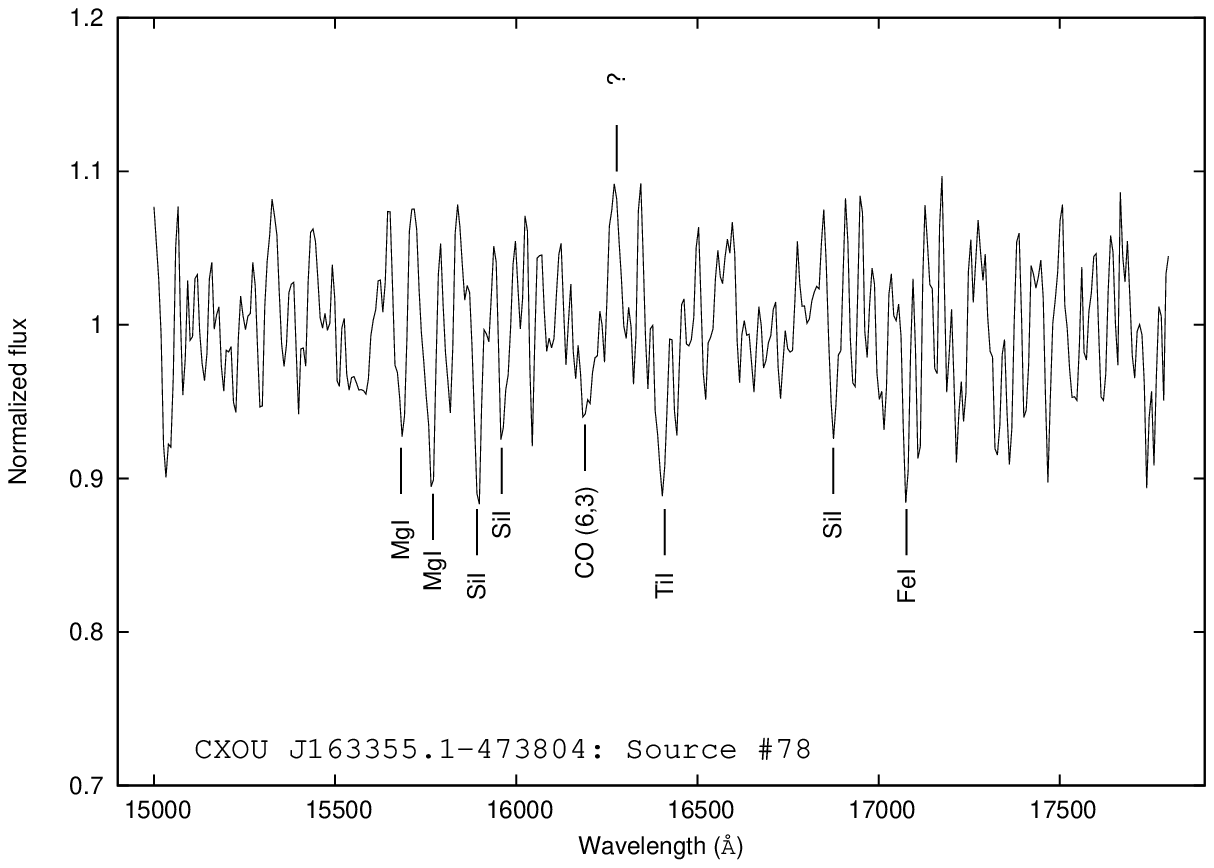}&\includegraphics[width=8.5cm]{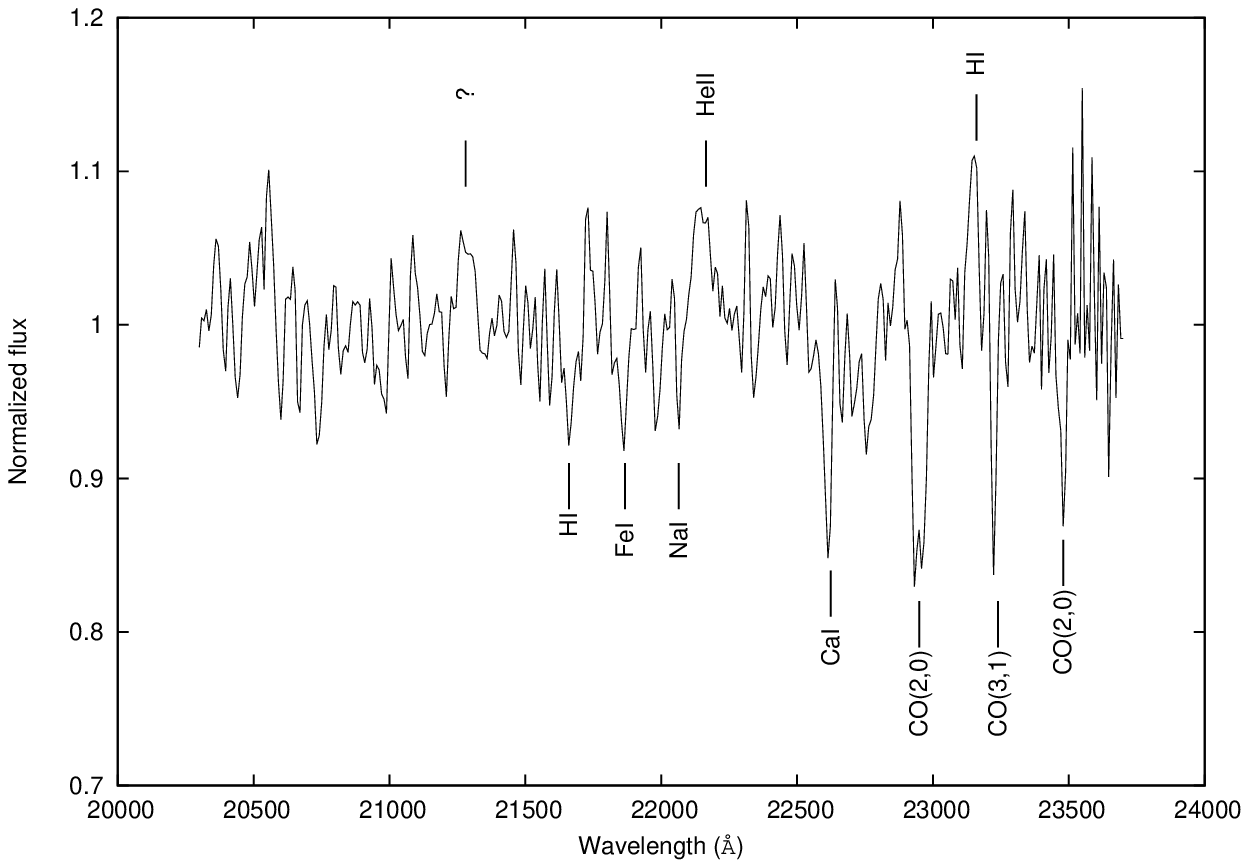}\\
\includegraphics[width=8.5cm]{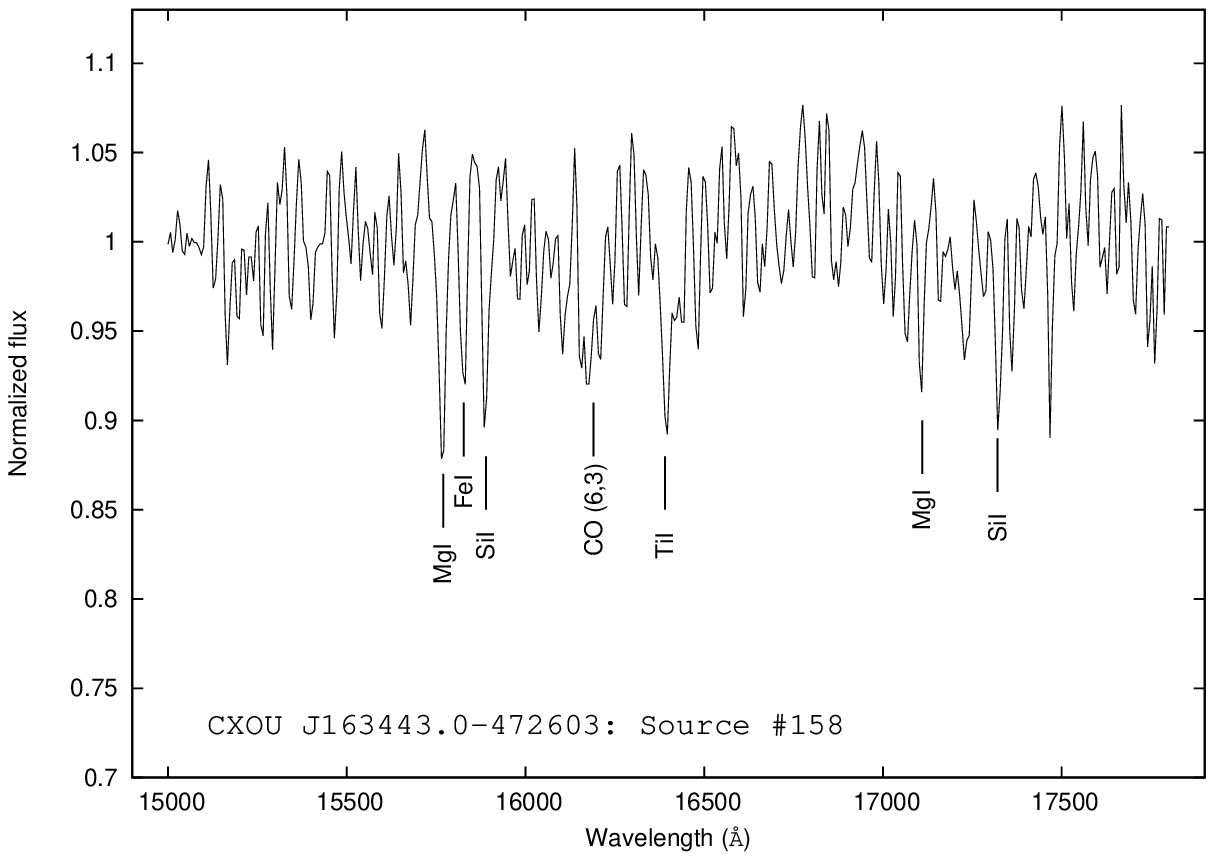}&\includegraphics[width=8.5cm]{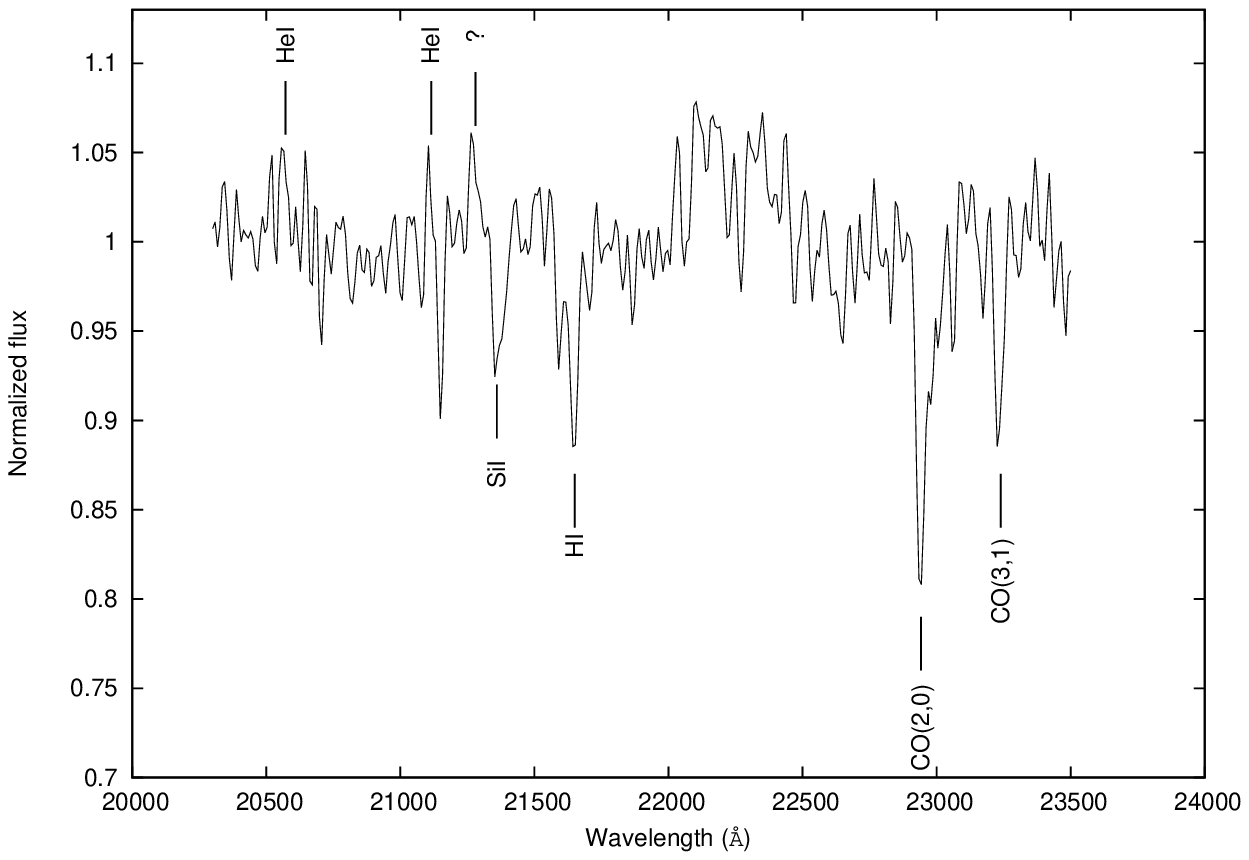}\\

\end{tabular}
\caption{Near-IR \textit{H}-band (left) and \textit{K}-band (right)
spectra of Source 38 (top), Source 78 (middle), and Source 158 (bottom).}
\label{irsp}
\end{figure*}

%% file: Figures/sp_ir2.figures
\begin{figure*}
\begin{tabular}{cc}
\includegraphics[width=8.5cm]{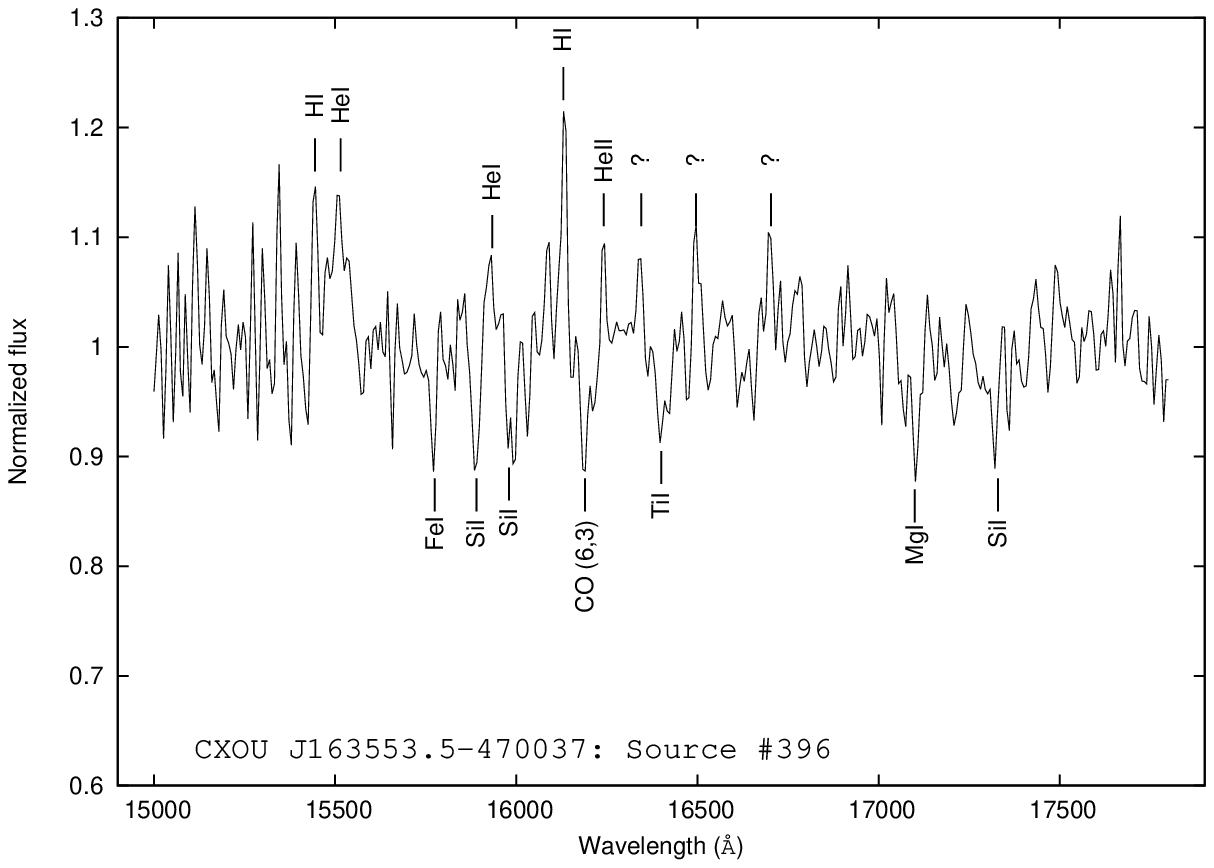}&\includegraphics[width=8.5cm]{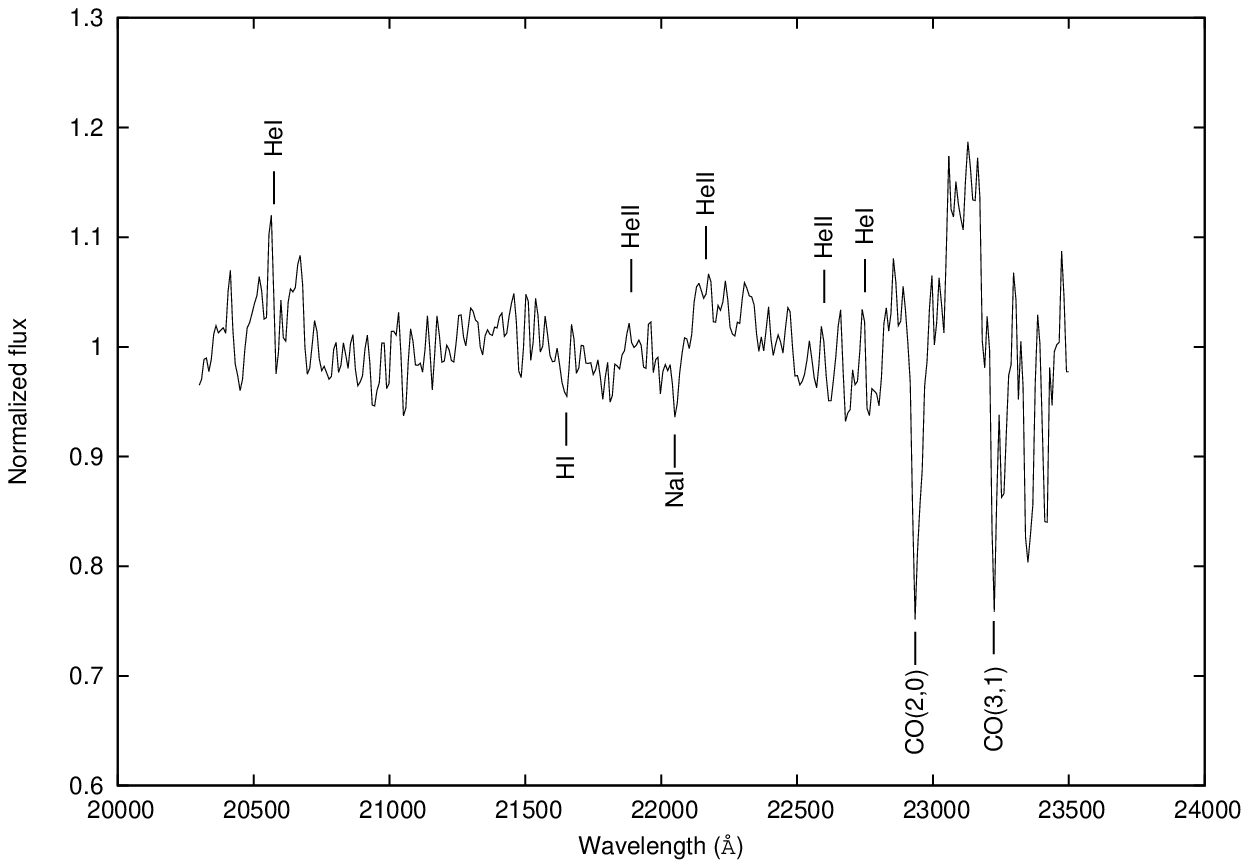}\\
\includegraphics[width=8.5cm]{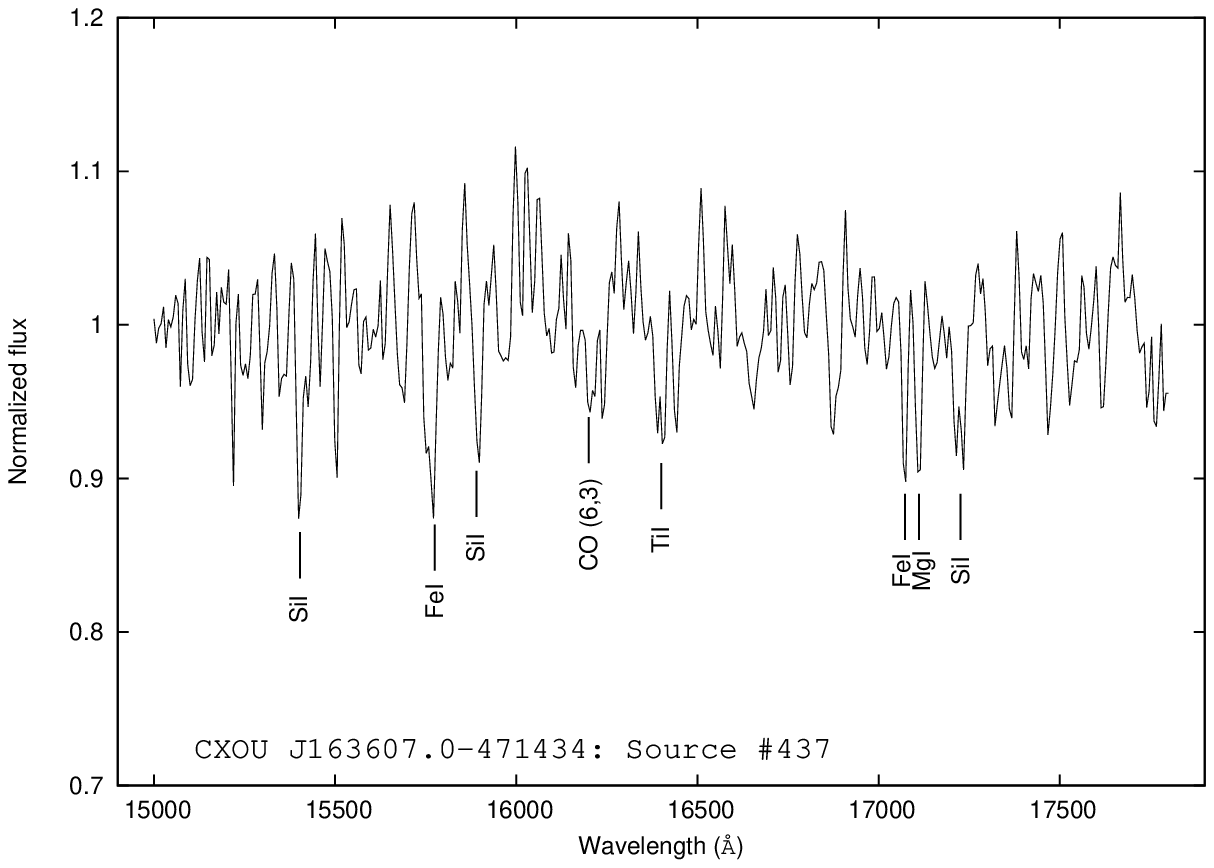}&\includegraphics[width=8.5cm]{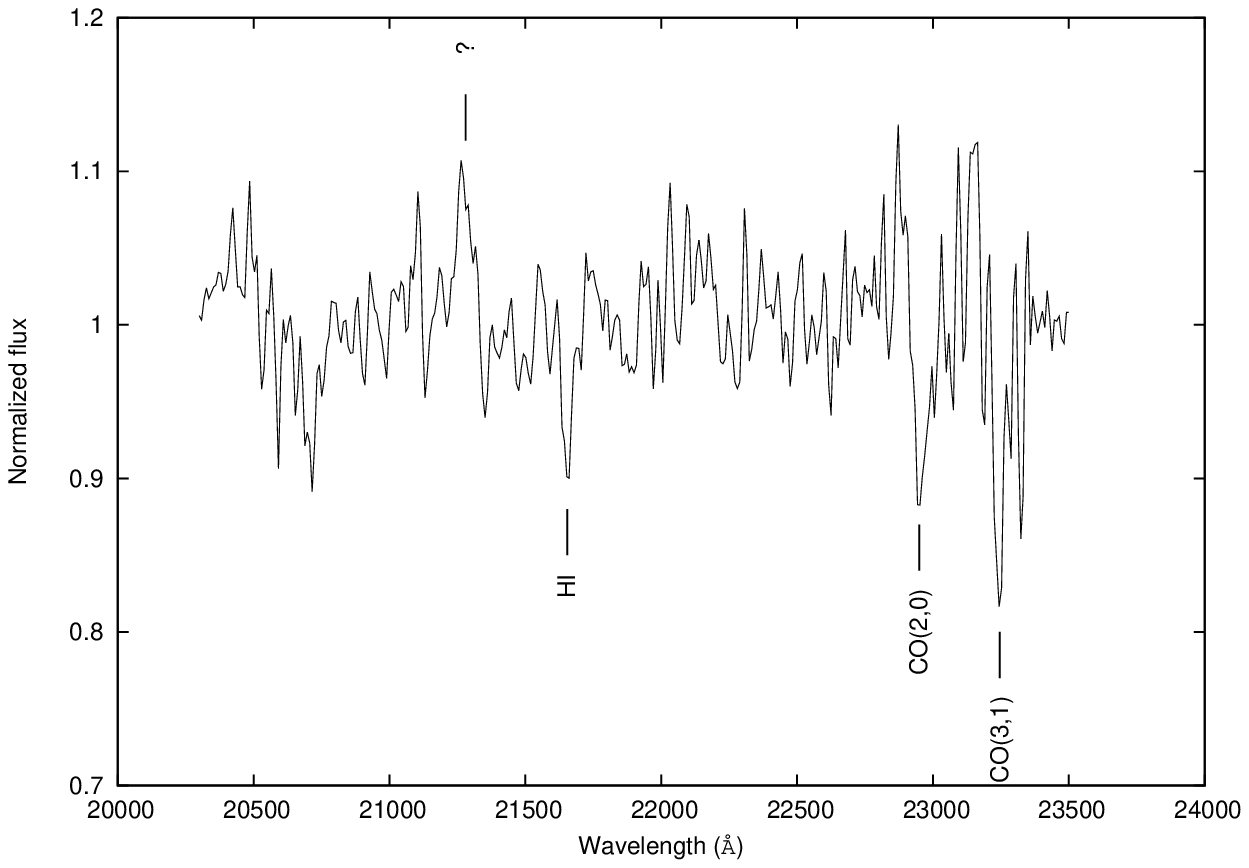}\\
\includegraphics[width=8.5cm]{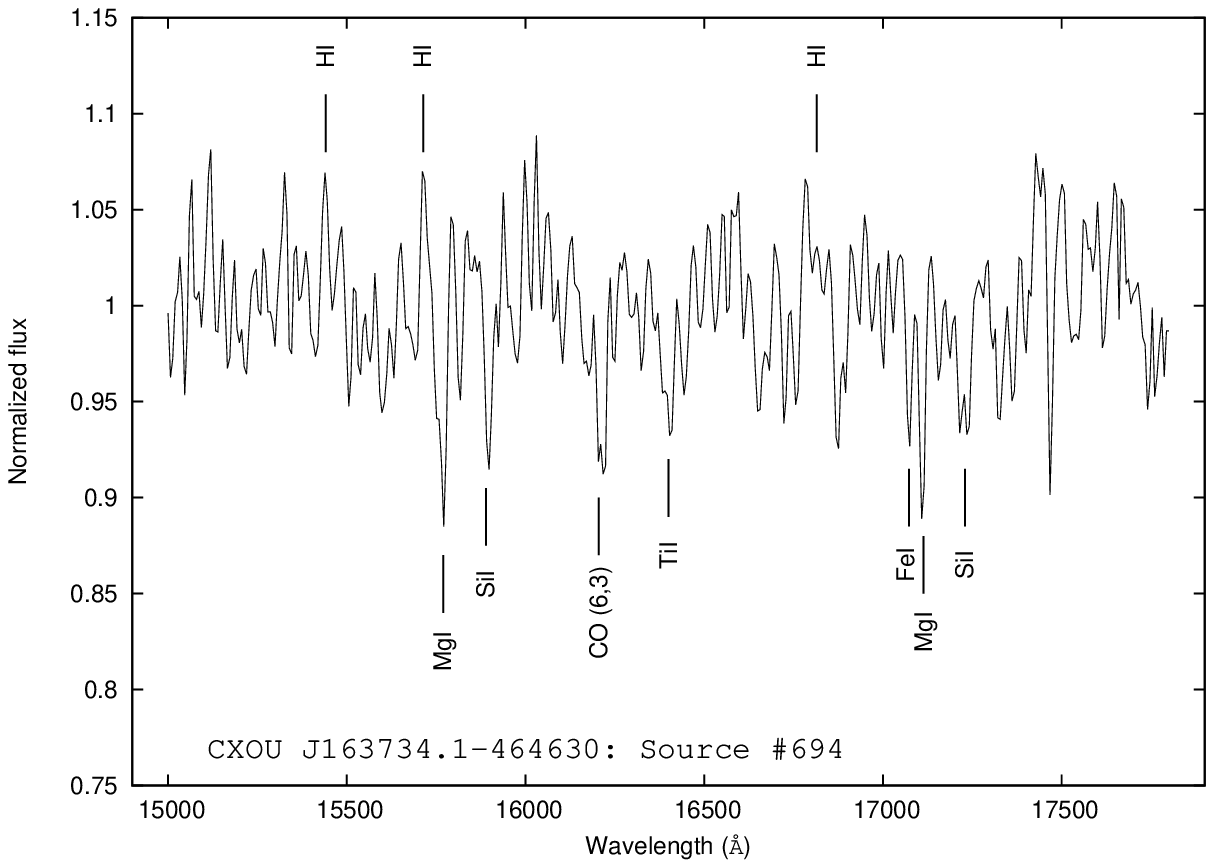}&\includegraphics[width=8.5cm]{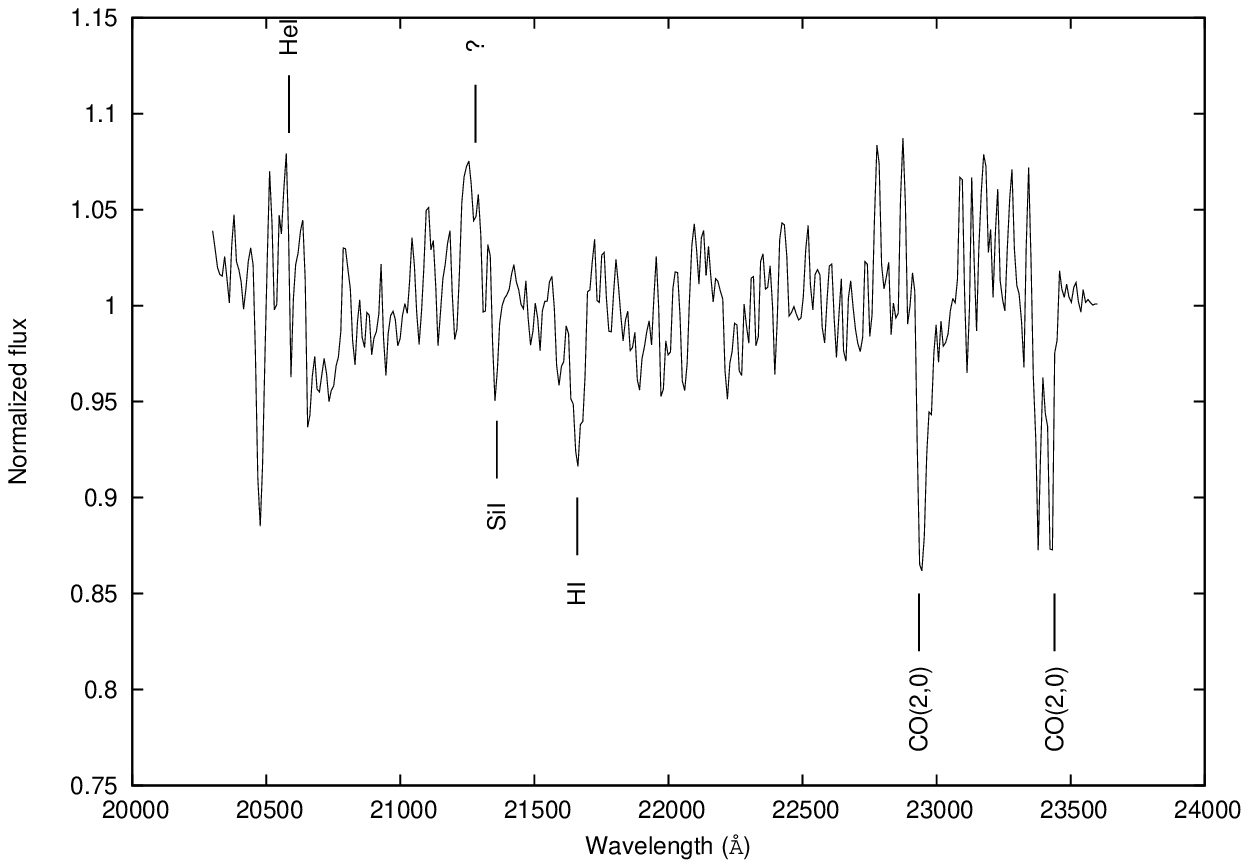}\\

\end{tabular}
\caption{Near-IR \textit{H}-band (left) and \textit{K}-band (right)
spectra of Source 396 (top), Source 437 (middle), and Source 694
(bottom).}
\label{irsp2}
\end{figure*}

%% file: Figures/sp_ir3.figures
\begin{figure*}
\begin{tabular}{cc}

\includegraphics[width=8.5cm]{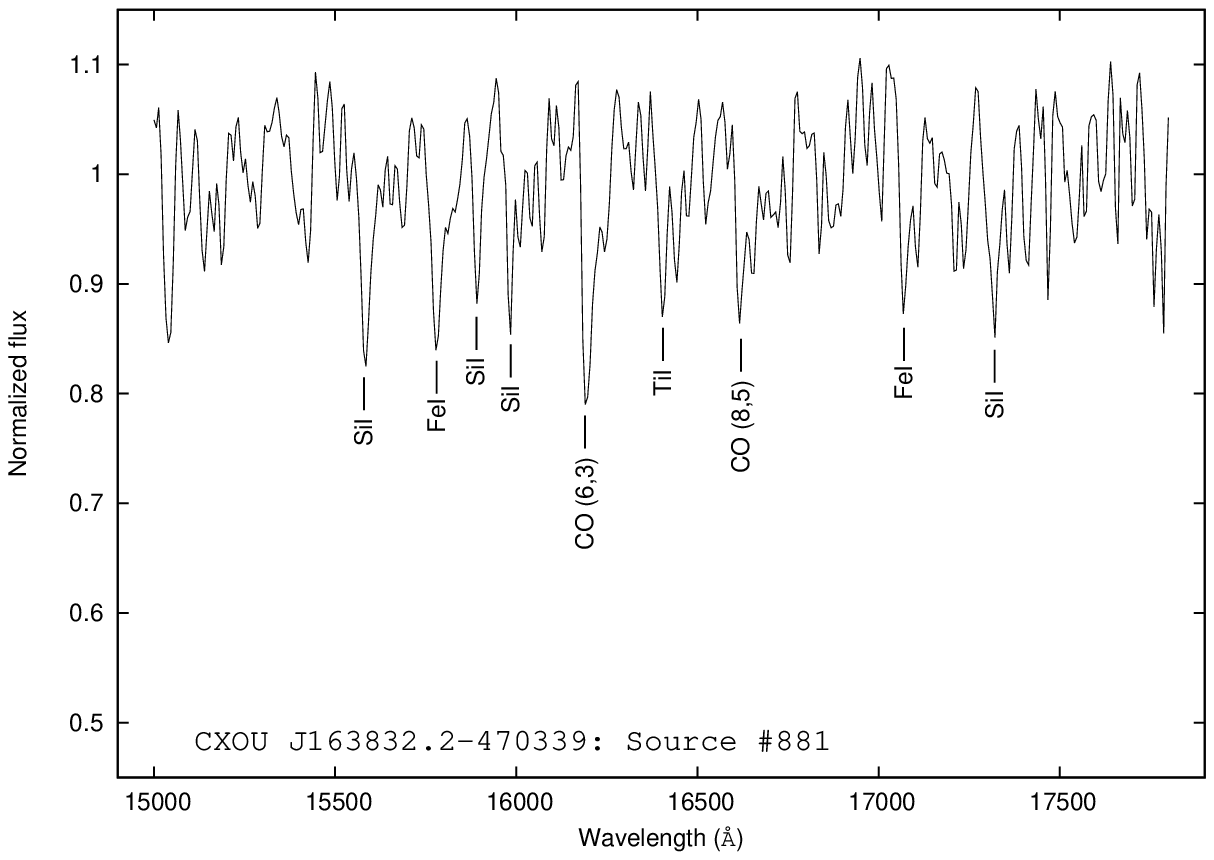}&\includegraphics[width=8.5cm]{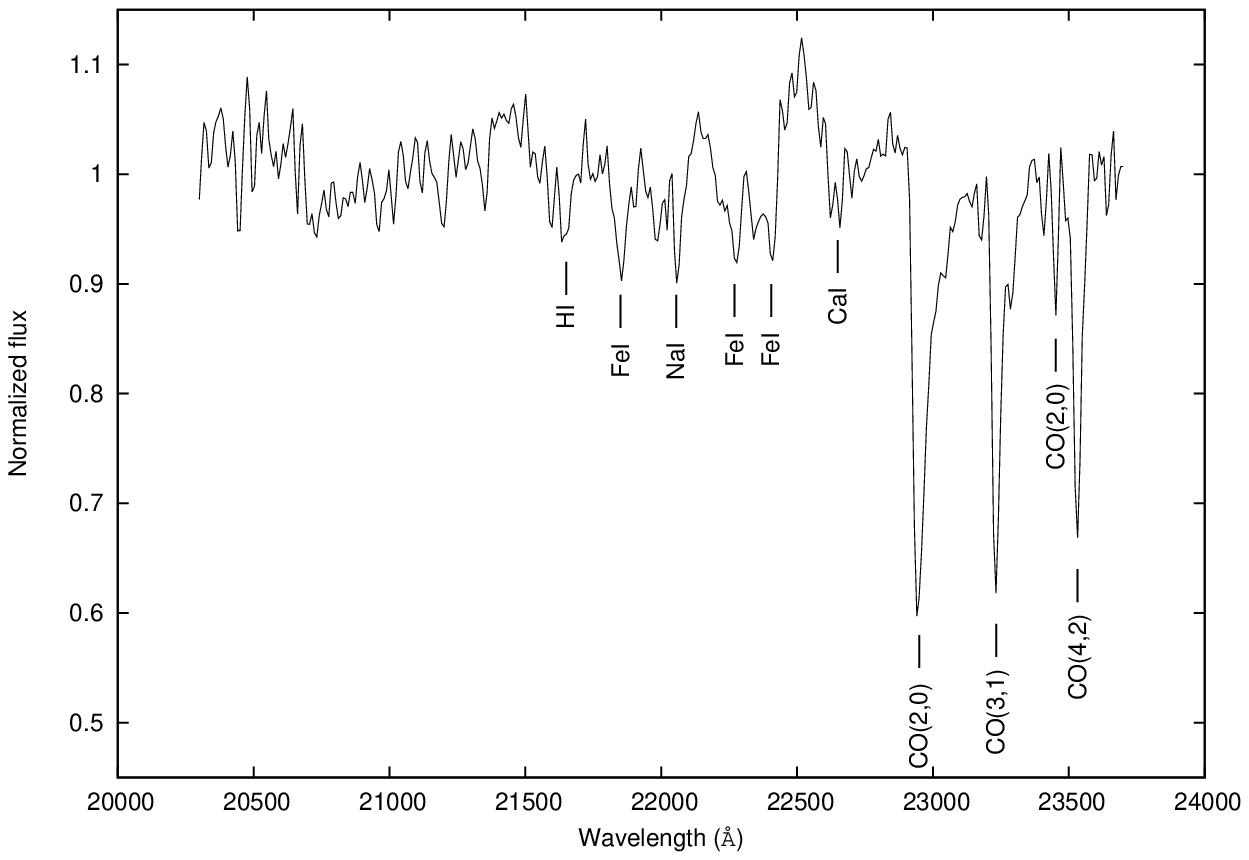}\\
\includegraphics[width=8.5cm]{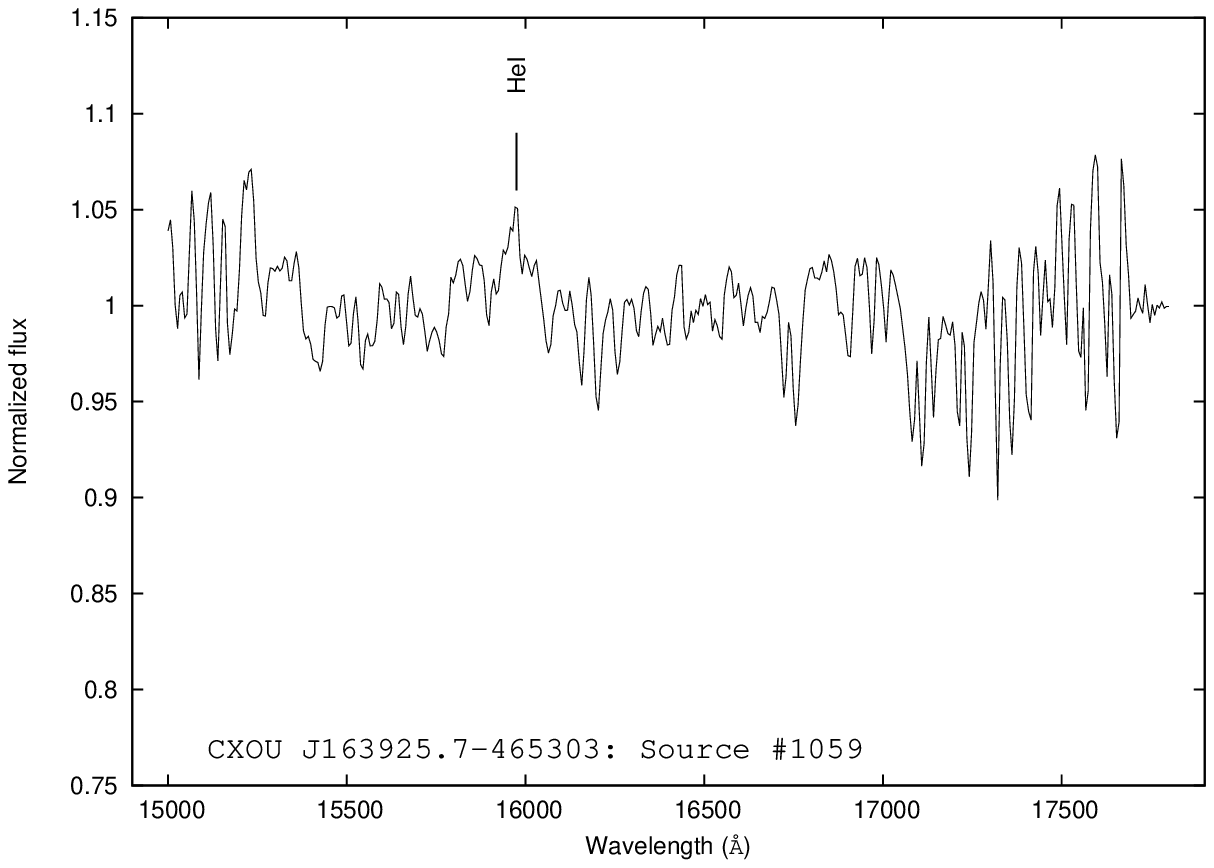}&\includegraphics[width=8.5cm]{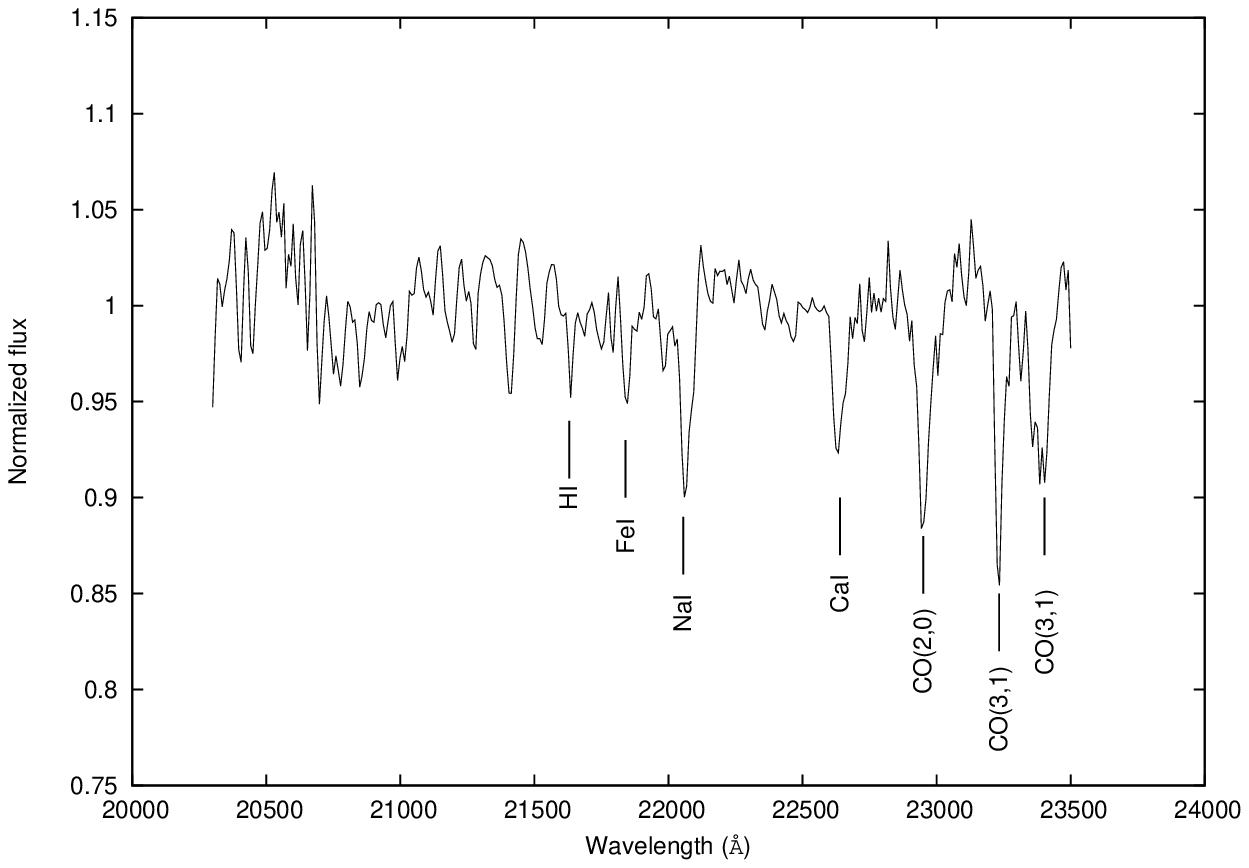}\\
\includegraphics[width=8.5cm]{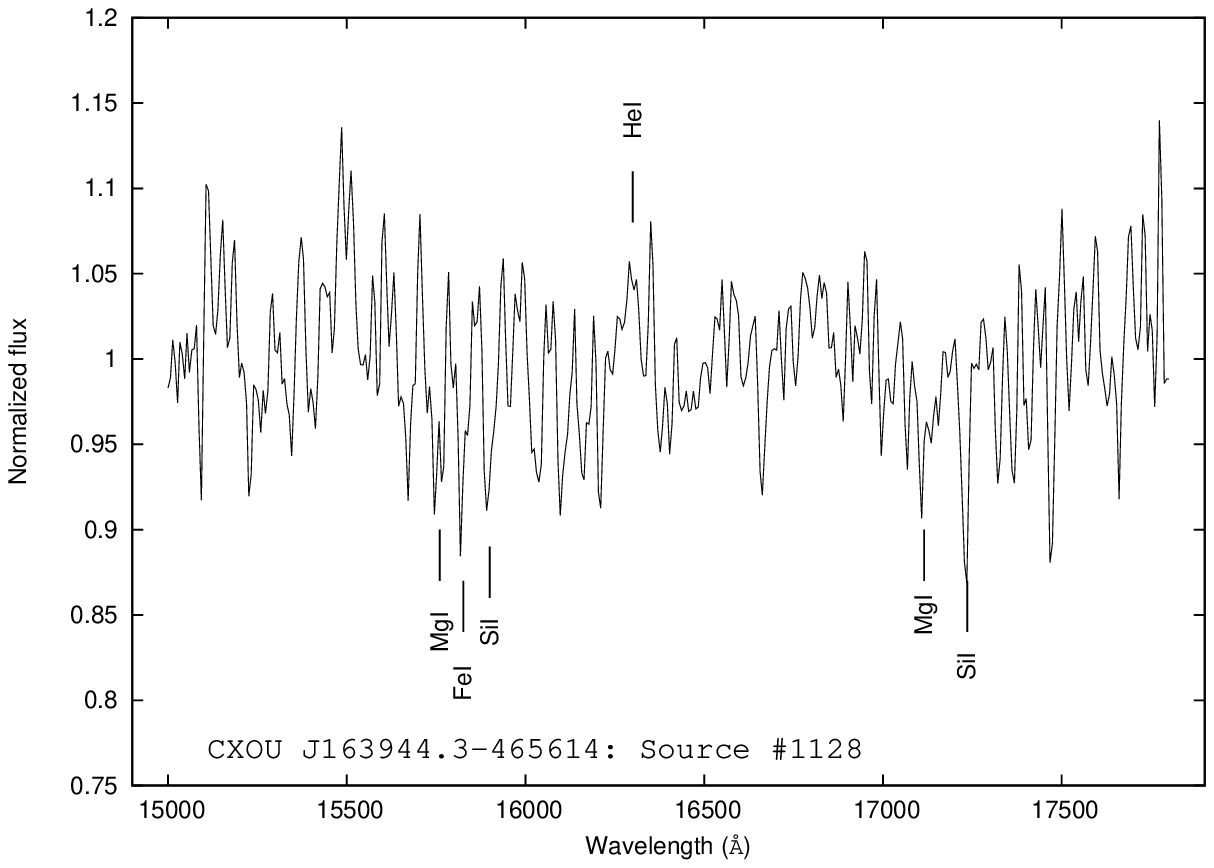}&\includegraphics[width=8.5cm]{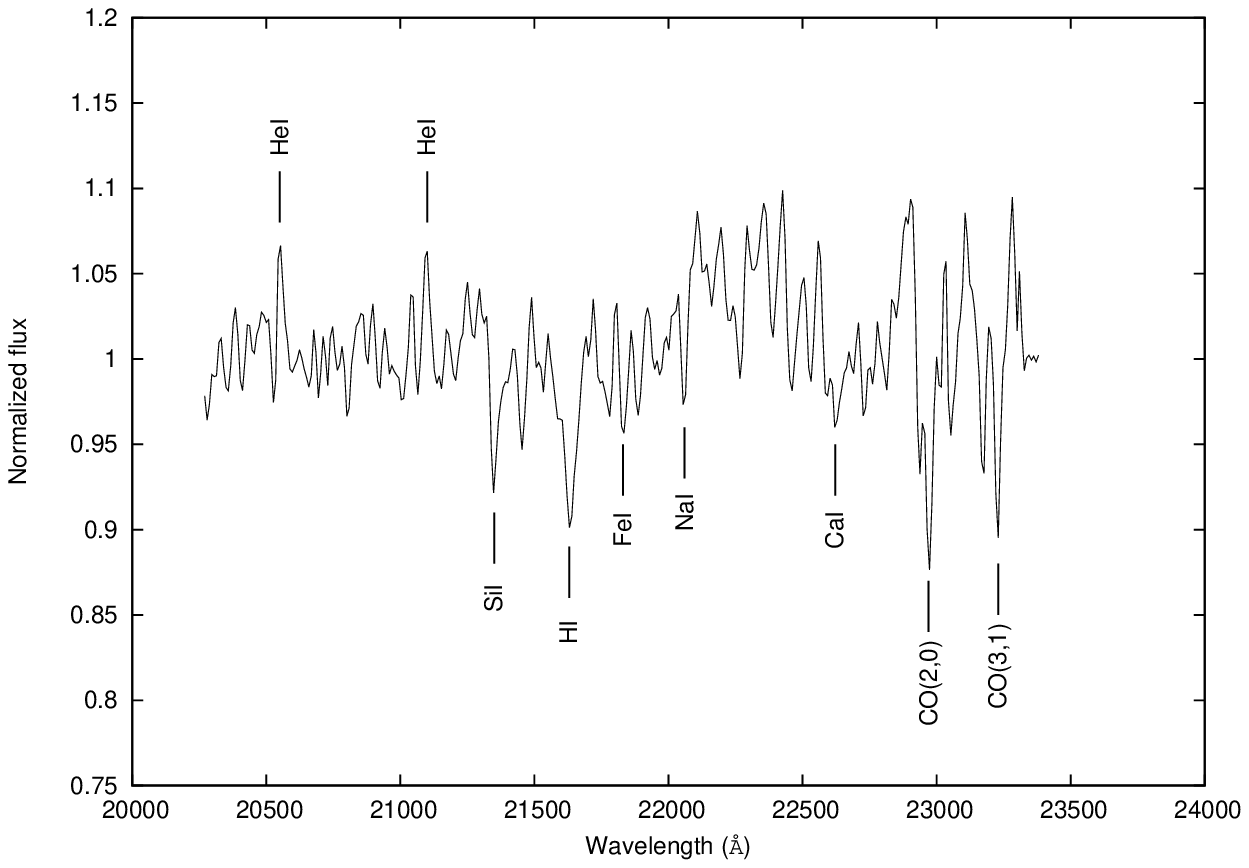}\\

\end{tabular}
\caption{Near-IR \textit{H}-band (left) and \textit{K}-band (right)
spectra of Source 881 (top), Source 1059 (middle), and Source
1128 (bottom).}
\label{irsp3}
\end{figure*}

%% file: Figures/sp_ir4.figures
\begin{figure*}
\begin{tabular}{cc}
\includegraphics[width=8.5cm]{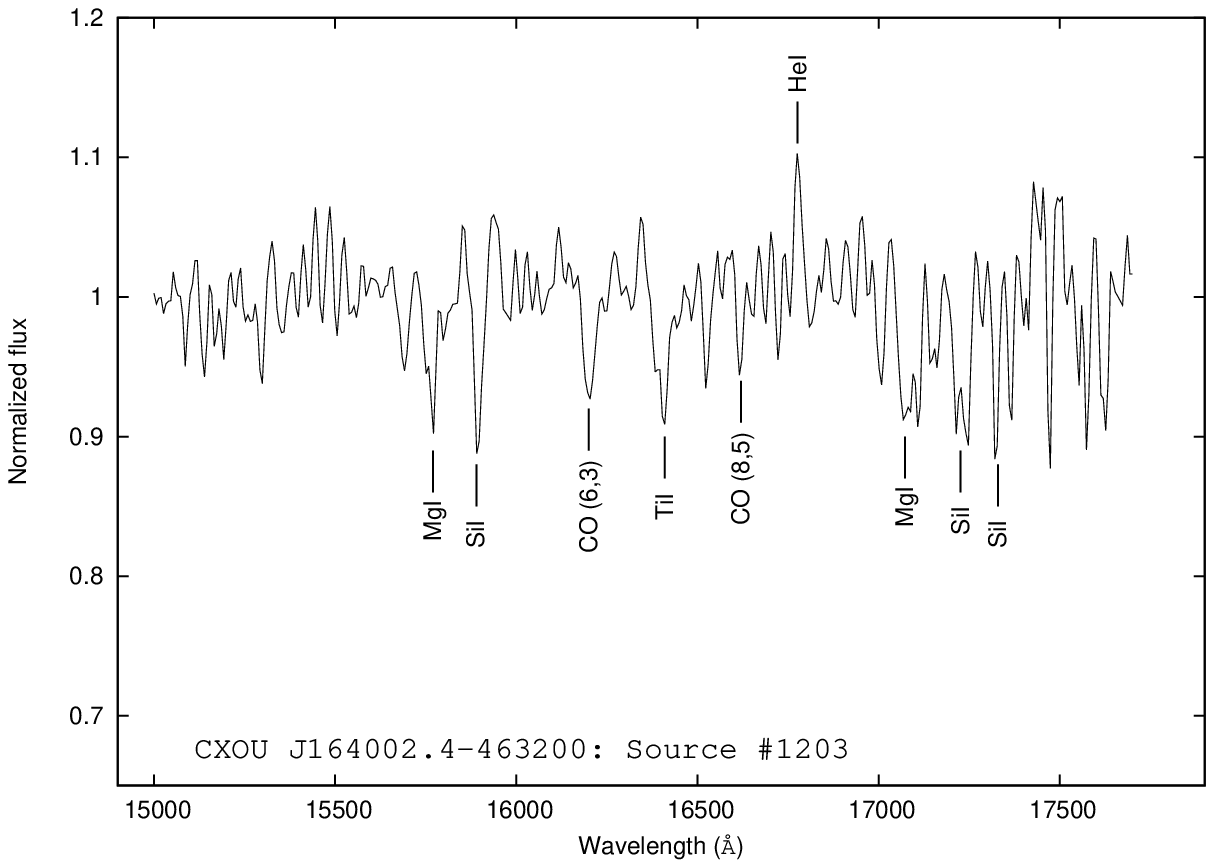}&\includegraphics[width=8.5cm]{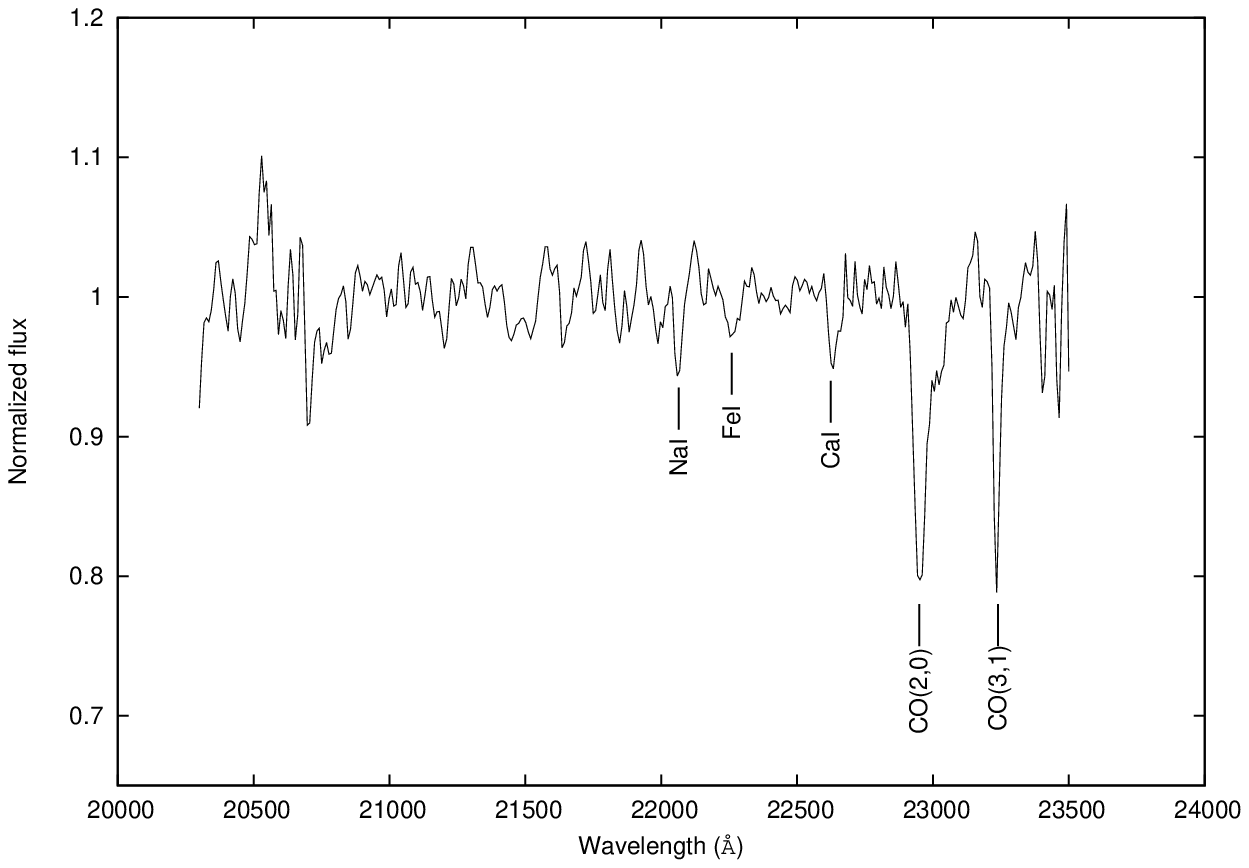}\\
\includegraphics[width=8.5cm]{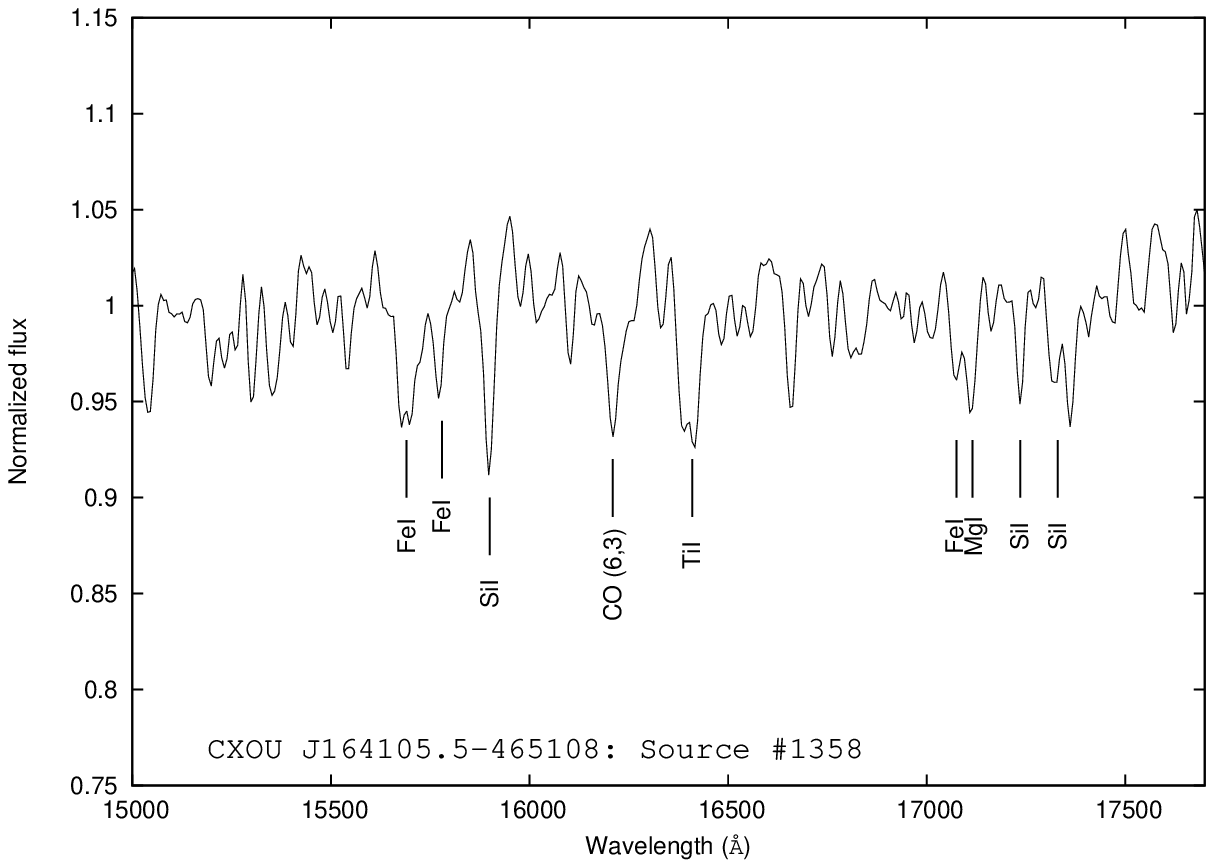}&\includegraphics[width=8.5cm]{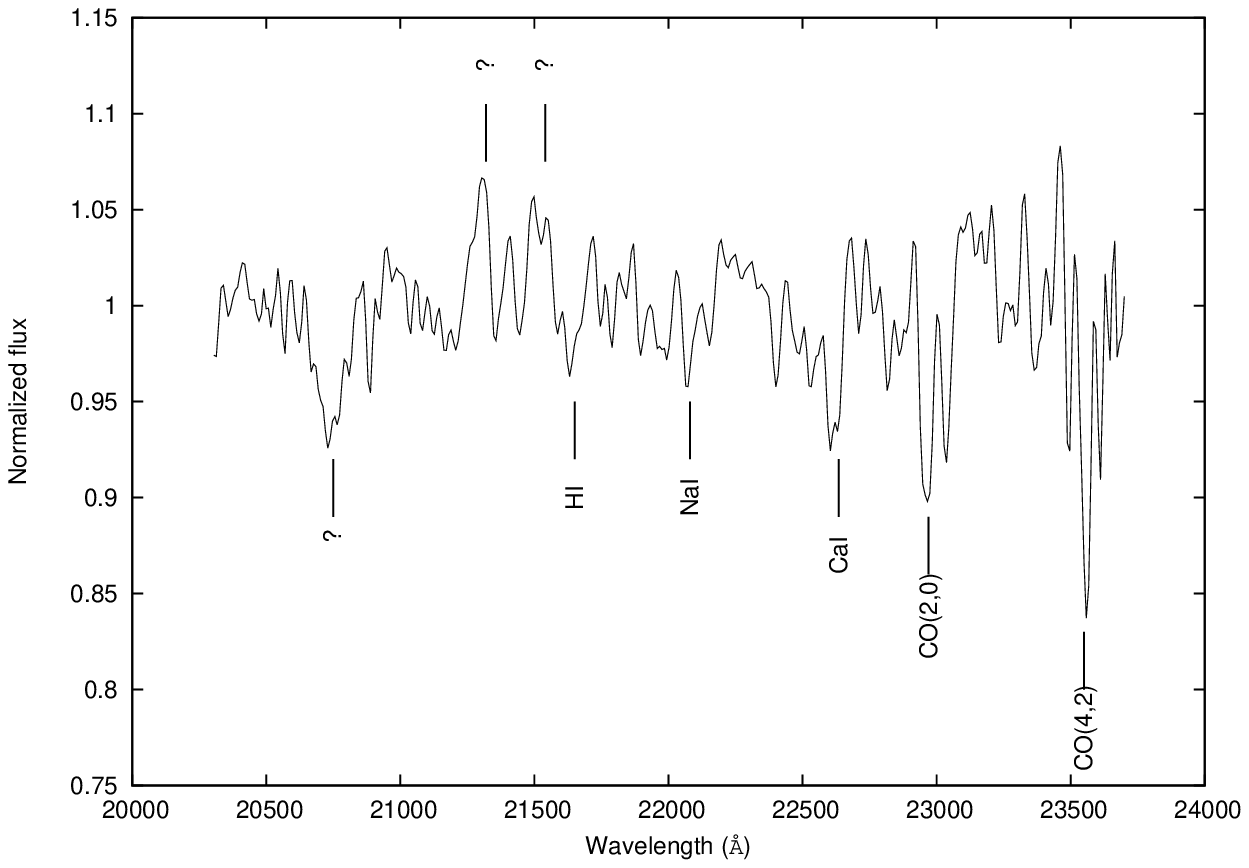}\\
\includegraphics[width=8.5cm]{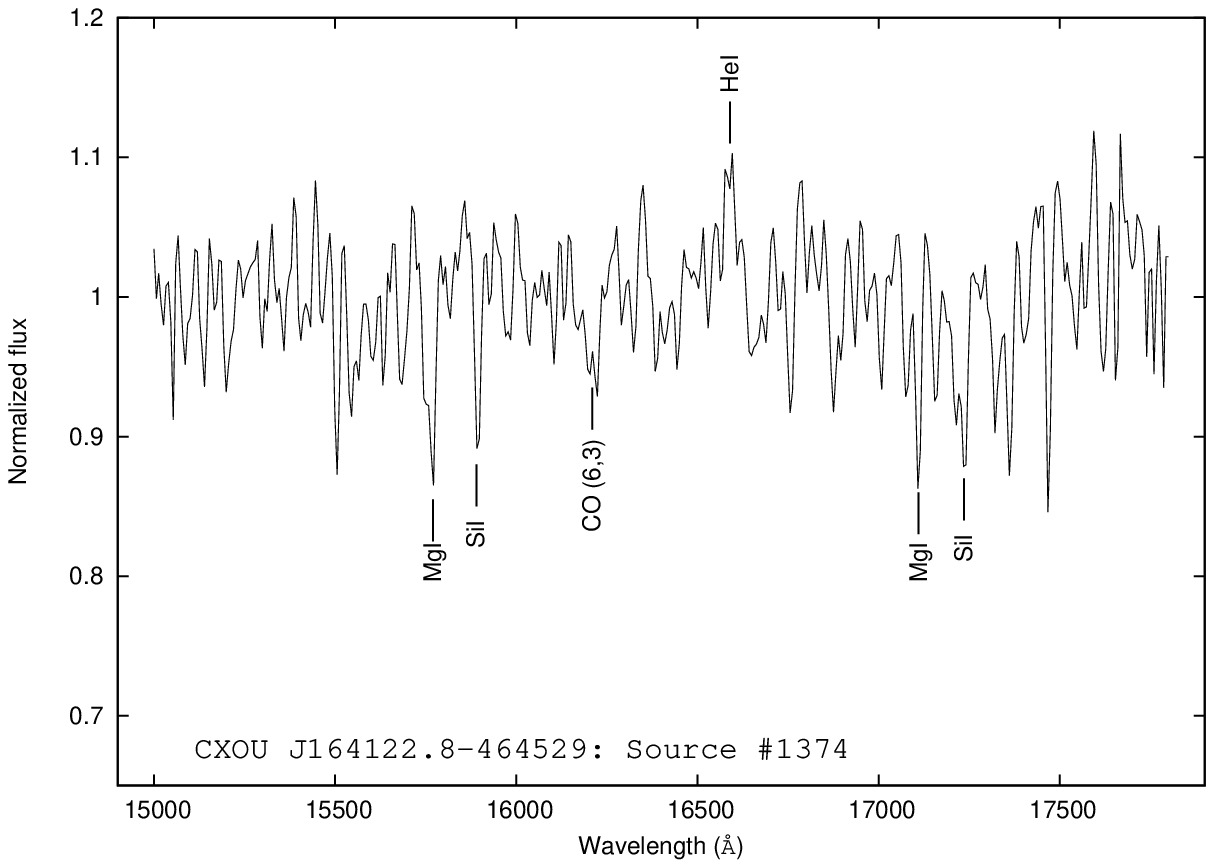}&\includegraphics[width=8.5cm]{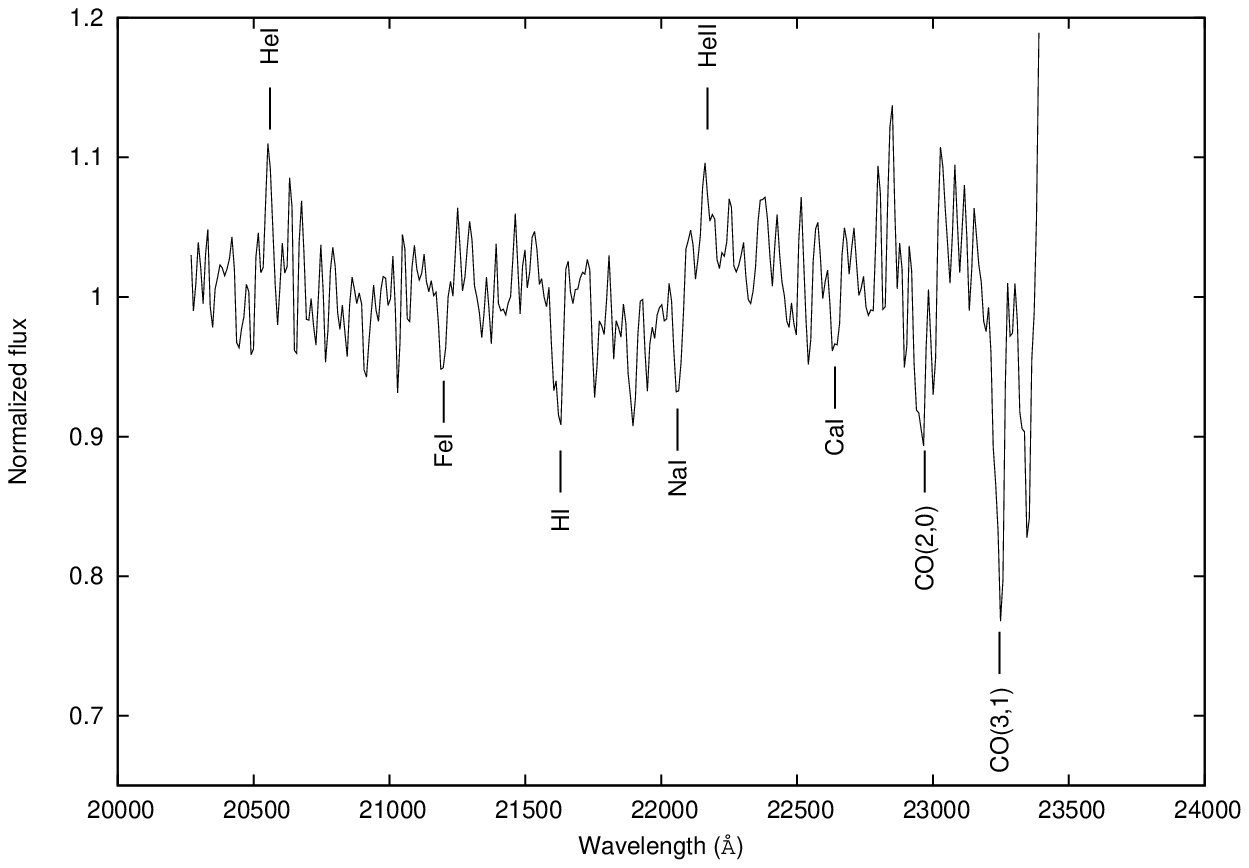}\\

\end{tabular}
\caption{Near-IR \textit{H}-band (left) and \textit{K}-band (right)
spectra of Source 1203 (top), Source 1358 (middle), and Source
1374 (bottom).}
\label{irsp4}
\end{figure*}

%% file: Figures/sp_ir5.figures
\begin{figure*}
\begin{tabular}{cc}
\includegraphics[width=8.5cm]{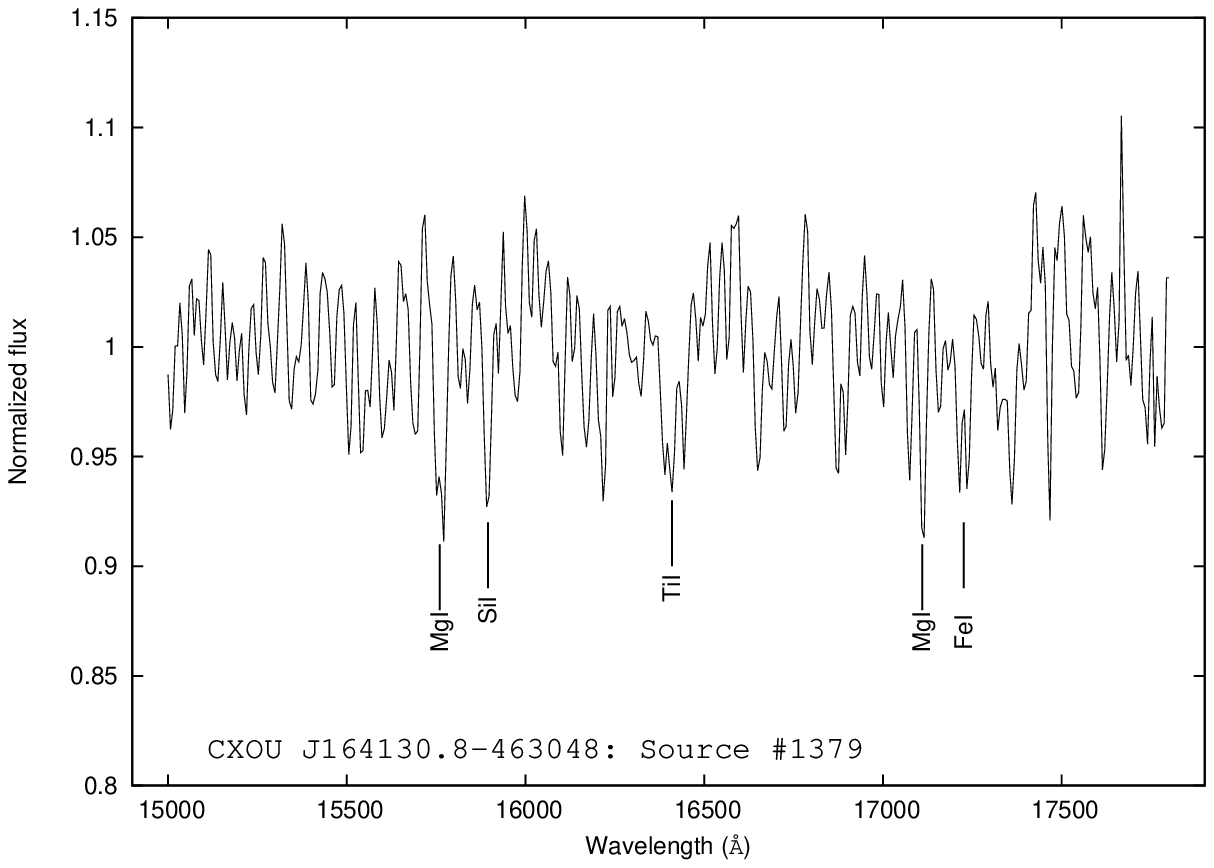}&\includegraphics[width=8.5cm]{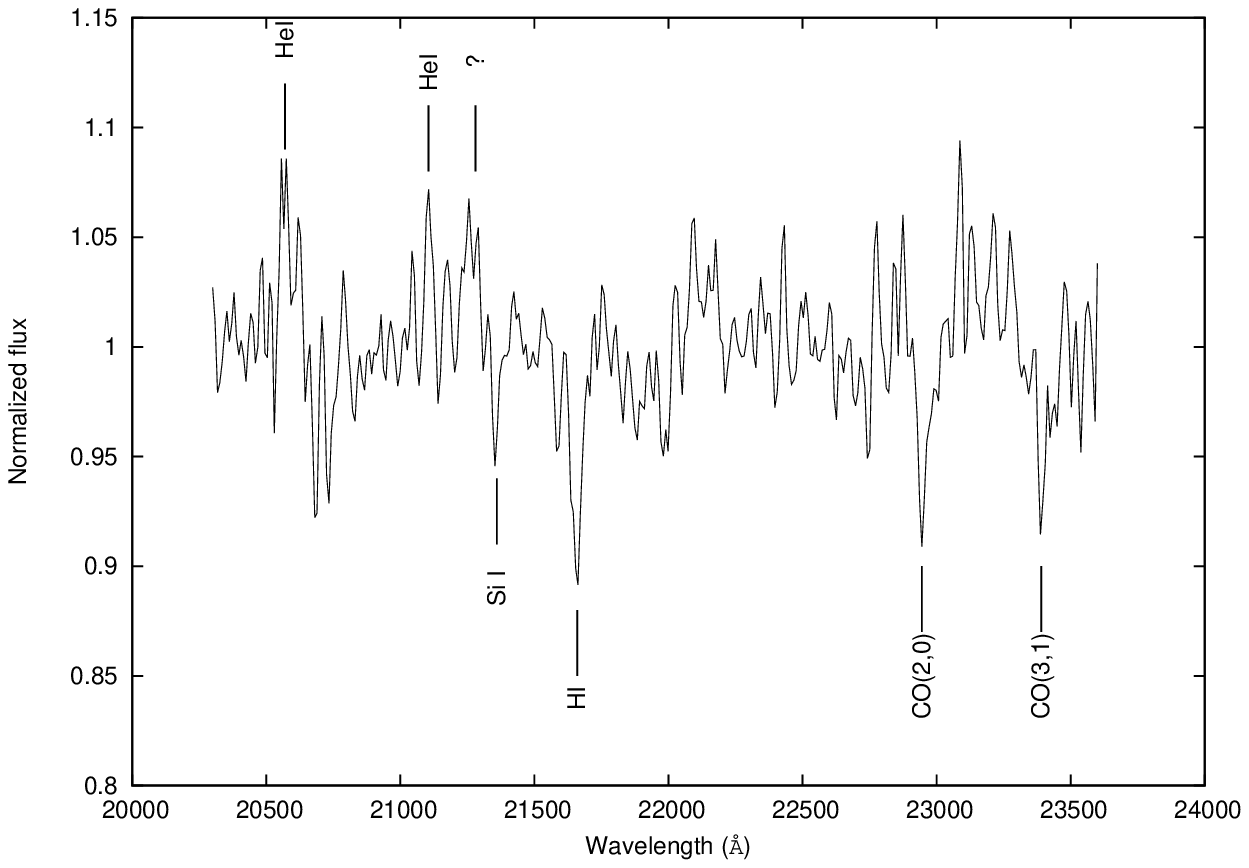}\\
\includegraphics[width=8.5cm]{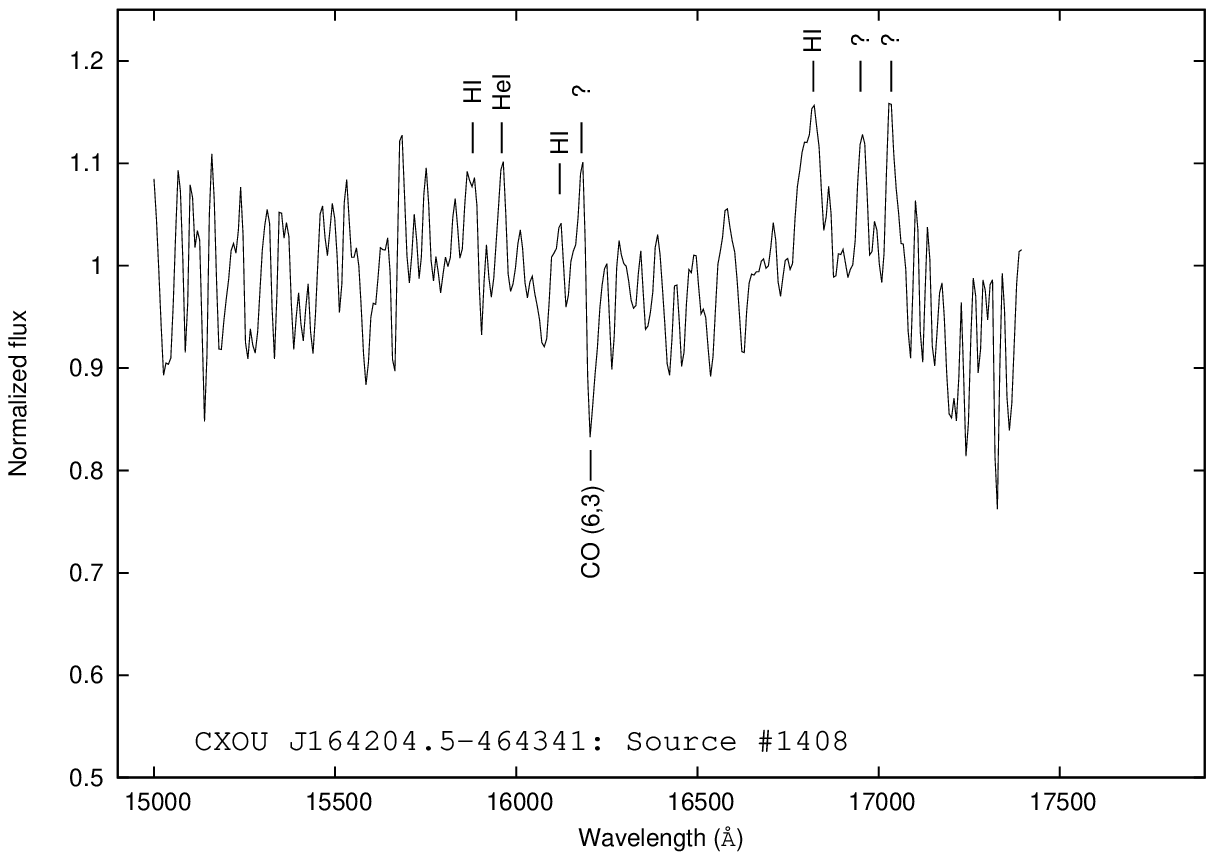}&\includegraphics[width=8.5cm]{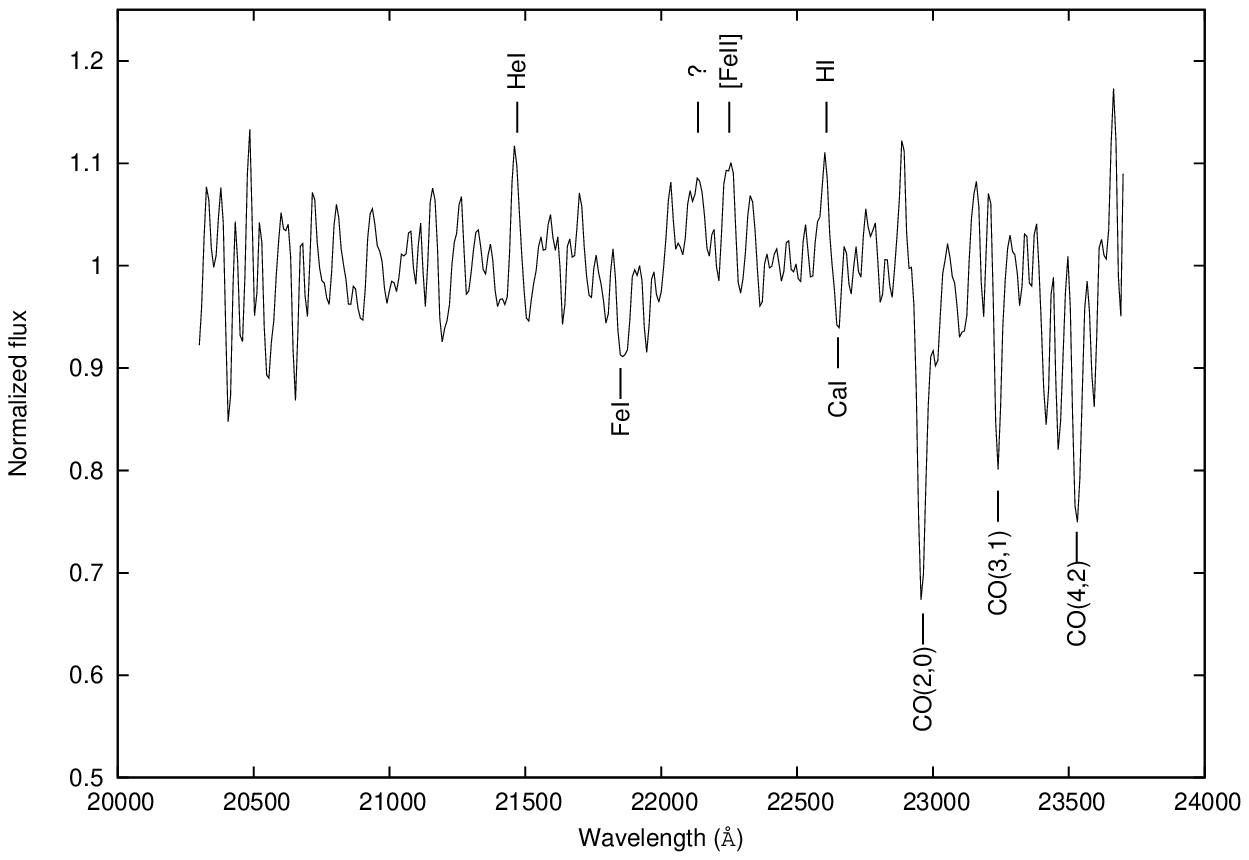}\\

\end{tabular}
\caption{Near-IR \textit{H}-band (left) and \textit{K}-band (right)
spectra of Source 1379 (top) and Source 1408 (bottom).}
\label{irsp5}
\end{figure*}

%% file: Tables/class_lm.table
\begin{table}[!ht]
\caption{\scriptsize Tentative spectral classification of the low
mass stars in our sample, based on the relative
strength of CO (6,3)~16198~\AA, Br{\small $\gamma$}, and CO~(2,0)~22957~\AA.}
\label{class}
\resizebox{0.71\textwidth}{!}{\begin{minipage}{\textwidth}
\begin{tabular}{lll}
\hline
\hline
No.&Relative strength&Spectral classification\\
\hline
38	&  Strong CO (6,3); No Br{\small $\gamma$}; Strong CO (2,0) & Early MIII\\
78	&  Weak CO (6,3); Weak Br{\small $\gamma$}; Strong CO (2,0) & Late G III\\
158	&  Weak CO (6,3); Strong Br{\small $\gamma$}; Strong CO (2,0) & Early/mid GI-III\\
396	&  Medium CO (6,3); No Br{\small $\gamma$}; Strong CO (2,0) & Mid/Late KI-III\\
437	&  Weak CO (6,3); Strong Br{\small $\gamma$}; Medium CO (2,0) & Early GI-III\\
694	&  Medium CO (6,3); Strong Br{\small $\gamma$}; Medium CO (2,0) & Early KI-III\\
881	&  Strong CO (6,3); Weak Br{\small $\gamma$}; Strong CO (2,0) & Late K/Early MI-III\\
1059	&  No CO (6,3); No Br{\small $\gamma$}; Medium CO (2,0) & Early MV\\
1128	&  No CO (6,3); Strong Br{\small $\gamma$}; Medium CO (2,0) & GV\\
1203	&  Medium CO (6,3); No Br{\small $\gamma$}; Strong CO (2,0) & Mid/Late KI-III\\
1358	&  Weak CO (6,3); Weak Br{\small $\gamma$}; Medium CO (2,0) & Early KV \\
1374	&  Weak CO (6,3); Medium Br{\small $\gamma$}; Weak CO (2,0) & Late G/Early KV\\
1379	&  No CO (6,3); Strong Br{\small $\gamma$}; Weak CO (2,0) & Mid GIII \\
1408	&  Strong CO (6,3); No Br{\small $\gamma$}; Strong CO (2,0) & KIII\\
\hline
\end{tabular}
\end{minipage}}
\end{table}